%% file: eq_eff_shared_effort_game.tex
\documentclass[preprint,12pt]{elsarticle}

\input{macros}
\usepackage{xcolor}

\journal{Discrete Applied Mathematics}

\begin{document}

\begin{frontmatter}

\title{When Effort May Fail: Equilibria of Shared Effort with a Threshold\tnoteref{label1}}
\tnotetext[label1]{A very preliminary version of this paper appears in~\protect\cite{PolevoyTrajanovskideWeerdt14} as a conference publication.
We recently got back to actively work on it and have obtained additional results that are incorporated in this manuscript.}

\author[label3]{Gleb Polevoy\fnref{label2}}
\ead{gpolevoy@mail.uni-paderborn.de}
\ead[url]{https://www.hni.uni-paderborn.de/alg/mitarbeiter/158866889309402/}
\fntext[label2]{The research was started when Gleb Polevoy and Stojan Trajanovski were with Delft University of Technology.}
\affiliation[label3]{organization={Paderborn University},
					city={Paderborn},
					postcode={33102}, 
					state={North Rhine-Westphalia},
					country={Germany}}

\author[label4]{Stojan Trajanovski\fnref{label2}}
\ead{sttrajan@microsoft.com}
\ead[url]{https://tstojan.github.io/}
\affiliation[label4]{organization={Microsoft},
					addressline={2 Kingdom Street, Paddington}, 
					city={London},
					postcode={W2 6BD}, 
					country={United Kingdom}}

\author[label5]{Mathijs de Weerdt\corref{cor2}}
\ead{M.M.deWeerdt@tudelft.nl}
\cortext[cor2]{The corresponding author: M.M.deWeerdt@tudelft.nl.}
\affiliation[label5]{organization={Delft University of Technology},
					addressline={Van Mourik Broekmanweg 6}, 
					city={Delft},
					postcode={2628XE}, 
					country={The Netherlands}}


\input{abstractNE}



\begin{keyword}
Shared effort games \sep Contribution threshold \sep Equilibrium \sep Price of anarchy \sep Price of stability 
\sep fictitious play \sep narrowing down/broadening (a solution concept) \sep cyclically strong equilibrium
\end{keyword}

\end{frontmatter}



\input{IntroductionNE}

\input{Model}
\input{NEandCharacterizeExistPotential}
\input{cyc_sne}
\input{mix_ne}
\input{RelatedWork}

\input{ConclusionNE}

\bibliographystyle{elsarticle-harv}
\bibliography{library}


\newpage
\appendix

\input{OmitProof}
\input{Simulation}

\end{document}

%% file: macros.tex
\usepackage[ruled]{algorithm2e} 

\SetAlFnt{\small}
\SetAlCapFnt{\small}
\SetAlCapNameFnt{\small}
\SetAlCapHSkip{0pt}
\IncMargin{-\parindent}

\usepackage{url}

\usepackage{latexsym,amsmath,amssymb}
\usepackage{amsthm}
\usepackage{graphicx}

\usepackage{ifthen}
\usepackage{mdwlist,paralist}

\usepackage{comment}
\excludecomment{longversion}

\usepackage{epsfig}
\usepackage{epsf}

\usepackage[geometry]{ifsym}
\usepackage{rotating}

\usepackage{float}
\usepackage[caption=false]{subfig} 

\usepackage{tikz}
\usetikzlibrary{shapes,snakes}
\usetikzlibrary{arrows.meta}

\tikzstyle{process} = [rectangle, minimum width=1cm, minimum height=1cm, text centered, draw=black, fill=gray!30]
\tikzstyle{star_process} = [star, minimum width=1.4cm, minimum height=1.4cm, text centered, draw=black, fill=gray!30]
\tikzstyle{line} = [latex, very thick,-,>=stealth]
\tikzstyle{arrow} = [thick,-latex]

\newboolean{Longversion} \setboolean{Longversion}{false}
\newcommand{\version}{Not_AAMAS_02_2014} 

\newcommand*{\eqns}[1]{Eq.~#1}
\newcommand*{\eqnsref}[1]{\eqns{\eqref{#1}}} 
\newcommand*{\forms}[1]{Formula(s)~#1}
\newcommand*{\formsref}[1]{\forms{\eqref{#1}}} 
\newcommand*{\chapt}[1]{Chapter~#1}
\newcommand*{\sectn}[1]{Section~#1}
\newcommand*{\sectnref}[1]{\sectn{\ref{#1}}} 
\newcommand*{\fig}[1]{Figure~#1}
\newcommand*{\figref}[1]{\fig{\ref{#1}}} 

\newcommand{\set}[1]{\left\{ #1 \right\}} 
 
\newcommand{\paren}[1]{\left( #1 \right)} 

\newcommand{\inv}[1]{\frac{1}{#1}} 
\newcommand{\abs}[1]{\left| #1 \right|} 
\newcommand{\ceil}[1]{\left\lceil {#1} \right\rceil}

\newcommand{\naturals}{\mathbb{N}} 

\newcommand{\reals}{\mathbb{R}}
\newcommand*{\realsP}{{\ensuremath{\mathbb{R}_{+}}}}

\newcommand*{\harder}[1]{#1-hard}

\newcommand*{\NP}{\textrm{NP}}

\newcommand*{\NPH}{\mbox{\textrm{\harder{\NP}}}}

\DeclareMathOperator{\cross}{\text{\rm \Cross}} 

\newcommand*{\defas}{\ensuremath{\stackrel{\rm \Delta}{=}}}

\theoremstyle{definition}

\newtheorem{theorem}{Theorem}
\newtheorem{defin}{Definition}
\newtheorem{remark}{Remark}
\newtheorem{lemma}{Lemma}
\newtheorem{proposition}{Proposition}

\newtheorem{corollary}{Corollary}

\newtheorem{example}{Example}

\newcommand{\defined}[1]{\emph{#1}}

\newcommand{\NE}{\text{NE}\xspace} 

\newcommand{\FP}{\ensuremath{\text{FP}}} 
\newcommand{\ISFP}{\ensuremath{\text{ISFP}}} 

\newcommand{\thetaEqTheta}[1]{\ensuremath{M_{\text{eq}}^{#1}}} 
\newcommand{\thetaEq}{\thetaEqTheta{\theta}} 
\newcommand{\thetaProp}{\ensuremath{M_{\text{prop}}^\theta}} 

\DeclareMathOperator{\poa}{PoA} 
\DeclareMathOperator{\pos}{PoS} 

\DeclareMathOperator{\sw}{SW} 
\DeclareMathOperator{\BR}{BR} 
\DeclareMathOperator{\br}{br} 

\renewenvironment{proof}[0]{{\it Proof. }}{\qed} 

%% file: abstractNE.tex
\begin{abstract}
People, robots, and companies mostly divide time and effort among projects, and
\defined{shared effort games} model people investing resources in public endeavors and sharing the generated values.
In linear $\theta$ sharing (effort) games, a project's value is linear in the total contribution, 
thus modelling predictable, uniform, and scalable activities.
The threshold $\theta$ for effort defines which contributors win and receive their share,
equal share modelling standard salaries, equity-minded projects, etc. 
Thresholds between 0 and 1 model games such as paper co-authorship and shared assignments,
where a minimum positive contribution is required for sharing in the value.
We constructively characterise the conditions for the existence of a pure equilibrium for $\theta\in\{0,1\}$, and for two-player games with a general threshold,%
\ifthenelse{\equal{\version}{AAMAS_02_2014}}{
with
close budgets and
}{
}%
and find the prices of anarchy and stability.
We also provide existence and efficiency results for more than two players, and use generalised fictitious play simulations to show when a pure equilibrium exists and what its efficiency is.
We propose a  method for studying solution concepts by
refining a solution concept 
and finding a large natural subclass of games where
the refinement coincides with the original solution concept (Nash, in this case).
This means that the original concept 
narrows down to a more demanding concept on certain games, providing 
new insights for comparing both concepts.
We also prove mixed equilibria always exist and bound
their efficiency.
%
\end{abstract}



%% file: IntroductionNE.tex
\section{Introduction}\label{Sec:introduct}
Many real-world situations include a group of players investing resources across multiple projects. 
Examples of such situations include financial investments, contributions to online communities~\cite{HarperLiChenKonstan07}, Wikipedia~\cite{ForteBruckman05}, political campaigns~\cite{Siegel2009}, paper co-authorship~\cite{KleinbergOren2011}, social exchange networks~\cite{KleinbergTardos2008}.
Naturally, also hobbies and attention spread fall under this category.
In the formal analyses of these type of games, it is often assumed that the obtained revenue from such projects is linear in the total contribution and is shared equally~\cite{Roberson06,KleinbergOren2011,BachrachSyrgkanisVojnovic12}. Basically, performing standard predictable activities, which are scalable and homogeneous,
often yields linear value (and we also tend to linear thinking~\cite{deLanghePuntoniLarrick2017}). Equal division takes place with standard salaries, fundraising for community projects, equal distribution of recognition to all the contributors, equity-minded projects (the same reward regardless the contribution), non-rivalrous values, like public parks, which can be enjoyed equally by all those who have access to them, etc.
However, in most of the above examples, revenue is only shared among those who contribute at least a certain amount of \emph{effort}. 
In this work we model this as a threshold on a player's contribution. 
In particular, we analyse the situation when the considered threshold is relative to the investments of the other players.
The abstract model of this paper is the first theoretical study of such shared effort games with a \defined{threshold}, assuming linear project values. This provides the foundations for modelling contributions in various kinds of projects, basically parallel games, such as considering splitting time among projects, each of which is a reciprocal interaction~\cite{PolevoydeWeerdt2017b}.

An example with a relative threshold of $1$ is an all pay-auction. Another example is the Colonel Blotto game (see e.g.~\cite{Roberson06}), where there are several battlefields over which each of the $2$ players distributes her soldiers, and the winner at each battlefield is the one who has put more people there. A player's payoff is the number of projects (battlefields) where she has won, the value of each project being fixed. Thus, when our model is reduced to $2$ players with a fixed value of each project, we obtain the Colonel Blotto competition, whereas we typically assume a project has a value that increases as a function of the total contribution, modelling constructive collaboration, rather than countenance.
These examples are ``highly thresholded" because only the player whose effort per project  is maximum collects the complete revenue.  
%
%
%
%
%

We now present examples of roughly linear growing project values,
each being equally shared among all those who contribute at least a certain threshold.
\begin{enumerate}
  \item	Consider a start-up where developers are building a new tool to sell
    and divide the revenue equally, or any firm paying equal salaries to everyone who works enough to keep the job. Here, the resulting value is assumed to grow approximately linearly in the contributed time, when the idea being implemented is clear and no surprises are expected. 

	\item Assigning points for an exercise, where a percentage of the perfect work is required to 
	obtain the (equal) homework’s credits~\cite{grad_policy_2016} is an absolute threshold example
	from a course at the University of Maryland.
	Linearity is a reasonable model for predictable tasks. Although our threshold is relative, 
	while in the Maryland example it is absolute, still, professors often relate the individual grades 
	to the average level of the students.
	
	\item	As a recreational example, think of kids selling lottery tickets for their sports club where a significant part of the income is used to improve the club’s facilities; kids that sell at least a certain number (threshold) of tickets, relatively to the others, are equally rewarded by the club with the same symbolic present (e.g., club memorabilia, like a hat, a scarf, or a jersey).
\end{enumerate}

\begin{figure}
\centering
\begin{tikzpicture}[node distance=1cm and 0.8cm]

\node (text0) [xshift=3cm] {{\LARGE Projects}};
\node (text1) [xshift=6cm] {{\Large $P(x) = 4 x$}};
\node (text2) [xshift=12cm] {{\Large $P(x) = 2 x$}};

\node (pro0) [below of=text0] {};
\node (pro1) [process,below of=text1] {};
\node (pro2) [process,below of=text2] {};
\node (text3) [below of=pro0, yshift=-2.5cm] {{\LARGE Authors}};
\node (auth1) [star_process,below of=pro1, yshift=-2.5cm] {};
\node (auth2) [star_process,below of=pro2, yshift=-2.5cm] {};

\draw [arrow] (auth1) -- (pro1) node [pos=.5, above, sloped] (TextNode1) {{\Large \textbf{$4$} hours}};
\draw [arrow] (auth1) -- (pro2) node [pos=.25, above, sloped] (TextNode2) {{\Large \textbf{$1$} hour}};
\draw [arrow] (auth2) -- (pro1) node [pos=.3, above, sloped] (TextNode3) {{\Large \textbf{$10$} hours}};
\draw [arrow] (auth2) -- (pro2) node [pos=.27, above, sloped] (TextNode4) {{\Large \textbf{$10$} hours}};

\draw [blue,-latex, very thick] (pro1) to [bend left=-45] node [pos=.2, above, sloped]  (TextNode5) {{\Large $28$}} (auth1);
\draw [blue,-latex, very thick] (pro1) to [bend left=70] node [pos=.2, above, sloped]  (TextNode5) {{\Large $28$}} (auth2);
\draw [blue,-latex, very thick] (pro2) to [bend left=65] node [pos=.2, above, sloped]  (TextNode5) {{\Large $22$}} (auth2);

\end{tikzpicture}
\caption{The co-authors invest what is shown in the arrows that go up,
every project's value is defined as the $P$ function of the total
contribution, and it is equally shared among the contributors who
contribute above the relative threshold of $0.2$. The obtained shares are
denoted by the arrows that go down.
}%
\label{fig:paper_coauthor_2}%
\end{figure}
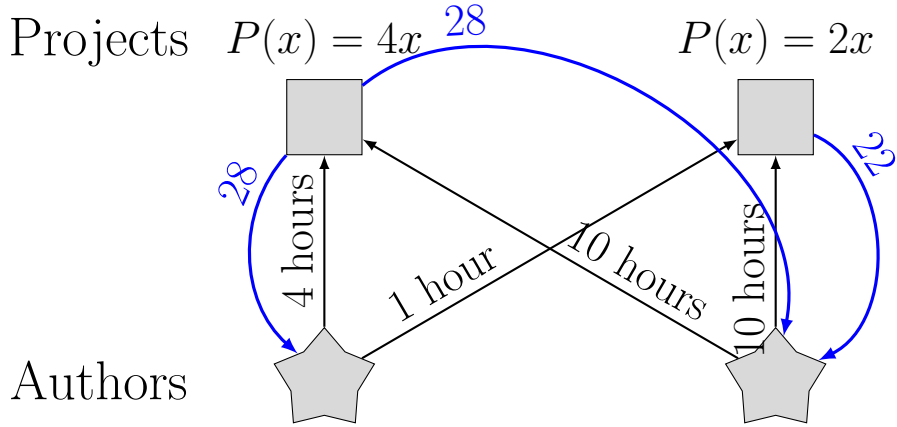

The following example with a smaller threshold is used later to further illustrate the model.
\newcounter{ex_collab_scientists}
\setcounter{ex_collab_scientists}{\value{example}}
\begin{example}\label{ex:collab_scientists}
Consider two collaborating scientists in a narrow research field. They can
work on their papers alone or together. When they collaborate on a paper
(a project), an
author has to contribute at least $0.2$ of the work of the other one, in order
to be considered a co-author. The value of a paper, being the recognition,
is equally divided among the authors.
Author $1$ has the time budget of $5$ hours to work, and author $2$ has
$20$ hours.
The value of the first paper is $4$ times the total
time the authors put in, while the less ``hot'' second paper rewards
the contributors with a value of only twice the contributed time.
This is illustrated
in \figref{fig:paper_coauthor_2}. In the figure, the first paper
receives the total contribution of $4 + 10 = 14$, creating the value
of $4 \cdot 14 = 56$. Both contributors are authors, since
$4 \geq 0.2 \cdot 10$, and the value is equally divided between them.
The second paper receives $1 + 10 = 11$, and yields the value of
$2 \cdot 11 = 22$. Here, only the contributor of $10$ is an author,
since $1 < 0.2 \cdot 10$, and he, thus, receives the whole value of $22$.
This is not a Nash equilibrium, since the second contributor would benefit
from moving the $1$ hour contribution to the first paper, increasing her
share from the first paper by 2.
On the other hand, if both authors invest all their time in paper $1$,
the situation is stable. Indeed, moving a part to paper $2$ would benefit
nobody, since the paper is twice less valuable than paper $1$, so
sharing the value of paper $1$ is as good as contributing alone
to paper $2$. The social welfare in this equilibrium is maximum possible,
since everyone contributes to the most valuable project.
In general, 
we would like to find stable contributions, and whether
they will be efficient for both authors, relatively to the maximum
possible divisions of the authors' time budgets.
\end{example}

It is important to understand and predict 
the stable contribution strategies, and suggest\footnote{Further influence methods are discussed in \sectnref{Sec:conclus_further}.} the efficient ones in situations where
players invest resources in several projects and share the generated values. 
The classical stability model is
the Nash equilibria, and the efficiency model of stable situations
is the ratio of the least total utility of the players in an equilibrium to the optimum, called the price of anarchy ($\poa$)~\cite{KoutsoupiasPapadimitriou2009},
and the ratio of the largest total utility in equilibrium to the optimal total utility, called the price of stability ($\pos$)~\cite{SchulzMoses2003,AnshelevichDasGuptaKleinbergTardosWexlerRoughgarden04}. If the price of anarchy is close to $1$, then all equilibria are good, and we may follow any equilibrium profile.
If the price of anarchy is low, while the price of stability is high, then we have to regulate the play by suggesting the efficient equilibria,
while if even the price of stability is low, the only way to make the play socially efficient is changing the game through, for example, subsidising~\cite{MondererTennenholtz2004} or subsidies and fines~\cite{PolevoyDziubinski2022}.

This price of anarchy for shared effort games was bounded in~\cite{BachrachSyrgkanisVojnovic12}, but assuming 
that a player obtains at least $1/k$th of her marginal contribution,
which fails to hold in a positively thresholded model.
Threshold was introduced in~\cite{PolevoyTrajanovskideWeerdt14}, and 
pure equilibria for two players with close budgets and projects with
linear utility functions were characterised.
There is no analysis of the existence of Nash Equilibria (\NE)
and their efficiency in \emph{general} linear shared effort games and
nothing about mixed Nash equilibria is known. 
%
This paper fills this gap with the following contributions.%
\footnote{Relatively to \protect\cite{PolevoyTrajanovskideWeerdt14}, we extend the theory
also for budgets not within a threshold factor from each other, 
characterise the non-suppressed equilibria for threshold $1$, 
provide the novel concept of \defined{narrowing down} Nash to
the newly defined \defined{cyclically strong equilibrium},
prove the existence of an \NE{} in the mixed case, answering this natural question, 
and substantially extend the simulations.} 

\paragraph{Our Contributions}
\begin{description}
	\item[For $2$ players,]	\label{do:pure:exist_2}
	a complete constructive characterization of the existence of pure \NE{}
	in Theorem~\ref{the:NE_charac_2players},
	and
	exact values of the price of anarchy and stability.
	\item[For more players,] \label{do:pure:0_1_suff_sim}
	analysing the cases of $0$ threshold
	and $1$ threshold (Theorem~\ref{the:one_threshold_char}),
	sufficiency results on existence of pure \NE{}
	and efficiency bounds,
	and exploring this domain through simulation of fictitious play to find pure \NE{} and their efficiency in 2-project multi-player games.
	The simulations need to be engineered based on the case in question, so 
	the importance of the simulations is mainly in the proof of concept of implementing 
	the infinite strategy fictitious play for 2 projects.
	This involves designing
	an $O(n \log n)$ best response algorithm for $2$-project multi-player games
	and proving that for more projects, best-responding is \NPH, even for the same project functions and $2$ players.
	
	\item[For any number of players,]	
	we introduce the concept of narrowing down (or broadening) a solution concept.
	In our case, we strengthen Nash equilibrium by adding a requirement of
	no profitable cyclical deviations of whole budgets, alluding to the
	famous top trading cycles mechanisms~\cite{AbdulkadirogluSonmez1999}.
	We then demonstrate that when no budget can make another one fall below
	the threshold, both solution concepts coincide.
	
	\item[For any number of players,]	\label{do:mix}
	a proof of existence of mixed \NE{} (it is an infinite game, so Nash's Theorem~\cite{Nash51} does not apply)
	and extending efficiency bounds to mixed equilibria.
\end{description}
%


%

%
After defining
shared effort games in the next section,
we first concentrate on pure strategies. 
We theoretically treat the existence and efficiency of \NE{}
for thresholds $0$ and $1$ and then
for games with two players in Section~\ref{Sec:char_two}.
Afterwards, we study games with any number of players.
In \sectnref{Sec:cyc_sne}, we define a new solution concept and narrow down
Nash equilibria to that concept on a natural subclass of games, whetting the appetite for defining further 
solution concepts and studying their relationships by analogy.
Finally, we treat the mixed extension in Section~\ref{Sec:mix_ne}, proving the
existence of a mixed \NE{} and showing some efficiency bounds
of the pure case generalise to the mixed case.
We conclude and discuss the future work in
\sectnref{Sec:conclus_further}.
When a proof is mostly technical, we provide its sketch and intuition in the body of the paper and
defer the complete proof to
\ref{Sec:omit_proof}. 
We conduct further investigation of existence and efficiency of pure \NE{} for more than $2$ players by employing fictitious play
simulations in~\ref{Sec:Sim}.
Concretely, we define an Infinite-Strategy
Fictitious Play and simulate it by computing a best response, till and if it finds an equilibrium within some time. We
check whether we have found an \NE{} and if that is the case,
what its efficiency is.

We introduce and study a threshold on the basic model, motivating more refinements of the model.
This is a major impact, alongside the mothodological implications of the notion of narrowing down 
of equilibria from \sectnref{Sec:cyc_sne}, which motivates defining other appropriate 
solution concepts and seeing when a narrowing down or a broadening occurs.

%% file: Model.tex
\section{Model}\label{Sec:model}

To model investing effort in shared projects,
we define \defined{shared effort games}, extending the model
by~\cite{BachrachSyrgkanisVojnovic12} with thresholds.
The games consist of players who contribute to projects,
and share the values the projects generate. Henceforth, 
we use the general term ``value'' instead of the monetary
``revenue''. Formally,
players
$N = \set{1, \ldots, n}$ contribute to projects~$\Omega$. 
The contribution of player~$i$ to project~$\omega \in \Omega$ is
denoted by~$x^i_\omega \in \realsP$.
Each player~$i$ has a budget~$B_i > 0$, and the strategy
space of player~$i$ ({i.e., the set of her possible actions) is%
\footnote{We denote ``defined as'' by $\defas$.}
$S^i \defas \set{x^i = (x^i_\omega)_{\omega \in \Omega} \in \reals_{+}^{\abs{\Omega}} \mid \sum_{\omega \in \Omega}{x^i_\omega} \leq B_i}$.
Assume w.l.o.g.~(without loss of generality) that $B_n \geq \ldots \geq B_2 \geq B_1 > 0$.
Being an $\abs{\Omega} - 1$-dimensional simplex, this set is
\emph{compact} and \emph{convex}.
Denote the vector of all the contributions by
$x = (x^i_\omega)^{i\in N}_{\omega \in \Omega}$
and the strategies of all the players except $i$ by $x^{-i}$.

To define the utilities, each
project~$\omega \in \Omega$ is associated with its (really or perceivably) linear (allows
concentrating on cumulative scalable cooperation
across multiple ``projects'', like manufacturing, technical writing, 
Wikipedia or environmental activism)
\defined{project function}, which determines its \defined{value}, based
on the total contribution $x_\omega \defas \sum_{i \in N}{(x_\omega^i)}$
that the project receives; formally,
$P_\omega(x_\omega) \colon \realsP \to \realsP$,
and $P_j(y) = \alpha_j y$. W.l.o.g.,
$\alpha_m \geq \alpha_{m - 1} \geq \ldots \geq \alpha_1 > 0$.
We denote the number of
projects with the largest coefficient of project functions by $k \in \naturals$, i.e.~%
$\alpha_m = \alpha_{m - 1} = \ldots = \alpha_{m - k + 1} > \alpha_{m - k} \geq \alpha_{m - k - 1} \geq \ldots \geq \alpha_1$.
We call those $k$ projects \defined{steep}.
The project's value is distributed among the players according to the
following rule. From each project~$\omega \in \Omega$, each player~$i$ that
contributes enough gets an equal
share, denoted $\phi_\omega^i(x_\omega) \colon \realsP \to \realsP$.
%
Such sharing is relevant to many applications where a minimum contribution
is required to share the value, such as paper co-authorship and homeworks,
and we study predominantly such games.
For any $\theta \in [0, 1]$, define the players who get a share as
those who bid at least a $\theta$ fraction of the maximum bid size
to~$\omega$,
\begin{equation*}
N^{\theta}_\omega \defas \set{i \in N_\omega
\mid x^i_\omega \geq \theta \cdot \max_{j\in N_\omega}{x^j_\omega}}.
\end{equation*}
The $\theta$-equal sharing mechanism equally divides
the project's value among all the
users who contribute at least $\theta$ of the maximum bid to the project,
as we now formally define.

\begin{defin}
The \defined{$\theta$-equal sharing mechanism},
denoted by~$\thetaEq$, is
\begin{equation*}
\phi^i_\omega(x_\omega) \defas
	\begin{cases}
	\frac{P_\omega(x_\omega)}{\abs{N^{\theta}_\omega}} & \mbox{if } i \in N^{\theta}_\omega,\\
	0 & \mbox{otherwise}.\\
	\end{cases}
\end{equation*}
\end{defin}

The utility of a player~$i \in N$ is defined to be
\begin{equation*}
u^i(x) \defas \sum_{\omega \in \Omega}{\phi^i_\omega(x_\omega)}.
\end{equation*}
The social welfare is defined as the total utility, i.e.~%
$\sw(x) \defas \sum_{i=1}^n {u^i(x)}$.

We finally define the games formally.
\begin{defin}[Shared-effort game with $\theta$-equal sharing]
A \defined{shared-effort game with $\theta$-equal sharing} is a tuple $\Gamma = (N, (B_i)_{i \in N}, \Omega, (P_\omega)_{\omega \in \Omega}, \theta, M^\theta_{eq})$, where $N = \{1, \ldots, n\}$ is a set of players, $B_i > 0$ is the resource budget of player $i$, $\Omega$ is a set of projects, $P_\omega(x_\omega) = \alpha_\omega \cdot x_\omega$ is the linear project function of project $\omega$, $\theta \in [0,1]$ is the contribution threshold, and $M^\theta_{eq}$ is the $\theta$-equal sharing mechanism (Definition~1). The strategy space of player $i$ is the simplex $S_i = \{x^i \in \mathbb{R}^{|\Omega|}_+ \mid \sum_\omega x^i_\omega \leq B_i\}$, and the utility is $u_i(x) = \sum_{\omega \in \Omega} \varphi^i_\omega(x_\omega)$.
\end{defin}

We assume linear project functions and $\theta$-equal sharing
unless explicitly said otherwise.

We do not model costs of resources, allowing to model
splitting resources that cannot be used otherwise, like 
in Blotto games; applications include splitting externally 
provided funds, human resources, or military equipment.
Our model can also represent investing effort that has been 
determined to be invested, thus being practically a sunk cost 
(e.g., employers at work or soldiers).
The cost being sunk indeed has repercussions on the structure 
of the equilibria, such as the existence of suppressed players 
in equilibria, and the similarity to all-pay auction is 
incomplete, because unlike all-pay auctions or public good 
provision models, where players incur a direct  cost of exerting 
effort, our framework does not include a negative payoff 
component for effort.

Reconsider the example from \sectnref{Sec:introduct} to illustrate the above model.
\newcounter{tmp}
\setcounter{tmp}{\value{example}}
\setcounter{example}{\value{ex_collab_scientists}}
\begin{example}[Continued]
The scientists $N$ invest in
the papers (projects) $\Omega$.
Assume that a paper's total value for the reputation of its
authors is proportional to the total investment in the paper. That is,
the project's functions $P_\omega$ are linear.
In order to be considered an author, a minimum threshold $\theta$ of the maximum
contribution is required, and
a paper's total contribution to the authors' reputation is
equally divided among all its authors.
This is a shared effort game with a threshold $\theta \in (0, 1)$ and
equal sharing.
\end{example}
\setcounter{example}{\value{tmp}}

%% file: NEandCharacterizeExistPotential.tex
\section{Pure Nash Equilibrium}\label{Sec:NECharachPotential}

\input{NashEquilibria}

%% file: NashEquilibria.tex
We begin with general results on equilibria in shared effort games.
Then, we completely characterise the existence of Nash equilibria
for two players, 
following with existence results for any number of players in \sectnref{Sec:NECharachPotential:exist}.
Finally, we provide the prices of anarchy and stability in \sectnref{Sec:NECharachPotential:eff}.
%
%
%
%
We study pure \NE{} in this section, which are natural to play, since a pure
shared effort game already has uncountably infinite strategy spaces and non-continuous
utilities.

Without a threshold ($\theta = 0$), or when $\theta = 1$, things would be simple,
as we now describe.
\begin{theorem}\label{the:zer_threshold_pot}
A game with $\thetaEqTheta{0}$ admits a potential function,
 a pure \NE{} exists and $\poa = \pos = 1$.
\end{theorem}
\begin{proof}
The strategy space of player~$i$ is $S^i$, and denote
$S \defas S^1 \times \ldots \times S^n$. Define~$P \colon S \to \reals$ by
$P(x) \defas \sum_{\omega\in \Omega}{\frac{P_\omega(x_\omega)}{\abs{N_\omega}}}$.
This is a potential function, because it is equal to the utility of any
player, and therefore, when player~$i$ changes her strategy, her utility
changes exactly as the potential does. 

The game possesses a pure \NE, whenever the potential function
attains its maximum. As the linear project functions are continuous and the
spaces are non-empty compact, being simplexes, so the potential function always attains its maximum (see
Lemma~$4.3$ in~\cite{MondererShapley1996PG}).

Moreover, a profile is an \NE{}
if and only if all the players arbitrarily divide their budgets among the most valuable
projects.
Therefore, being an \NE{} is equivalent to maximising the
social welfare, implying $\poa = \pos = 1$.
\end{proof}

To characterise existence and efficiency of \NE{} in other cases, we need some definitions.
Given a strategy profile, we call a project that receives no contribution
a \defined{vacant} project.
We define players that do not obtain a share from a given project as
\defined{dominated} at that project.
We call them \defined{suppressed}
if they also contribute to that project. Formally,
\begin{defin}\label{def:domin_suppress}
The \defined{dominated} players at a project~$\omega$ are
$D_\omega \defas N_\omega \setminus N^{\theta}_\omega$, and
the \defined{suppressed} players at a project~$\omega$ are
$S_\omega \defas \set{i \in N_\omega : x_\omega^i > 0} \setminus N^{\theta}_\omega$.
If a player is dominated/suppressed at all the projects where he contributes,
we simply call him \defined{dominated/suppressed}.
\end{defin}

In an \NE{}, a player is suppressed at a project if and only if it
is suppressed at all the projects where it contributes. This holds since
if a player is suppressed at project $p$ but it also contributes to
project $q \neq p$ and is not suppressed there, then it would like to move
its contribution from $p$ to project $q$.

Consider distinct projects functions 
$\alpha_{m = m_1} > \alpha_{m_2} > \alpha_{m_3} > \ldots > \alpha_{1 = m_l}$, 
denoting $m_1 = m$ and $m_l = 1$, where $l \geq 1$ is the number of distinct
project functions. We choose $m_j$ as an arbitrary $j'$ such that 
$\alpha_{j'} = \alpha_{m_j}$.
For each $j \in \set{1, 2, \ldots, l}$, we call every project~$\omega$ 
with $\alpha_\omega = \alpha_{m_j}$ a \defined{$j$-level project}.
In the same spirit, consider distinct budgets 
$B_{n = n_1} > B_{n_2} > B_{n_3} > \ldots > B_{1 = n_p}$,
denoting $n_1 = n$ and $n_p = 1$, $p \geq 1$ being the number of distinct
budgets. We choose $n_i$ as an arbitrary $i'$ such that 
$\alpha_{i'} = \alpha_{n_i}$.
For each $q \in \set{1, 2, \ldots, p}$, call every player~$i$ 
with $B_i = B_{n_q}$ a \defined{$q$-level player}, 
and call $B_i$ the \defined{$q$-level budget}.
Thus, given a level, one can talk about projects and budgets on that level.

For each level $j \in \set{1, 2, \ldots, l}$, 
let $r_j$ be the number of projects
of level $j$, namely $r_j \defas \abs{\set{\omega \in \Omega : \alpha_\omega = \alpha_{m_j}}}$.
Similarly, for each level $j \in \set{1, 2, \ldots, p}$, let $s_j$ be the number of players
of level $j$, namely $s_j \defas \abs{\set{i \in N : B_i = B_{n_i}}}$.
We are now ready to characterise all the equilibria without suppressed players
for threshold $\theta = 1$.
\begin{theorem}\label{the:one_threshold_char}
A game with $\thetaEqTheta{1}$ possesses \NE{} where nobody is suppressed
if and only if all the following conditions hold:
\begin{enumerate}\label{theta_1_cond}
	\item \label{theta_1_cond:num}
	The number of distinct project functions~$l$ is at least 
	the number of distinct budgets~$p$, namely $l \geq p$.
	
	\item	\label{theta_1_cond:enough_budget}
	For each level $j \in \set{1, 2, \ldots, p - 1}$,
	there either exists a single $j$-level budget such that 
	$B_{n_j} > r_j B_{n_{j + 1}}$, or there exist at least $r_j$ budgets of level $j$.
	
	\item	\label{theta_1_cond:enough_alpha}
	For each level $j \in \set{1, 2, \ldots, p - 1}$ and level $d \in \set{j + 1, j + 2, \ldots, p}$, 
	$\alpha_{m_j} \geq (1 + \ceil{s_d / r_d}) \alpha_{m_d}$.
\end{enumerate}
\end{theorem}
\begin{proof}
In one direction, we assume the existence of $x \in \NE$ where no player is 
suppressed and prove that all the conditions above hold.
The key observation is that since exactly the maximum contributors equally 
share the project's value, contributing to a project yields the same value, 
regardless how many others contribute there too, if nobody is suppressed.
Therefore, in an equilibrium without suppression, everyone contributes 
only to the most profitable project, where that is not suppressed.
Thus, all the $1$-level budgets go to the $1$-level projects. The $2$-level
budgets would be suppressed if they went where a $1$-level budget is,
violating the assumption of not being suppressed, and otherwise such
a suppression would be a profitable deviation for any $1$-level player,
violating the equilibrium assumption, so they all go to the $2$-level 
projects. Inductively, $j$-level projects contribute all to $j$-level 
projects.
%
Therefore, $l \geq p$, yielding condition~\ref{theta_1_cond:num}.

Next, consider any level $j \in \set{1, 2, \ldots, p - 1}$.
If there exists just one $j$-level budget $B_{n_j}$, then the only
way to render the deviation of any $j + 1$-level to a $j$-level project
non-profitable is to have more than $B_{n_{j+ 1}}$ contributed to 
every $j$-level project, implying $B_{n_j} > r_j B_{n_{j + 1}}$.
On the other hand, if there are multiple $j$-level budgets, then
no splitting of these budgets is possible, to avoid a suppression from
another $j$-level budget being a profitable deviation. Thus, the only
way to render the deviation of any $j + 1$-level to a $j$-level project
non-profitable is to have a full $j$-level budget contribute to 
every $j$-level project. This implies 
condition~\ref{theta_1_cond:enough_budget}.

Finally, consider again any level $j \in \set{1, 2, \ldots, p - 1}$
and any greater level $d \in \set{j + 1, j + 2, \ldots, p}$.
At the $d$ level, at least one project obtains at least
$\ceil{s_d / r_d}$ contributions from $d$-level players. Therefore,
making the suppression of the total $d$-level contribution to 
a $d$-level project by a $j$-level player being non-profitable requires
$\alpha_{m_j} x \geq \alpha_{m_d} (x + \ceil{s_d / r_d} B_{n_d}), \forall x \in (B_{n_d}, B_{n_j})$,
namely $\alpha_{m_j} \geq (1 + \ceil{s_d / r_d}) \alpha_{m_d}$,
implying condition~\ref{theta_1_cond:enough_alpha}.

Conversely, we now assume all the conditions above 
and demonstrate an equilibrium profile without suppressed players.
For each level $j \in \set{1, \ldots, p}$, let all the players of
level $j$ contribute all their budgets to the projects of level $j$.
There exist enough project levels because of 
condition~\ref{theta_1_cond:num}.
If there exists a single such player, she should equally split her
budget among all the $j$-level projects. If multiple such players exist, 
they should not split their respective budgets, and every $j$-level 
project should receive at least one such player's contribution, which 
is possible, because of condition~\ref{theta_1_cond:enough_budget}.
No project of level $d$ should have more than $\ceil{s_d / r_d}$
players contribute there. Intuitively, this means balancedness.

We now prove that the defined profile is an equilibrium without
suppression. Suppression does not exist, since a player
contributes to the same project only together with equally
sized contributions. 

As for the non-existence of profitable
deviations, consider the possible deviations. Contributing
to a more profitable project (a smaller level) would result 
in suppression, because every project of level $j$ receives
a contribution greater than $B_{n_{j + 1}}$.
Contributing to another project on the same level would either
result in suppression if contributing less than the full budget
with a full budget contribution, or yield the same utility. 
Last, contributing to a least profitable (higher level) project 
is non-profitable because of condition~\ref{theta_1_cond:enough_alpha}, 
and since no project of level $d$ has more than $\ceil{s_d / r_d}$
players contribute there. This implies the profile is an \NE.
\end{proof}

Next, we use the structural insights from the previous theorem to
analyse the efficiency of equilibria for $\theta = 1$.
\begin{proposition}\label{prop:one_threshold_eff}
If a game with $\thetaEqTheta{1}$ possesses \NE{} where nobody is suppressed,
then the social welfare of any such \NE{} is
$\sum_{q \leq p}{\alpha_{m_q} (\sum_{i : B_i = B_{n_q}}{B_i})}$, whereas the social optimal is
$\alpha_{m_1} \sum_{i \in N}{B_i}$.
\end{proposition}
\begin{proof}
As we showed in the proof of Theorem~\ref{the:one_threshold_char},
in any such equilibrium, all the players of level $j$ contribute 
everything to the projects of the level $j$. 
This implies the statement.
\end{proof}

\ifthenelse{\equal{\version}{AAMAS_02_2014}}{
\subsection{Characterization for Two Players with Close Budgets}
}{
}%
\label{Sec:char_two}

\subsection{Existence of Nash Equilibrium}\label{Sec:NECharachPotential:exist}

All Nash equilibria are invariant to multiplication.
\begin{proposition}\label{prop:inv_mult}
Consider a game with project functions
$P_j (x) = \alpha_j \cdot x$. If profile $x = \paren{x^1, \ldots, x^n}$
is an \NE, then the following hold:
\begin{description}
	\item[Multiplying projects:]	$x$ is also an equilibrium in the game
	obtained by multiplying all the project coefficients by some positive
	$p$.
	\item[Multiplying budgets and profile:]	$p \cdot x$ is an \NE{} in the
	game obtained by multiplying all the budgets by some positive $p$.
\end{description}
\end{proposition}
The proof is deferred to the appendix.

Having dealt with $\theta = 0$ and $\theta = 1$, we may assume $\theta \in (0, 1)$.
We first assume $2$ players, i.e., $n = 2$, and completely characterise
this case.
We introduce Lemmas~\ref{lemma:NEchar}, \ref{lemma:NEchar:close_B},
and \ref{lemma:NEchar:far_B}, before 
characterizing the existence of \NE. These lemmas describe what must hold in any \NE.
Their proofs appear in the appendix.

\begin{lemma}\label{lemma:NEchar}
Consider
an equal $\theta$-sharing game with two players with 
$0 < \theta < 1$. 
%
Then the following hold in any \NE.
\begin{enumerate}
	\item	\label{lemma:NEchar:list:at_least_1_steep}
	At least one player contributes to a steep project.
	\item	\label{lemma:NEchar:list:non_suppr_non_steep}
	Suppose that a non-suppressed player, contributing to a steep
	project, contributes to a non-steep project as well.
	Then, it
	contributes either alone or precisely the least amount it should
	contribute to achieve a portion in the project's value.
\end{enumerate}\label{lemma:NEchar:list}
\end{lemma}

The following lemma treats budgets that are close to each other.
\begin{lemma}\label{lemma:NEchar:close_B}
Consider
an equal $\theta$-sharing game with two players with 
$0 < \theta < 1$. 
%
If $B_1 \geq \theta B_2$,
the following hold in any \NE.
\begin{enumerate}
	\item Each player contributes to every steep project.
	
	\item	A non-steep project receives the contribution of at most
	one player.
\end{enumerate}
\end{lemma}

We need another definition.
\begin{defin}
A \defined{$2$-steep} project is a maximally valuable non-steep project.\footnote{Several such projects may exist, thus ``a''.}
\end{defin}

The following lemma treats budgets that are far from each other.
\begin{lemma}\label{lemma:NEchar:far_B}
Consider
an equal $\theta$-sharing game with two players with 
$0 < \theta < 1$. 
%
If $B_1 < \theta B_2$, 
then the following hold in any \NE{} where no player is suppressed.
\begin{enumerate}
	\item \label{lemma:NEchar:far_B:list:contr_1}
	Player $1$ contributes only to non-steep projects.
	
	\item	\label{lemma:NEchar:far_B:list:each_pos_util}
	Each player receives a (strictly) positive utility, unless all projects are the same.%
	\footnote{That is, unless $k = m$.}
	
	\item	\label{lemma:NEchar:far_B:list:2_all_steep_and_thresh_others}
	Player $2$ contributes alone to every steep project,
	and perhaps to a non-steep project together with $i$, the
	threshold amount.
		
	\item	\label{lemma:NEchar:far_B:list:2_steep}
	The projects that are non-steep and also non-$2$-steep receive zero contribution.
	
	\item	\label{lemma:NEchar:far_B:list:B_ratio}
	If player $2$ contributes to a $2$-steep project,
	then there exists only a single $2$-steep project.
	
\end{enumerate}\label{lemma:NEchar:far_B:list}
\end{lemma}

We are finally ready to characterise the existence of an \NE{}
for two players. The characterisation depends on the threshold,
the ratio between the budgets and the highest coefficients.
\begin{theorem}\label{the:NE_charac_2players}
Consider an equal $\theta$-sharing game with two players with budgets $B_1, B_2$. W.l.o.g.,
$B_2 \geq B_1$. Assume $0 < \theta < 1$, and project functions $P_j(x) = \alpha_j \cdot x$ with coefficients
$\alpha_m = \alpha_{m - 1} = \ldots = \alpha_{m - k + 1} > \alpha_{m - k} \geq \alpha_{m - k - 1} \geq \ldots \geq \alpha_1$
(ordered w.l.o.g.).
\ifthenelse{\equal{\version}{AAMAS_02_2014}}{
For $B_1 \geq \theta B_2$, this game
has a pure strategy \NE{} if and only if
the following both hold.%
\footnote{If $\alpha_{m - k}$ does not exist,
consider the containing condition to be vacuously true.}
\begin{enumerate}
	\item $\frac{1}{2}\alpha_{m} \geq \alpha_{m - k}$,\label{NE_charac_2players:conds:close_B:steep}
	\item	$B_1 \geq k\theta B_2$;\label{NE_charac_2players:conds:close_B:not_steep}
\end{enumerate}
}{
A pure \NE{} exists if and only if
one of the following holds.%
\footnote{If $\alpha_{m - k - 1}$ (and) or $\alpha_{m - k}$ does not exist,
consider the containing condition to be vacuously true.}

\begin{enumerate}
\item	$B_1 \geq \theta B_2$ and the following both hold.
\begin{enumerate}
	\item $\frac{1}{2}\alpha_{m} \geq \alpha_{m - k}$,\label{NE_charac_2players:conds:close_B:steep}
	\item	$B_1 \geq k\theta B_2$;\label{NE_charac_2players:conds:close_B:not_steep}
\end{enumerate}\label{NE_charac_2players:conds:close_B}

\item $B_1 < \theta B_2$ and also at least one of the following holds.
\begin{enumerate}
	\item \label{NE_charac_2players:conds:far_B:no_suppress:no_inter}
	$B_1 < \frac{\theta B_2}{k}$ and $\frac{\alpha_{m - k}}{\alpha_m} \leq \min\set{\frac{1}{1 + \theta}, \frac{2\theta}{1 + \theta}}$,
	
	\item	\label{NE_charac_2players:conds:far_B:no_suppress:inter}
	$B_1 < \frac{\theta B_2}{k + \theta^2}$ and $\alpha_{m - k} \geq 2 \alpha_{m - k - 1}$ and $\frac{2\theta}{1 + \theta} \leq \frac{\alpha_{m - k}}{\alpha_m} \leq \frac{2(1 - \theta)}{2 - \theta}$ and $m - k$ is the only $2$-steep project,
	
	\item	\label{NE_charac_2players:conds:far_B:suppress}
	$B_1 < \frac{\theta}{\abs{\Omega}} B_2$ and all the
	project functions are equal, i.e.~$\alpha_m = \alpha_1$.
\end{enumerate}\label{NE_charac_2players:conds:far_B}
\end{enumerate}\label{NE_charac_2players:conds}
}%
\end{theorem}
The idea of the proof is as follows. To show existence of an equilibrium
under the assumptions of the theorem,
we provide a strategy profile and prove that no unilateral deviation is
profitable. We show the other direction by assuming that a given profile
is an \NE{} and deriving the asserted conditions, employing
Lemmas~\ref{lemma:NEchar}, \ref{lemma:NEchar:close_B} and~\ref{lemma:NEchar:far_B}
that describe what holds in an equilibrium.

The full proof appears in~\ref{Sec:omit_proof}.

Consider some structural insights from the proof of 
Theorem~\ref{the:NE_charac_2players}. When the ratio between the
budgets is small, namely $B_1 \geq \theta B_2$, both players can be 
in equilibrium by equally investing in every steep project.
When the budgets are further apart, the level-$2$ player has
to invest in a level-$2$ project, while the level-$1$ player divides
her budget equally among the steep projects. This fits the case of
$\theta = 1$, analysed in Theorem~\ref{the:one_threshold_char}, where
the players of each level spread their budgets approximately evenly
among the projects of the corresponding level.
If the budgets are even further apart and the level-$2$ projects are 
a bit more profitable than in the former case, then the level-$1$ player 
can obtain her part of the level-$2$ project, while spreading the rest 
of her budget evenly among the steep projects. 
Finally, if the level-$1$ budget allows complete domination, then 
level-$1$ player can spread her budget evenly, thereby guaranteeing 
an equilibrium.
Thus, there is a tendency to spread budgets in an even manner, 
though the players with larger budgets often ``push'' the other players 
from the more profitable projects. Budgets serve as keys to the more 
profitable projects.

We conclude that besides the equilibria with $\alpha_m = \alpha_{m - k}$,
there exists an \NE{} if and only if $\alpha_{m - k}$ is at most a constant
fraction of $\alpha_m$.
\begin{corollary}\label{cor:NE_exist_proj_rat_smaller_2players}
Assume the conditions of Theorem~\ref{the:NE_charac_2players}
and that there exist two projects.
Then, once all the parameters besides ${\alpha_{m- k}}$ and ${\alpha_m}$
are set and all the conditions involving the budgets are fulfilled,
there exists a $C > 0$, such that
an \NE{} exists if and only if $\frac{\alpha_{m- k}}{\alpha_m} \leq C$,
and, perhaps, if and only if $\frac{\alpha_{m- k}}{\alpha_m} = 1$.
\end{corollary}
\begin{proof}
First, for $2$ projects, the condition involving $\alpha_{m - k - 1}$
in \ref{NE_charac_2players:conds:far_B:no_suppress:inter} is vacuously true.

Consider the bounds on the possible values of
$\frac{\alpha_{m- k}}{\alpha_m}$ such that at least one \NE{} exists,
besides case~\ref{NE_charac_2players:conds:far_B:suppress}.
From Theorem~\ref{the:NE_charac_2players},
the only way that this corollary could be wrong would require
the upper bound on $\frac{\alpha_{m- k}}{\alpha_m}$
from~\ref{NE_charac_2players:conds:far_B:no_suppress:no_inter}
to be strictly smaller than the lower bound
from~\ref{NE_charac_2players:conds:far_B:no_suppress:inter},
while the two bounds on $\frac{\alpha_{m- k}}{\alpha_m}$
from~\ref{NE_charac_2players:conds:far_B:no_suppress:inter}
gave a non-empty segment.
These conditions mean that both
$\frac{1}{1 + \theta} < \frac{2\theta}{1 + \theta}$
and $\frac{2\theta}{1 + \theta} \leq \frac{2(1 - \theta)}{2 - \theta}$
should hold.
The first inequality means $\theta > 0.5$, while the second one
means $\theta \leq 0.5$. Since these conditions cannot hold
simultaneously, the corollary is never wrong.
\end{proof}

The sufficiency conditions~\ref{NE_charac_2players:conds:close_B}
and~\ref{NE_charac_2players:conds:far_B:suppress}
extend for a general $n$ as follows.

\begin{theorem}\label{the:NE_charac_nplayers}
Consider an equal $\theta$-sharing game with $n \geq 2$ players with budgets
$B_n \geq \ldots \geq B_2 \geq B_1$ (the order is w.l.o.g.), $0 < \theta < 1$, and project functions $P_j(x) = \alpha_j \cdot x$ with coefficients
$\alpha_m = \alpha_{m - 1} = \ldots = \alpha_{m - k + 1} > \alpha_{m - k} \geq \alpha_{m - k - 1} \geq \ldots \geq \alpha_1$
(the order is w.l.o.g.).

This game
has a pure strategy \NE{} if%
\ifthenelse{\equal{\version}{AAMAS_02_2014}}{
$B_{n - 1} \geq \theta B_n$ and the following both hold.%
\footnote{If $\alpha_{m - k}$ does not exist,
consider the containing condition to be vacuously true.}
\begin{enumerate}
	\item $\frac{1}{n}\alpha_{m} \geq \alpha_{m - k}$,\label{NE_charac_nplayers:conds:close_B:steep}
	\item	$B_1 \geq k\theta B_n$;\label{NE_charac_nplayers:conds:close_B:not_steep}
\end{enumerate}
}
{
one of the following holds.%
\footnote{If $\alpha_{m - k}$ does not exist,
consider the containing condition to be vacuously true.}

\begin{enumerate}
\item	$B_{n - 1} \geq \theta B_n$ and the following both hold.
\begin{enumerate}
	\item $\frac{1}{n}\alpha_{m} \geq \alpha_{m - k}$,\label{NE_charac_nplayers:conds:close_B:steep}
	\item	$B_1 \geq k\theta B_n$;\label{NE_charac_nplayers:conds:close_B:not_steep}
\end{enumerate}\label{NE_charac_nplayers:conds:close_B}

\item $B_{n - 1} < \theta B_n$ and also the following holds.
\begin{enumerate}	
	\item	\label{NE_charac_nplayers:conds:far_B:suppress}
	$B_{n - 1} < \frac{\theta}{\abs{\Omega}} B_n$ and all the
	project functions are equal, i.e.~$\alpha_m = \alpha_1$.
\end{enumerate}\label{NE_charac_nplayers:conds:far_B}
\end{enumerate}\label{NE_charac_nplayers:conds}
}%
\end{theorem}
\begin{proof}%
\ifthenelse{\equal{\version}{AAMAS_02_2014}}{
It is analogous to the proof for $n = 2$.
All the players equally
divide their budgets among all the steep projects.
}{
It is analogous to the proof for $n = 2$, noticing the following.
In case~\ref{NE_charac_nplayers:conds:close_B}, everyone equally
divides her budget among all the steep projects.
In case~\ref{NE_charac_nplayers:conds:far_B}, player $n$ dominates
everyone else.
}
\end{proof}

\subsection{Efficiency}\label{Sec:NECharachPotential:eff}

In order to facilitate decisions\footnote{Influence methods are discussed in \sectnref{Sec:conclus_further}.}, it is important to analyse
the efficiency of the various Nash Equilibria.
We aim to find the price of anarchy ($\poa$), which is the ratio of a
worst \NE's efficiency to the optimum possible one, and the price
of stability ($\pos$), which is the ratio of a
best \NE's efficiency to the optimum possible one.
We first completely resolve the two-player case.
\begin{theorem}\label{the:NE_effic_2players}
Consider an equal $\theta$-sharing game with two players with budgets $B_1, B_2$. W.l.o.g.,
$B_2 \geq B_1$. Assume $0 < \theta < 1$, and project functions $P_j(x) = \alpha_j \cdot x$ with coefficients
$\alpha_m = \alpha_{m - 1} = \ldots = \alpha_{m - k + 1} > \alpha_{m - k} \geq \alpha_{m - k - 1} \geq \ldots \geq \alpha_1$
(the order is w.l.o.g.).%
\footnote{If $\alpha_{m - k}$ does not exist,
consider the containing condition to be vacuously true.}

\ifthenelse{\equal{\version}{AAMAS_02_2014}}{
Assume that $B_1 \geq \theta B_2$ and the following both hold.
\begin{enumerate}
	\item $\frac{1}{2}\alpha_{m} > \alpha_{m - k}$,\label{NE_effic_2players:conds:close_B:steep}
	\item	$B_1 \geq k\theta B_2$;\label{NE_effic_2players:conds:close_B:not_steep}
\end{enumerate}\label{NE_effic_2players:conds:close_B}

Then, there exists a pure strategy \NE{}
and there holds:
$\poa = \pos = 1$.
}
{
\begin{enumerate}
\item	Assume that $B_1 \geq \theta B_2$ and the following both hold.
\begin{enumerate}
	\item $\frac{1}{2}\alpha_{m} \geq \alpha_{m - k}$,\label{NE_effic_2players:conds:close_B:steep}
	\item	$B_1 \geq k\theta B_2$;\label{NE_effic_2players:conds:close_B:not_steep}
\end{enumerate}\label{NE_effic_2players:conds:close_B}

Then, there exists a pure strategy \NE{}
and there holds:
$\pos = \poa = 1$.

\item Assume that $B_1 < \theta B_2$ and also at least one of the following holds.
\begin{enumerate}
	\item \label{NE_effic_2players:conds:far_B:no_suppress:no_inter}
	$B_1 < \frac{\theta B_2}{k}$ and $\frac{\alpha_{m - k}}{\alpha_m} \leq \min\set{\frac{1}{1 + \theta}, \frac{2\theta}{1 + \theta}}$.
	
	Then, there exists a pure strategy \NE{}
	and the following holds.
	$\pos = \frac{\alpha_m B_2 + \alpha_{m - k} B_1}{\alpha_m (B_1 + B_2)}$.
	If the case~\ref{NE_effic_2players:conds:far_B:no_suppress:inter}
	holds as well, then
	$\poa = \frac{\alpha_m (B_2 - \theta B_1) + \alpha_{m - k} (B_1 (1 + \theta))}{\alpha_m (B_1 + B_2)}$;
	otherwise, $\poa = \pos$.
	
	\item	\label{NE_effic_2players:conds:far_B:no_suppress:inter}
	$B_1 < \frac{\theta B_2}{k + \theta^2}$ and $\alpha_{m - k} \geq 2 \alpha_{m - k - 1}$ and $\frac{2\theta}{1 + \theta} \leq \frac{\alpha_{m - k}}{\alpha_m} \leq \frac{2(1 - \theta)}{2 - \theta}$ and $m - k$ is the only $2$-steep project.
	
	Then, there exists a pure strategy \NE{}
	and the following holds.
	If the case~\ref{NE_effic_2players:conds:far_B:no_suppress:no_inter}
	holds as well, then
	$\pos = \frac{\alpha_m B_2 + \alpha_{m - k} B_1}{\alpha_m (B_1 + B_2)}$;
	otherwise, $\pos = \frac{\alpha_m (B_2 - \theta B_1) + \alpha_{m - k} (B_1 (1 + \theta))}{\alpha_m (B_1 + B_2)}$.
	In any case,
	$\poa = \frac{\alpha_m (B_2 - \theta B_1) + \alpha_{m - k} (B_1 (1 + \theta))}{\alpha_m (B_1 + B_2)}$.
	
	\item	\label{NE_effic_2players:conds:far_B:suppress}
	$B_1 < \frac{\theta}{\abs{\Omega}} B_2$ and all the
	project functions are equal, i.e.~$\alpha_m = \alpha_1$.
	
	Then, there exist pure \NE{}
	and 
	$\pos = 1, \poa = \frac{B_2}{B_1 + B_2}$.
\end{enumerate}\label{NE_effic_2players:conds:far_B}
\end{enumerate}\label{NE_effic_2players:conds}
}%
\end{theorem}

We derive the exact lower bound (infimum) and the maximum of the
price of anarchy and stability.
\begin{corollary}\label{cor:NE_effic_2players_tight_bound}
Consider an equal $\theta$-sharing game with two players with budgets $B_1, B_2$. W.l.o.g.,
$B_2 \geq B_1$. Assume $0 < \theta < 1$, and project functions $P_j(x) = \alpha_j \cdot x$ with coefficients
$\alpha_m = \alpha_{m - 1} = \ldots = \alpha_{m - k + 1} > \alpha_{m - k} \geq \alpha_{m - k - 1} \geq \ldots \geq \alpha_1$
(the order is w.l.o.g.).%
\footnote{If $\alpha_{m - k}$ does not exist,
consider the containing condition to be vacuously true.}
Then, the infimum of $\pos$ over all the cases is
$\frac{k}{k + \theta} (> 0.5)$, and the maximum is $1$.
The same holds for $\poa$.
\end{corollary}

We next prove some efficiency results 
for
a general $n \geq 2$.

\begin{theorem}\label{the:NE_effic_nplayers}
Consider an equal $\theta$-sharing game with $n \geq 2$ players with budgets
$B_n \geq \ldots \geq B_2 \geq B_1$, $0 < \theta < 1$ (the order is w.l.o.g.), and project functions $P_j(x) = \alpha_j \cdot x$ with coefficients
$\alpha_m = \alpha_{m - 1} = \ldots = \alpha_{m - k + 1} > \alpha_{m - k} \geq \alpha_{m - k - 1} \geq \ldots \geq \alpha_1$
(the order is w.l.o.g.).%
\footnote{If $\alpha_{m - k}$ does not exist,
consider the containing condition to be vacuously true.}

\ifthenelse{\equal{\version}{AAMAS_02_2014}}{
Assume that $B_{n - 1} \geq \theta B_n$ and the following both hold.
\begin{enumerate}
	\item $\frac{1}{n}\alpha_{m} \geq \alpha_{m - k}$,\label{NE_effic_nplayers:conds:close_B:steep}
	\item	$B_1 \geq k\theta B_n$;\label{NE_effic_nplayers:conds:close_B:not_steep}
\end{enumerate}

Then, there exists a pure strategy \NE{}
and there holds:
$\pos = 1$.
}{
\begin{enumerate}
\item	Assume that $B_{n - 1} \geq \theta B_n$ and the following both hold.
\begin{enumerate}
	\item $\frac{1}{n}\alpha_{m} \geq \alpha_{m - k}$,\label{NE_effic_nplayers:conds:close_B:steep}
	\item	$B_1 \geq k\theta B_n$;\label{NE_effic_nplayers:conds:close_B:not_steep}
\end{enumerate}\label{NE_effic_nplayers:conds:close_B}

Then, there exist pure \NE{}
and 
$\pos = 1$, $\poa \geq \frac{(1 + (n - 1)\theta) (B_{n - 1} + B_n)}{n (\sum_{i \in \set{1, 2, \ldots, n}}{B_i})}$.

\item Assume that $B_{n - 1} < \theta B_n$. \label{NE_effic_nplayers:conds:far_B:suppress}
	Then, $\poa \geq \frac{B_n}{\sum_{i\in \set{1, 2, \ldots, n}}{B_i}}$.
	
	If we also have
	$B_{n - 1} < \frac{\theta}{\abs{\Omega}} B_n$ and all the
	project functions are equal, i.e.~$\alpha_m = \alpha_1$,	
	then there exist pure \NE{}
	and 
	$\pos = 1, \poa = \frac{B_n}{\sum_{i\in \set{1, 2, \ldots, n}}{B_i}}$.
\label{NE_effic_nplayers:conds:far_B}
\end{enumerate}\label{NE_effic_nplayers:conds}
}%
\end{theorem}

To prove more bounds on the price of anarchy, we employ the
so-called \defined{smoothness argument} from Roughgarden~\cite{Roughgarden2015}.
This means showing that for a socially optimal profile
$x^*$ and any profile $x$, there exist $\lambda > 0$ and $\mu > -1$
such that
\begin{equation}
\sum_{i = 1}^n{u^i((x^*)^i, x^{-i})} + \mu \cdot \sw(x) \geq \lambda \cdot \sw(x^*).
\label{eq:smooth_game}
\end{equation}
Roughgarden proves that this implies that each pure \NE{} has a social
welfare of at least $\frac{\lambda}{1 + \mu}$ of the optimal
social welfare, i.e.~$\poa \geq \frac{\lambda}{1 + \mu}$.
\begin{remark}
We employ the relaxation from~\cite[Remark~$2.3$]{Roughgarden2015} that $x^*$ does
not have to be any profile but we may pick any socially optimal one, and
we use the payoff-maximisation version of the
argument (see~\cite[\sectn{$2.3.2$}]{Roughgarden2015}).
\end{remark}

Equipped with this tool, we prove the following.
\begin{theorem}\label{the:NE_effic_n_smooth}
Consider an equal $\theta$-sharing game with $n \geq 2$ players with budgets
$B_n \geq \ldots \geq B_2 \geq B_1 > 0$ (the order is w.l.o.g.), $0 \leq \theta \leq 1$, and project functions $P_j(x) = \alpha_j \cdot x$ with coefficients
$\alpha_m = \alpha_{m - 1} = \ldots = \alpha_{m - k + 1} > \alpha_{m - k} \geq \alpha_{m - k - 1} \geq \ldots \geq \alpha_1 > 0$
(the order is w.l.o.g.).
%
Let $l \in \set{1, 2, \ldots, n}$ be the smallest integer such that
	$B_1 \leq B_2 \leq \ldots \leq B_{l - 1} < \theta B_n$, but 
	$B_l \geq \theta B_n$.
	Then, $$\poa \geq \frac{1 + (n - 1)\theta}{n} \frac{\sum_{i = l}^{n - 1}{B_i}}{\sum_{i = 1}^{n}{B_i}}
	+ \frac{1 + (n - l) \theta}{n - l + 1} \frac{B_n}{\sum_{i = 1}^{n}{B_i}}.$$
\end{theorem}
\begin{proof}
We show the smoothness argument for some $\lambda > 0$ and $\mu > -1$.
%
Setting $\mu$ to zero
implies that the left hand side of \eqnsref{eq:smooth_game} is minimised
as follows. The utility of players less than $l$ is zero, since $n$
can suppress them.
To minimise the utility of a player $i$ that $l \leq i \leq n - 1$, every
other player contributes exactly $\theta B_i$ (if her budget allows), and
to minimise the utility of $B_n$, every player from $l$ till $n - 1$
contributes $\theta B_n$, while the players less than $l$ contribute
nothing. This yields the following lower bound on the left hand side:
\begin{equation*}
\sum_{i = l}^{n - 1}{ \frac{\alpha_m B_i (1 + (n - 1)\theta)}{n} }
+ \frac{\alpha_m B_n (1 + (n - l)\theta) }{n - l + 1} + 0 \cdot \sw(x).
\end{equation*}
The right hand side of \eqnsref{eq:smooth_game}, equal to
$\lambda \cdot \alpha_m \sum_{i = 1}^n{B_i}$, is bounded by the
left hand side if and only if
$$\lambda \leq \frac{1 + (n - 1)\theta}{n} \frac{\sum_{i = l}^{n - 1}{B_i}}{\sum_{i = 1}^{n}{B_i}}
	+ \frac{1 + (n - l) \theta}{n - l + 1} \frac{B_n}{\sum_{i = 1}^{n}{B_i}}.$$
Combined with $\mu = 0$, this implies the lower bound on the price of
anarchy.
\end{proof}

If $B_{n - 1} < \theta B_n$, 
this theorem with $l = n$
implies that the price of anarchy is at least
$\frac{B_n}{\sum_{i = 1}^ n {B_i}}$, matching
part~\ref{NE_effic_nplayers:conds:far_B:suppress} of
Theorem~\ref{the:NE_effic_nplayers}.

To simplify the theorem, we draw a simpler, though
a looser, bound.
\begin{corollary}\label{cor:NE_effic_n_smooth_simpler}
Under the conditions of Theorem~\ref{the:NE_effic_n_smooth},
the  following bound holds.
%
	Let $l$ be the first integer such that
	$B_1 \leq B_2 \leq \ldots \leq B_{l - 1} < \theta B_n$, but 
	$B_l \geq \theta B_n$.
	Then $\poa \geq \theta \frac{\sum_{i = l}^n {B_i}}{\sum_{i = 1}^n {B_i}}.$
\end{corollary}

To summarise, we first characterise the cases of zero threshold and threshold~$1$. 
We then provide a complete characterisation of the existence and efficiency 
of equilibria for two players with a general threshold, and extend selected 
existence and efficiency results to $n \geq 2$ players with a general threshold.
Notice that our (pure) model is \emph{not} the mixing
of the model where a player may invest
in at most one project,
because the mixing would extend the utilities
linearly in the mixing coefficients~\cite[Definition~{32.1}]{OsborneRubinstein94}, which is not the case
in our (pure) model with a positive threshold.

%% file: cyc_sne.tex
\section{Narrowing down Nash to Cyclically Strong Equilibrium}\label{Sec:cyc_sne}

We now introduce a  generic method to study solution concepts.
Having introduced a new solution concept by relaxing or strengthening some properties of an existing solution concept, imagine it subsequently turns out that both concepts coincide on a subclass of games. 
Finding such a subclass of games, especially a maximal such subclass,
where the two solution concepts coincide, involves understanding 
the distinction between the two solution concepts. This also allows 
studying the limits between the solution concepts.

Now, if the new concept generally strengthens the previous one, then we conclude that on the subclass where both concepts are equivalent, the requirements of the original concept imply the new concept, implying a \defined{narrowing down} of the original concept to the new one. On the other hand, if the new concept relaxes the original one, then their equivalence on a certain subset of games means that already the lighter requirements imply the original concept, demonstrating a \defined{broadening} of the original concept.
This also allows characterising one solution using the equivalent one, and to transfer the efficiency bounds.

Here, we are going to (apparently) strengthen the Nash equilibria on thresholded shared effort games
by adding the requirement that no cyclical deviation of whole budgets is allowed. Cyclical moves
are practically and theoretically important, such as the famous top trading cycles mechanisms%
~\cite{AbdulkadirogluSonmez1999}.
Subsequently, we will prove that the resulting solution concept is actually equivalent to 
Nash equilibrium on thresholded shared effort games where no budget can suppress another one, 
thereby constituting a narrowing down of Nash equilibrium. In other words, having no profitable 
unilateral deviations already implies possessing no cyclical deviation of whole budgets. 
This exemplifies a new useful technique we hope will serve many researchers in the future.

Formally, given a shared effort game, we define
\begin{defin}\label{def:cyc_sne}
A \defined{cyclically strong equilibrium} is a Nash equilibrium~$x$,
where there exists no coalitional deviation of players~$i_1, i_2, \ldots, i_p$ 
such that
\begin{enumerate}
	\item for each $l = 1, 2, \ldots, p$, there exists a project~$\omega \in \Omega$,
	such that $x^{i_l}_{\omega} = B_{i_l}$, namely, all $i_l$'s budget is in one project;
	\item	if each player~$i_l$ above deviates to invest all $B_{i_l}$
	into the cyclically next player's project, thereby creating profile~$x'$,
	then no deviating player loses and at least one of them 
	strictly benefits w.r.t.~$x$.
\end{enumerate}
\end{defin}
Cyclically strong equilibrium is a restriction of Nash and a relaxation of
strong Nash equilibrium~\cite{Aumann1959}.

Interestingly, the new concept often coincides with Nash equilibrium.
\begin{theorem}
In any thresholded shared effort game where $B_i \geq \theta B_j, \forall i, j \in N$, 
any Nash equilibrium is also cyclically strong.
\end{theorem}
\begin{proof}
Consider an $x \in \NE$, and consider a cyclic coalitional deviation
by players~$i_1, i_2, \ldots, i_p$ from Definition~\ref{def:cyc_sne}, 
assuming by contradiction that no player loses
and at least one strictly benefits. 
Since no unilateral deviation is profitable, while now no player loses
and at least one strictly benefits, then either some 
$B_{i_l} < \theta B_{i_{l + 1}}$ (modulo $p$), which is assumed not to be the case,
or each deviation on this cycle has to be of a budget of size at most 
the size of the previous deviation, thus 
$B_{i_1} \geq B_{i_2} \geq \ldots \geq B_{i_p} \geq B_{i_1}$,
with at least one inequality being strict. However, this cycle implies
$B_{i_1} > B_{i_1}$, in contradiction to the assumption of a cyclic
deviation above.
\end{proof}

This property allows for a better theoretical study of Nash equilibria,
as well as providing a practical property they all possess.

The described approach analyses the \emph{refinement} by finding 
\emph{subclasses of games} where it coincides with 
the original solution concept. Similarly, one can study 
a \emph{subclass of games} by analysing the \emph{refinements} 
that coincide with the original solution concept on that subclass.

%% file: mix_ne.tex
\section{Mixed Nash Equilibrium}\label{Sec:mix_ne}

As we have seen, a game may not possess a pure \NE. Therefore,
we naturally turn to mixed extensions%
\footnote{A mixed extension has strategies that are distributions on the
pure strategies and the respective utilities are the expected utilities
under these distributions.} and ask whether a mixed extension always has an \NE. At first, this
is unclear.
As the game is infinite, the 
theorem by Nash~\cite{Nash51} about the existence of a mixed \NE{}
in finite games
is irrelevant. Since the game is not continuous, even
the theorem by Glicksberg~\cite{Glicksberg1952} about the existence of
a mixed \NE{} in continuous games is not applicable. Fortunately, we
answer affirmatively employing a more general existence theorem
by Maskin and Dasgupta~\cite[Theorem~{$5^*$}]{DasguptaMaskin1986}, which 
requires some definitions. 
Let the strategy sets be $A_i \subset \reals^m$.
For each pair of players $i, j \in {1, \ldots, n}$, let $D(i)$ be
a positive natural, and for a $d \in \set{1, \ldots, D(i)}$, let
$f^d_{i, j} \colon \reals \to \reals$ be one-to-one and continuous, such that
$(f^d_{i, j})^{-1} = f^d_{j, i}$. For every player~$i$, we define
\begin{eqnarray}
A^*(i) \defas
\{(a_1, \ldots, a_n) \in A \mid \exists j \neq i, \exists k \in \set{1, \ldots, m}, \exists d \in \set{1, \ldots, D(i)}, \nonumber\\
\text{ such that } a_{j, k} = f^d_{i, j}(a_{i, k})\}.
\end{eqnarray}
They define weakly lower semi-continuity, which intuitively means that
there is a set of directions, such that approaching a point from any of
these directions gives values at least equal to the function at the point. 
Find the precise definition in appendix \sectnref{Sec:omit_proof}.

Theorem~{$5^*$} from~\cite{DasguptaMaskin1986} states that if the
strategy sets $A_i \subset \reals^m$ are non-empty, convex, and compact,
the utility function $u^i$ are continuous, except for 
$A^{**}(i) \subseteq A^*(i)$, the sum of all the utilities
is upper semi-continuous, and for every player $i$, $u^i(a_i, a_{-i})$
is bounded and weakly lower semi-continuous in $a_i$, then, there
exists a mixed Nash equilibrium in this game.

Using Theorem~{$5^*$} from~\cite{DasguptaMaskin1986}, we prove
(see the proof in \sectnref{Sec:omit_proof})
\begin{theorem}\label{the:mix_NE_exist}
Any (linear) shared effort game with 
$\theta$-equal sharing has a mixed Nash equilibrium.
\end{theorem}

The existence result automatically extends to the solution concepts that
include mixed Nash equilibria, such as correlated~\cite{Aumann1974} and
coarse correlated~\cite{MoulinVial1978} equilibria.
Luckily, not only existence results but also some efficiency bounds 
extend to other equilibria, as we describe next.

First, the invariance to multiplication holds for the mixed,
correlated and coarse correlated equilibria as well, since
the proof Proposition~\ref{prop:inv_mult} extends from 
linearity of expectation.

We are about to show that some bounds on the
social welfare of solution concepts extend to the other equilibria
as well.
An important preliminary observation is that the maximum social welfare
stays the  same even when (correlated) randomization is allowed; it is
always $\alpha_m \sum_{i = 1}^n {B_i}$.

Consider the results of Theorem~\ref{the:NE_effic_nplayers}.
Its lower bounds on the price of anarchy stem from the utility that
certain players can always achieve, and the bounds, therefore, hold for mixed,
correlated and coarse correlated equilibria as well. Since any pure \NE{}
is also a mixed/correlated/coarse-correlated \NE, the rest of the
efficiency results, based on presenting an \NE{}, also extend to the
other solution concepts.

As for Theorem~\ref{the:NE_effic_n_smooth} and its corollary,
Roughgarden~\cite[Theorem~$3.2$]{Roughgarden2015} proves that the lower
bounds on the price of anarchy that are proven using the
smoothness argument go over to the mixed, correlated and
coarse correlated equilibria.

To conclude,
\NE{} exist in the mixed case, and
the efficiency bounds from Theorems~\ref{the:NE_effic_nplayers}
and~\ref{the:NE_effic_n_smooth} apply there too.
%

%% file: RelatedWork.tex
\section{Related Work}\label{Sec:rel_work}

Since
understanding motivation is necessary to implement recommendations
about contribution,
we first discuss why people contribute
to projects and how such contributions have been modeled. Then, we
provide the basic game-theoretic background and present the existing work
on existence and efficiency of \NE{} for sharing
effort, concluding that no analysis of the general setting has taken
place, a gap which we partially fill.
%

\paragraph{Contribution to Projects}

\ifthenelse{\equal{\version}{AAMAS_02_2014}}{
}{
Motivation to contribute to public projects can be
both extrinsic, like a payment or a record for the CV, and intrinsic,
such as exercising one's favorite skills~\cite{KaufmannSchulzeVeit11}
or, mostly, conservation citizen science projects%
~\cite{MaundIrvineLawsonSteadmanRiselyCunninghamDavies2020}.
\citeauthor{AllahbakhshBenatallahIgnjatovicMotahariBertinoDustdar2013}~\cite{AllahbakhshBenatallahIgnjatovicMotahariBertinoDustdar2013} 
discuss rewards for contributing to crowdsourcing, based on quality, where contributors have to fulfill given
requirements and build reputation (extrinsic) over time.
Wang et al.~\cite{WangFesenmaier2003} model motivation to contribute to
online travelling communities and conclude the importance of both the practical motives,
such as supporting travellers, building relationships, and hoping for a future
repay (extrinsic), as well as of internal drives to participate.
Forte and Bruckman~\cite{ForteBruckman05} study why people contribute
to Wikipedia, by asking contributors, and conclude that the reasons
are similar to those of scientists and include the desire to publish
facts about the world (intrinsic).
Bagnoli and Mckee~\cite{BagnoliMckee1991} empirically check when people
contribute to a public good, like building a playground. They find that
if people know the threshold for the project's success and benefit
from collective contributing, then they will contribute, in agreement with
the theory of~\cite{BagnoliLipman1989}. The work argues that knowing such
information is realistic, suggesting the real cases of hiring a lobbyist
and paying to a ski club as evidence.
This conclusion supports our rationality assumption.

The concrete ways
to motivate such contributions have been studied too. For instance, Harper et al.~\cite{HarperLiChenKonstan07}
find that explicitly comparing a person's contribution to the contribution
of others helps focusing
on the desired features of the system, but does not change the interest
in the system per se.
The influence of revealing how much people contribute to a movie
rating community is experimentally studied
in~\cite{ChenHarperKonstanLi2010}.
Initiating participation in online communities is experimentally studied
in~\cite{LudfordCosleyFrankowskiTerveen04} on the example of the influence
of similarity and uniqueness of ratings on participation.
}

We now present the required theoretical background.
\paragraph{Background: Equilibria and Efficiency}

Given a general non-cooperative game in strategic form%
~$(N, (S_i)_{i \in N}, (u_i)_{i \in N})$, a \defined{Nash equilibrium}
is a profile~$s \in S$, such that no unilateral deviation is
beneficial, namely $\forall i \in N, \forall s_i' \in S_i$,
$u_i(s_i', s_{-i}) \leq u_i(s)$.

\ifthenelse{\equal{\version}{AAMAS_02_2014}}{
}{
An \NE{} can be inefficient, such as in the famous example of the
prisoner's dilemma~\cite[Example~$16.2$]{OsborneRubinstein94}. 
%
The ratios of the objectives in the least or the most efficient
\NE{} and in the optimum, called \defined{price of anarchy} ($\poa$)~\cite{KP99}
and \defined{price of stability} ($\pos$)~\cite{AnshelevichDasGuptaKleinbergTardosWexlerRoughgarden04}, respectively,
constitute the most popular approaches to quantifying this inefficiency%
~\cite[\chapt{17}]{Nisanetal07}.
The price of anarchy measures the best guarantee on
an \NE, while the price of stability measures the cost of leading the
game to a specific equilibrium.
Roughgarden and Tardos~\cite[\chapt{17}]{Nisanetal07} discuss inefficiency of equilibria in
non-cooperative games and consider the examples of network,
load balancing and resource allocation games.
This work argues that understanding exactly when selfish behaviour is socially
profitable is important, since in many applications, implementing control
is extremely difficult.

We combinatorially study existence and efficiency of \NE, and then turn
to simulations.
}

\ifthenelse{\equal{\version}{AAMAS_02_2014}}{
}{
\paragraph{Background: Classical Fictitious Play}\label{fict_play}
In our simulations,
we generalise and employ the widely studied fictitious play,
introduced by Brown~\cite{Brown1951},
to find equilibria.
In this play, each player best-responds to the product
of cumulative marginal
histories of the others' actions at every time step.
It is a myopic learning process.
If the game is finite, then if a fictitious play converges,
then the distribution in its limit is an \NE{}~\cite{Manor2008}.
Conversely, a game is said to possess the fictitious play property
if every fictitious play approaches equilibrium in this game~\cite{MondererShapley1996PGFPP}. Many
researchers show games that possess this property, for example, finite two-player
zero-sum games~\cite{Robinson1951}
and finite weighted potential games~\cite{MondererShapley1996PGFPP}.
A famous example for a game without such property is a $3 \times 3$
game by Shapley~\cite{Shapley1963}. In this game, there is a cyclic
fictitious play that plays each strategy profile for at least an
exponentially growing number of times, and therefore, does not converge
at all. Moreover, even its subsequences do not converge to an \NE.
}

\paragraph{Background: Mixed Equilibria and their Existence}
Since pure equilibria may not exist, we also consider
the \defined{mixed extensions} of a game, where the strategies
are the probability distributions over the original (pure)
strategies and the utilities are defined as the expected 
utilities under the played distributions. Then, a 
\defined{mixed Nash equilibrium} is defined as an 
equilibrium of the mixed extension of a game.
Regarding the existence of a mixed \NE, Nash proved in his classical
paper~\cite{Nash51} that a mixed extension of a finite game always
possesses an \NE. Glikcsberg~\cite{Glicksberg1952} showed the existence
of a mixed \NE{} for
continuous games. However, in shared effort games with a positive
threshold, the threshold creates discontinuity. Dasgupta and Maskin
prove the existence of a mixed equilibrium for a subclass of
possibly discontinuous games~\cite{DasguptaMaskin1986}. We show that
shared effort games can be cast to fit Dasgupta and Maskin in
Theorem~\ref{the:mix_NE_exist}.

\paragraph{Extant Models of Sharing Effort}

The models that resemble ours, but expressed in the terms of
cooperative games, where every coalition of players has a value, include
the works by Zick, Elkind and Chalkiadakis~\cite{ZickElkind2011, ZickChalkiadakisElkind2012}.
Contributing to
a coalition can be considered as contributing to a project in
a shared effort game, and in both cases players have budgets
and obtain shares. 
%
%
Another coalitional model, where participating in a coalition can be seen 
as contributing to projects appears in~\cite{GalNguyenTranZick2020},
though in addition to being a different kind of game, 
this model allows no participation in multiple coalitions.
Despite these similarities, cooperative games are not 
games in our sense, since they do not define utilities. Moreover, even if we
considered a shared effort game as a particular case of a cooperative game,
a positive threshold would vitiate the individual rationality of the core, since a
player who obtains a positive share when she is the only contributor to a project may obtain nothing when
others contribute to the same project.

We now move to non-cooperative games modelling sharing effort.
Shared effort games where only the biggest contributor obtains the project's value, while everyone pays, are called all-pay auctions,
and their equilibria are studied, for instance, by Baye, Kovenock and de~Vries~\cite{BayeKovenockdeVries96}. That work shows cases where each player obtains the expected payoff of zero,
and cases where the winner obtains the difference between the two highest valuations, while the rest obtains zero.
All-pay auctions model lobbying, single-winner contests, political campaigns, striving for a job promotion (see e.g.~\cite{Siegel2009}) and Colonel Blotto games with two~\cite{Roberson06} or more~\cite{KovenockRoberson2021,BoixAdseraEdelmanJayanti2021} players.
In a Colonel Blotto game, the colonels divide their armies across battlefields, and at every battlefield, the larger force wins. The number of the won battlefields defines the utility of a colonel. Namely, the winner of a battlefield takes it fully to her utility.
For two players, Roberson~\cite{Roberson06} analyses the equilibria of this game and their expected payoffs. Any outcome is socially optimum, since this is a constant-sum game,
modelling a confrontation. The $n$-player Blotto game has been studied too~\cite{KovenockRoberson2021}. Shared effort games can model these games, but  typically, shared effort models cooperation, 
where a project's value increases with the total contribution.

The models below feature no thresholds required to share a project's value.
Anshelevich and Hoefer~\citep{AnshelevichHoefer2012} consider an undirected graph model, where the nodes are the players and each player divides its budget among its adjacent edges in minimum effort games (where the edges are $2$-player projects), each of which equally rewards both sides by the value of the project's success (i.e., duplication instead of division). Anshelevich and Hoefer prove the existence of equilibria, find the complexity of finding an \NE{}, and find that the $\poa$ is at most $2$. A related setting of \emph{multi-party computation games} appears in Smorodinsky and Tennenholtz~\cite{SmorodinskyTennenholtz06}. There, the players are computing a common function that requires them to compute a costly private value, motivating free-riding. The work suggests a mechanism, where honest computation is an \NE. This differs from our work, since they consider cost minimisation, and the choice of the players is either honestly computing or free riding, without choosing projects.
For shared effort games
with specific conditions (obtaining at least a constant share of one's marginal contribution to the project's value and no contribution threshold, i.e.~$\theta = 0$), Bachrach et al.~\citep{BachrachSyrgkanisVojnovic12} show that the price of anarchy ($\poa$) is at most the number of players. That work also upper bounds the $\poa$ for the case of convex project functions, where each player receives at least a constant share of its marginal contribution to the project's value.
The academic game~\cite{AckermanBranzei2017} considers discrete weights,
limits the collaboration to at most two agents per project, and the utility
functions model synergy and various ordering of the authors, which effect
is studied. This model employs the pairwise stability 
approach~\cite{JacksonWolinsky1996}, which in addition to guarding against 
unilateral deviations, guards against $2$ agents deviating to 
a common project too.
Gollapudi et al.~\cite{GollapudiKolliasPanigrahiPliatsika2017} allow an agent 
merely to choose which single team to join. They consider several 
profit sharing models, including equal sharing, but without any threshold.

There has been no research of the \NE{} of our problem with
a $\theta \in [0,1]$ sharing mechanism before the preliminary
version~\cite{PolevoyTrajanovskideWeerdt14} of our paper.
A positive $\theta \in (0,1]$ contradicts the condition
of receiving a constant share. We
provide precise conditions for existence of \NE, and find their efficiency. 
There exist two works that variate this model in various directions.
In~\cite{PolevoydeWeerdt2018a}, quotas and other requirements from
projects, such as paper and grants applications, are modeled by requiring
a project to be good enough in order to actually obtain its value.
They find equilibria and their efficiency and compare the efficiency of
equilibria for being within a quota and for attaining at least a certain
minimum value.
In~\cite{PolevoydeWeerdt2017b}, investing in reciprocal interactions,
like attending seminars and meeting friends, is modeled be assuming
each project to be a reciprocal interaction from which the interacting
parties obtain a value. They prove the convergence of such a process
and analyse the existence and efficiency of the equilibria.

%% file: ConclusionNE.tex
\section{Conclusions and Future Work}\label{Sec:conclus_further}

This paper considers shared effort games where the players contribute to the given projects, and subsequently share the linear values of these projects, conditionally on the allocated effort. We study existence and efficiency of the \NE{}. 
In the following, we describe when and how one can vary the parameters of the game
so as to improve the equilibrium efficiency; when that is impossible, we
can always apply tools like subsidising~\cite{MondererTennenholtz2004}
or other payment adjustments~\cite{PolevoyDziubinski2022}.

We discover that
multiplying all the  project functions or budgets by the same factor does not change the equilibria.
We first treat the equilibria for thresholds of $0$ and $1$.
For threshold $0$, pure equilibria always exist and are socially optimal, 
requiring no regulation. For threshold of $1$, on the other hand, there
exist non-trivial conditions for existence of equilibria without suppression,
and when they hold, the efficiency is uniquely determined and depends on 
the ratio between the sums of the budgets weighted by the project functions. 
Thus, striving for high efficiency would require making the projects similar 
to one another, but otherwise, regulation will not help, as the efficiency 
is the same across all such equilibria.
We then characterise the existence and efficiency of pure \NE{} for shared effort games with two players. 
When an \NE{} exists and the budgets are close to each other, 
all the \NE{} are socially optimal.%
\ifthenelse{\equal{\version}{AAMAS_02_2014}}{
}{
When the budgets are further apart, in the sense that smallest budget is less than threshold times the largest one, the efficiency depends on the ratio of the budgets and of the two or three
largest projects' coefficients
but is always greater than half of the optimum.
}%

When the budgets are close, we demonstrate an optimal \NE{} where everyone
equally spreads her budgets among the most valuable projects. This
motivates the organisers of any project to make their project most
valuable possible. Even second best project can receive no contributions in an optimal \NE.

For arbitrarily many players, we find socially optimal pure equilibria in some cases
and bound the efficiency from below.
%
We obtain further results on the existence and efficiency of \NE{} for more than two players by simulating fictitious play.
All Nash equilibria are invariant to multiplication of project coefficients and of budgets (Proposition~\ref{prop:inv_mult}). Therefore, simulation results obtained for normalized parameter values are representative of the general case.
Actually, the main contribution of the simulations resides in the presented methodology,
generalising fictitious play and computing a best response.
To this end, we generalise fictitious play to infinite strategy spaces and describe some of the
best responses of a player to the other players' strategies.
First, we corroborate all the theoretical predictions for the simulated cases.
\ifthenelse{\equal{\version}{AAMAS_02_2014}}{
For two players, an \NE{} is usually almost of the optimal efficiency.
}{
}%
The most important factor for existence and efficiency of an equilibrium is
the ratio of the largest to the second largest project function coefficients and of the largest to the
second largest budgets. Therefore, to influence the projects and the players to some extent,
one should influence the projects with the highest values and the players with the largest
budgets. 

Some efficiency bounds persist in the mixed extension, where we show that an \NE{} always exists. 

Our main conceptual contribution is narrowing down a Nash equilibrium
to a new solution concept, discovering that without suppression of whole budgets,
Nash equilibria have no profitable cyclical deviations. That inspires thinking about
other interactions and relevant solution concepts, which may coincide on important
classes of games, motivating much future work.

Further 
directions for future work follow.
Since the real value usually depends on the total contribution linearly up
to a point, we should study piecewise project functions.
Sometimes the value of the project grows in discrete steps; for example, 
in the times required to complete a product or to complete a certain part of an article.
In those cases, replacing our linear model with a discrete value function would require
us to check which properties still hold, and this more precise model would also inspire 
further realistic models in game theory.
More generally, we would like to extend our complete theoretical characterization of the
existence of (pure) \NE{} %
\ifthenelse{\equal{\version}{AAMAS_02_2014}}{
to far budgets, and also }{
}%
to more than two players and to various non-linear project functions. 
%
Next, we can try finding various \NE{} that are not yet analysed analytically
by approximately finding a best response and extending the simulations to more than two projects. 
Our simulations possess extra range representativeness because of 
the multiplication invariance, and we have tried to consider the practically 
realistic ranges parameters. Still, the parameter range of the simulations 
should be extended based on the studied scenario.
Another numerical approach is discretizing the game and applying 
off-the-shelf software to find the equilibria of the discretization, such
as Gambit~\footnote{\url{http://gambitproject.readthedocs.io/en/v15.1.1/intro.html\#}}
or Game Theory Explorer~\footnote{\url{http://banach.lse.ac.uk/}}.
Since randomization can be deliberately undertaken or 
describe beliefs or uncertain behaviour
~\cite[\sectn{3.2}]{OsborneRubinstein94},
we would like to find concrete mixed equilibria to be able to advise on playing them
or at least predict the outcomes, like we
do here for the pure equilibria.
Moreover, we would like to vary the model, studying
a relative threshold that depends on the median of the contributions, to dampen
the influence of extreme contributors, or studying an absolute threshold, 
and imposing further constraints on the players. 
Another option is introducing an explicit cost of effort, to model 
optional costs. This would change the set of equilibria, 
in particular eliminating suppressed players. This 
would also eliminate the continuum of equilibria when 
the suppressed player can do anything.
Converting important interactions, like
investment policies of political campaigns, to our model opens a promising
avenue, too.

We prove that Nash equilibrium narrows down to cyclically strong 
equilibrium on shared effort games where every two players' budgets 
are at most the threshold factor away. The next natural question 
is whether these are the only thresholded shared effort games 
where every Nash equilibrium is also cyclically strong, 
which we leave open. Answering that would shed additional light 
on the concept of cyclically strong equilibrium, by describing whether 
the budget ratios can characterise the distinction between 
general Nash and cyclically strong Nash equilibrium.

To conclude, we have analysed when contributions to public projects are
in equilibrium and what is lost in the equilibria relatively to the best
possible contribution profiles. 
The theoretical analysis of efficiency implies that
for two players with close budgets,
no coordination is needed, since the price of anarchy is $1$.
The price of anarchy is close to $1$ also for two players with budgets that
are far from each other, and it is
always more than half.
For three or more players, some coordination may substantially improve the total utility, though
we have seen many cases with efficiency above $0.75$.
We have also provided conditions for a general number of players where every
equilibrium is almost optimal, so no coordination is required.
In the scenarios where much is lost relatively to the optimum,
coordination may improve efficiency.

%% file: OmitProof.tex
\section{Omitted Proofs}\label{Sec:omit_proof}

Let us first prove Proposition~\ref{prop:inv_mult}.

\begin{proof}
The invariance to multiplying the projects stems from the fact that
multiplying all the project coefficients by $p$ multiplies all the utilities
by $p$. Since this happens to all the utilities, the
same relations keep holding between the various strategy profiles, and
thus, the same \NE{} remain.

We prove the second part by contradiction. If $p \cdot x$ is not an \NE,
then there exists a unilateral profitable deviation by player~$i$. Denote
the profile after such a deviation by $x'$. Then, $1/p \cdot x'$ is a
legal profile in the original game, and it is a unilateral 
deviation from profile~$x$ by player~$i$. This deviation is profitable
to~$i$, since the original deviation is profitable and all the utilities
are multiplied by multiplying the profile. This contradicts the assumption
that $x$ is an \NE{} of the original game.
\end{proof}

We now prove Lemma~\ref{lemma:NEchar}.

\begin{proof}
First, at least one player contributes somewhere, since otherwise any
positive contribution would be a profitable deviation for every player
(all $\alpha_i$s and $B_j$s are positive).
Moreover, at least one of the players contributes to a steep project, for the
following reasons. If only the
non-steep projects receive a contribution, then take any such project $p$.
If a single player contributes there, then this player would benefit
from moving to contribute to a vacant steep project.
If both players contribute
to $p$, then if one is suppressed, it would like to deviate to any steep
project where it would not be suppressed, and if no-one is suppressed,
then a player who contributes not less would like to contribute to a vacant
steep project instead.

We prove part~\ref{lemma:NEchar:list:non_suppr_non_steep} now.
Let $i \in N$ be any non-suppressed player among those who
contribute to a steep project, w.l.o.g., to project $m$.
Assume first that player $j \neq i$ is not suppressed.
Then, for any non-steep project where $i$ contributes, $i$
contributes either alone or precisely the least amount it should
contribute to achieve a portion in the project's value, because
otherwise $i$ would like to increase its contribution to $m$ on the
expense of decreasing its contribution to the considered
non-steep project.

Now, consider the case where $j$ is suppressed. Then, even if $j$
contributes to a non-steep project where $i$ contributes (and is
suppressed there), $i$ still will prefer to move some budget from
this project to $m$, since $i$ receives the whole value of $m$ as
well. Thus, this cannot be an \NE.

\end{proof}

The proof of Lemma~\ref{lemma:NEchar:close_B} appears now.

\begin{proof}
Since $B_1 \geq \theta B_2$, no player
is suppressed, because any player prefers not being suppressed,
and at any project,
a player who concentrates all its value there is not suppressed.

Every steep project receives a positive contribution from each player,
for the following reasons.
If only a single player contributes to a steep project, then the
player who does not contribute there will profit from
contributing there exactly the threshold value,
while leaving at least the threshold values at all the projects
where it contributed.
There is always a sufficient surplus to reach the
threshold because $B_1 \geq \theta B_2$.
If no player contributes to a steep project~$p$, then there exists another
steep project~$q$, where two players contribute, according to
part~\ref{lemma:NEchar:list:at_least_1_steep} of
Lemma~\ref{lemma:NEchar:list:at_least_1_steep} and what we have just described. The player who
contributes there strictly more than the threshold would profit from
moving some part of his contribution from $q$ to $p$, still remaining
not less than the threshold on $q$, contradictory to having
an \NE.

We next prove the second part of the lemma.
Since both players are non-suppressed contributors to steep projects,
then, according to
part~\ref{lemma:NEchar:list:non_suppr_non_steep} in Lemma~\ref{lemma:NEchar},
we conclude that there exist no non-steep projects
where $j$ and $i$ contribute together.
\end{proof}

\ifthenelse{\equal{\version}{AAMAS_02_2014}}{
}{
We now present the proof of Lemma~\ref{lemma:NEchar:far_B}.

\begin{proof}
We prove part~\ref{lemma:NEchar:far_B:list:contr_1} first.
Consider an \NE{} profile. Assume to the contrary that player $1$
contributes to a steep project, w.l.o.g., to project $m$. 
Since $B_1 < \theta B_2$ and no player is suppressed,
player $2$ could transfer to $m$ budget from other projects, such that
at each project, where $2$ was obtaining a share of the value,
$2$ still obtains a share,
and $2$ suppresses $1$ at $m$. This would increase $2$'s utility,
contrary to the assumption of an \NE.

Now, we prove parts~\ref{lemma:NEchar:far_B:list:each_pos_util} and~\ref{lemma:NEchar:far_B:list:2_all_steep_and_thresh_others}.
Since $1$ does not contribute to steep projects,
part~\ref{lemma:NEchar:list:at_least_1_steep} of Lemma~\ref{lemma:NEchar}
implies that $2$ contributes to a steep project
(say, $2$ contributes $y > 0$ to project $m$), and
part~\ref{lemma:NEchar:list:non_suppr_non_steep} of Lemma~\ref{lemma:NEchar}
implies that
if $2$ contributes to a non-steep project, it contributes there
either alone or precisely the least amount it should
contribute to achieve a portion in the project's value. Since contributing
alone is strictly worse than contributing this budget to a steep project,
$2$ may only contribute together with $1$, the threshold amount.
Therefore, $1$ receives a positive value in this profile,
and we have part~\ref{lemma:NEchar:far_B:list:each_pos_util}.
The only thing left to prove here is that $2$ contributes to each
steep project. If not, $1$ would prefer to move some of its contribution
there, in contradiction to the assumption of an \NE.

We prove part~\ref{lemma:NEchar:far_B:list:2_steep} now.
Assume to the contrary that a non steep and non $2$-steep project receives a
contribution. We proved in part~\ref{lemma:NEchar:far_B:list:2_all_steep_and_thresh_others} that $1$ contributes there,
alone or not.
For her, moving a small enough utility to a $2$-steep project
would increase her utility, regardless whether $2$ contributes to 
any of those projects. This is so because if $2$ contributes together
with~$1$, it contributes precisely the threshold amount, according
to part~\ref{lemma:NEchar:far_B:list:2_all_steep_and_thresh_others}.
This incentive to  deviate contradicts the assumption of an \NE.

We prove part~\ref{lemma:NEchar:far_B:list:B_ratio} now.
If $2$ contributes to a $2$-steep
project~$p$, then there may not exist another $2$-steep project, since
otherwise $1$ would like to transfer a small amount from $p$
to another $2$-steep project~$q$, such that without losing a share of the
value of~$p$, player $1$ gets the whole value of project~$q$.
}%
\ifthenelse{\equal{\version}{AAMAS_02_2014}}{
}{
\end{proof}
}%

We prepend the proof of the Theorem~\ref{the:NE_charac_2players} with the following technical lemma.
\begin{lemma}\label{lemma:NE_charac_2players:move_delta}
Consider an equal $\theta$-sharing game with two players with budgets $B_1, B_2$. W.l.o.g.,
$B_2 \geq B_1$. Assume $0 < \theta < 1$, and project functions $P_j(x) = \alpha_j \cdot x$ with coefficients
$\alpha_m = \alpha_{m - 1} = \ldots = \alpha_{m - k + 1} > \alpha_{m - k} \geq \alpha_{m - k - 1} \geq \ldots \geq \alpha_1$
(ordered w.l.o.g.).
Assume that no player is suppressed anywhere, and player $j$ does not
contribute to a non-steep project p. Consider player $i \neq j$.

Then, the following hold.
\begin{enumerate}
	\item \label{lemma:NE_charac_2players:move_delta:list:not_prof_to_p}
	If $\frac{1}{2}\alpha_{m} \geq \alpha_{p}$, then it is
not profitable for $i$ to move any budget $\delta > 0$ from any subset
of the steep projects to $p$ (or to a set of such projects).

	\item	\label{lemma:NE_charac_2players:move_delta:list:prof_to_steep}
	If $\frac{1}{2}\alpha_{m} > \alpha_{p}$, then it is
(strictly) profitable for $i$ to move any budget $\delta > 0$ from $p$
to any subset of the steep projects.
If $j$ is suppressed after such a move, then requiring
$\frac{1}{2}\alpha_{m} \geq \alpha_{p}$ is enough.

	\item	\label{lemma:NE_charac_2players:move_delta:list:prof_to_p}
	If $\frac{1}{2}\alpha_{m} < \alpha_{p}$ and it is
possible to move $\delta > 0$ from any subset of the  steep projects to
$p$, such that $i$ received and still receives half of the value of
these steep projects, then it is (strictly) profitable for $i$.
\end{enumerate}\label{lemma:NE_charac_2players:move_delta:list}
\end{lemma}
\begin{proof}
Before moving, player $i$'s utility is 
$\sum_{q \in \Omega}{(\inv{2} \text{or } 1)\alpha_q\cdot \paren{x^1_q+x^2_q}}$.

For part~\ref{lemma:NE_charac_2players:move_delta:list:not_prof_to_p},
assume $\frac{1}{2}\alpha_{m} \geq \alpha_{p}$.
If $i$ moves $\delta > 0$ from the steep projects to $p$, then
its utility from the steep projects decreases by at least
$0.5 \alpha_m \delta$, and its utility from $p$
increases by $\alpha_p \delta$. The total change is
$(-0.5 \alpha_m + \alpha_p) \delta$, and since
$\frac{1}{2}\alpha_{m} \geq \alpha_{p}$, this is
non-positive.

We prove part~\ref{lemma:NE_charac_2players:move_delta:list:prof_to_steep}
now. Moving $\delta$ from $p$ to a subset of the steep projects decreases
the utility of $i$ by $\alpha_p \delta$ and increases it by at least
$0.5 \alpha_m \delta$. Since $\frac{1}{2}\alpha_{m} > \alpha_{p}$,
the sum of these is (strictly positive).
If this move suppresses $j$, then the increase is more than
$0.5 \alpha_m \delta$, thus requiring
$\frac{1}{2}\alpha_{m} \geq \alpha_{p}$ is enough.

To prove part~\ref{lemma:NE_charac_2players:move_delta:list:prof_to_p},
assume that $\frac{1}{2}\alpha_{m} < \alpha_{p}$ and we can take $\delta > 0$
from some of the steep projects
where $i$ receives half of the value so as to keep receiving a half of the
new value. Then, moving this $\delta$ to $p$ decreases $i$'s utility from
the steep projects by $0.5 \alpha_m \delta$ and its utility from $p$
increases by $\alpha_p \delta$. The total change is
$(-0.5 \alpha_m + \alpha_p) \delta$, and since
$\frac{1}{2}\alpha_{m} < \alpha_{p}$, this is
(strictly) positive.
\end{proof}

And here is the proof of Theorem~\ref{the:NE_charac_2players}.

\begin{proof}
($\Rightarrow$) We prove the existence of \NE{} under the conditions of
the theorem.%
\ifthenelse{\equal{\version}{AAMAS_02_2014}}{
}{
We begin with case~\ref{NE_charac_2players:conds:close_B},
}%
supposing that $B_1 \geq k\theta B_2$ and
$\frac{1}{2}\alpha_{m - k + 1} \geq \alpha_{m - k}$.
Let both players allocate $1/k$th of their respective budgets to each of
the steep projects. We prove here that this is an \NE.
This profile provides each player with $k \cdot \inv{2}\alpha_m\cdot \frac{B_1+B_2}{k}
= \inv{2}\alpha_m\cdot (B_1+B_2)$.
For any player $i$, moving $\delta > 0$ to some non-steep projects is
not profitable, according to
part~\ref{lemma:NE_charac_2players:move_delta:list:not_prof_to_p}
of Lemma~\ref{lemma:NE_charac_2players:move_delta}.
Another possible deviation is reallocating budget among the
steep projects. Since $B_1 \geq k\theta B_2$, we conclude that
$B_2 \leq \frac{B_1}{k\theta}$, so $2$ is not able to suppress $1$
(and the other way around is clearly impossible, even more so)
and therefore, merely reallocating among the steep projects will not
increase the utility. The only deviation that remains to be considered is
simultaneously allocating $\delta > 0$ to some non-steep projects and
reallocating the rest of the budget among the steep ones. Any such
potentially profitable deviation can be looked at as
two consecutive deviations: first allocating $\delta > 0$ to
some non-steep projects, and then reallocating the rest of the budget
among the steep ones. Part~\ref{lemma:NE_charac_2players:move_delta:list:not_prof_to_p}
of Lemma~\ref{lemma:NE_charac_2players:move_delta}
shows that bringing back all $\delta > 0$ from non-steep projects to the
steep ones, without getting suppressed anywhere (which is possible since $B_1 \geq \theta B_2$)
will bear a non-negative profit. Therefore, we can
ignore the last form of deviations.
Therefore, this is an \NE.

\ifthenelse{\equal{\version}{AAMAS_02_2014}}{
}{
We now move to handle case~\ref{NE_charac_2players:conds:far_B}.
Case~\ref{NE_charac_2players:conds:far_B:no_suppress:no_inter}: suppose that
$B_1 < \frac{\theta B_2}{k}$ and $\frac{\alpha_{m - k}}{\alpha_m} \leq \min\set{\frac{1}{1 + \theta}, \frac{2\theta}{1 + \theta}}$.
Let player $1$ invest all its budget in $m - k$,
and let $2$ invest $\frac{B_2}{k}$ in each steep project.
We prove this is an \NE.
The only possibly profitable deviation for player $1$ is to invest in
steep projects. However, since $B_1 < \frac{\theta B_2}{k}$, player $1$
would obtain nothing from the steep projects.
Also player $2$ would not gain from a deviation, because
first, from our assumption,
\begin{eqnarray*}
\frac{\alpha_{m - k}}{\alpha_m} \leq \frac{1}{\theta + 1}
\iff \alpha_m (B_1 / \theta) \geq \alpha_{m - k}(B_1(1 + 1 / \theta)),
\end{eqnarray*}
and therefore, player $2$ would not profit from suppressing player $1$ at
project $m - k$.
Second, according to our assumption,
\begin{eqnarray*}
\frac{\alpha_{m - k}}{\alpha_m} \leq \frac{2\theta}{\theta + 1}
\iff \alpha_m (\theta B_1) \geq \frac{\alpha_{m - k} ((1 + \theta)B_1)}{2},
\end{eqnarray*}
and therefore, player $2$ would not profit from getting a half of the value of
project $m - k$.
Thus, no deviation is profitable.
Therefore this is an \NE.

Case~\ref{NE_charac_2players:conds:far_B:no_suppress:inter}: suppose that
$B_1 < \frac{\theta B_2}{k + \theta^2}$ and $\alpha_{m - k} \geq 2 \alpha_{m - k - 1}$ and $\frac{2\theta}{1 + \theta} \leq \frac{\alpha_{m - k}}{\alpha_m} \leq \frac{2(1 - \theta)}{2 - \theta}$ and $m - k$ is the only $2$-steep project.
Let player $1$ invest all its budget in $m - k$,
and let $2$ invest $\theta B_1$ in $m - k$
and $\frac{B_2 - \theta B_1}{k}$ in each steep project.
We prove that this is an \NE.
The possibly profitable deviations for player $1$ is to invest in
steep projects or in project $m - k - 1$. Here, we show them to be
non profitable.
First, since $B_1 < \frac{\theta B_2}{k + \theta^2} \iff B_1 < \theta \frac{B_2 - \theta B_1}{k}$,
there is no profit for $1$ from investing in a
steep project.
Second, according to our assumption,
\begin{eqnarray*}
 \alpha_{m - k} & \geq & 2 \alpha_{m - k - 1} \\
\iff \frac{\alpha_{m - k}(B_1 (1 + \theta))}{2}
& \geq & \frac{\alpha_{m - k}(B_1 (\theta^2 + \theta))}{2} + \alpha_{m - k - 1}(B_1(1 - \theta^2)),
\end{eqnarray*}
and therefore, player $1$ would not profit from investing in $m - k - 1$.
Next, we show that also player $2$ does not have incentives to deviate.
Since contributing to the non-$2$-steep projects would not increase $2$'s
utility, and since the way how the contribution is divided among
the steep projects does not influence the utility,
the possible deviations to increase player $2$'s utility are transferring
budget from the steep projects to $m - k$ or the other way around. We
show now that they are not profitable.
First, by our assumption,
\begin{eqnarray*}
\frac{\alpha_{m - k}}{\alpha_m} & \leq & \frac{2(1 - \theta)}{2 - \theta}\\
\iff \alpha_m (B_1 / \theta - \theta B_1) + \frac{\alpha_{m - k}(B_1(1 + \theta))}{2}
& \geq & \alpha_{m - k}(B_1 (1 + 1 / \theta)),
\end{eqnarray*}
and therefore, player $2$ would not profit from suppressing $1$ on project $m - k$.
Second, according to our assumption,
\begin{eqnarray*}
\frac{\alpha_{m - k}}{\alpha_m} \geq \frac{2\theta}{1 + \theta}
\iff \frac{\alpha_{m - k} (B_1(1 + \theta))}{2} \geq \alpha_m (\theta B_1),
\end{eqnarray*}
and therefore, player $2$ would not profit from moving $\theta B_1$ from $m - k$
to a steep project.
Thus, no deviation is profitable.
Therefore this is an \NE.

Case~\ref{NE_charac_2players:conds:far_B:suppress}: suppose that
$B_1 < \frac{\theta}{\abs{\Omega}} B_2$ and all the
project functions are equal. Then, player $2$ investing
$\frac{B_2}{\abs{\Omega}}$ in every project, and player $1$ using any
strategy is an \NE. To see this, notice that player $2$ obtains
$\alpha_m(B_2 + \sum_{\omega\in \Omega}{x^1_\omega})$, that is the maximum
possible utility. Player $1$ will be suppressed in any attempt to invest,
and therefore has no incentive to deviate.
Therefore this is an \NE.
}%

($\Leftarrow$)
We show the other direction now.
We assume that a given profile is an \NE{} and derive the conditions of the
theorem.
\ifthenelse{\equal{\version}{AAMAS_02_2014}}{
}{
We first suppose
that $B_1 \geq \theta B_2$ and we shall derive that the
conditions of~\ref{NE_charac_2players:conds:close_B} hold.
}%

Since $B_1 \geq \theta B_2$, then
according to Lemma~\ref{lemma:NEchar:close_B}, each player contributes to every
steep project.
Suppose to the contrary that $\frac{1}{2}\alpha_{m - k + 1} < \alpha_{m - k}$.
Let $i$ be a player who contributes to $m$ more than its threshold
there, and let $j$ be the other player.
Then, by part~\ref{lemma:NE_charac_2players:move_delta:list:prof_to_p}
of Lemma~\ref{lemma:NE_charac_2players:move_delta}, all non-steep
projects with coefficients larger than $0.5 \alpha_m$ must get a positive
contribution from $j$,
for otherwise $i$ would profit by transferring there part of its budget
from $m$. Therefore, the non-steep
projects with coefficients larger than $0.5 \alpha_m$ receive
no contribution from $i$, according to Lemma~\ref{lemma:NEchar:close_B}.

Therefore, at all the steep projects, player $j$ contributes
exactly its threshold value, while $i$ contributes above it. Also, $i$
contributes nothing to any non-steep project: we have shown this for the
non-steep
projects with coefficients larger than $0.5 \alpha_m$, now we show it for
the rest. If $i$
contributed to a non-steep project with coefficient
at most $0.5 \alpha_m$, he would benefit
from deviating to a steep one, by part~\ref{lemma:NE_charac_2players:move_delta:list:prof_to_steep}
of Lemma~\ref{lemma:NE_charac_2players:move_delta}
(when the coefficient is exactly $0.5 \alpha_m$, we use the fact that $j$
would be suppressed by such a deviation).

We assume that $B_1 \geq \theta B_2$, and thus, for any $i \neq j$ we have
\begin{eqnarray*}
\theta B_j \leq B_i
\iff B_j - \theta B_i \leq \frac{B_i}{\theta} - \theta B_i
\iff B_j - \theta B_i \leq \frac{B_i - \theta^2 B_i}{\theta}.
\end{eqnarray*}
Thus, a non-steep project with coefficients larger than
$0.5 \alpha_m$ receives from $j$ at most
$\frac{B_i - \theta^2 B_i}{\theta}$, and since $i$ can transfer to that
project $B_i - \theta^2 B_i$ without losing a share at the steep projects,
$i$ can transfer exactly $\theta$-share of $j$'s contribution there and
profit thereby,
by part~\ref{lemma:NE_charac_2players:move_delta:list:prof_to_p} of
Lemma~\ref{lemma:NE_charac_2players:move_delta} (that lemma assumes
$j$ does not contribute to those non-steep projects, but contributing
exactly the threshold to such a project is not worse than alone).
This profitable deviation contradicts our assumption and we conclude that
$\frac{1}{2}\alpha_{m} \geq \alpha_{m - k}$.

It is left to prove that $B_1 \geq k\theta B_2$.
Part~\ref{lemma:NE_charac_2players:move_delta:list:prof_to_steep}
of Lemma~\ref{lemma:NE_charac_2players:move_delta} implies
there are no contributions to non-steep projects, since they would
render the deviation to the steep projects profitable,
unless $\frac{1}{2}\alpha_{m - k + 1} = \alpha_{m - k}$, in which case
a $2$-steep project can get a positive investment from one player.
Thus, the players' utility is at most the same as when each steep
project obtains contributions from both players, and other projects
receive nothing. Thus, each player's utility is at most
$k\cdot (\alpha_m / 2)(\frac{B_1 + B_2}{k}) = (\alpha_m / 2)  (B_1 + B_2)$.
If $2$ could deviate to contribute
all $B_2$ to a steep project while suppressing $1$ there, player
$2$ would obtain $\alpha_m(B_2 + y)$,
for some $y > 0$. This is always profitable, since
\begin{eqnarray*}
B_2 \geq B_1
\Rightarrow B_2 + 2 y > B_1
\iff \alpha_m(B_2 + y) > (\alpha_m / 2)  (B_1 + B_2).
\end{eqnarray*}
Thus, since we are in an \NE, $2$ may not be able to suppress $i$ and therefore
$B_2 \leq \frac{B_1}{k} \inv{\theta}
\Rightarrow B_1 \geq k\theta B_2$.
Thus, we have proved that 
\ifthenelse{\equal{\version}{AAMAS_02_2014}}{
the conditions of the theorem hold.
}{
Conditions~\ref{NE_charac_2players:conds:close_B}
hold.

Suppose
that $B_1 < \theta B_2$ and we shall derive that
Conditions~\ref{NE_charac_2players:conds:far_B} hold.

We exhaust all the possibilities for an \NE, namely:
$1$ is suppressed,
$1$ is not suppressed and player $2$ does not contribute to non-steep projects,
and $1$ is not suppressed and player $2$ contributes to non-steep projects.
We show that each of this options entails at least one of the
sub-conditions of~\ref{NE_charac_2players:conds:far_B}.

First, assume $1$ is suppressed.\footnote{See definition~\ref{def:domin_suppress}.}
Then, $2$ invests more than $B_1 / \theta$
at each project. Therefore, $B_1 < \frac{\theta}{\abs{\Omega}} B_2$.
If not all the projects were steep, then $2$ would
profitably transfer some amount to a steep project from the non-steep
ones, while still dominating $1$ everywhere. This deviation would contradict the profile being an \NE.
Therefore, all the projects are steep and
condition~\ref{NE_charac_2players:conds:far_B:suppress}
holds.

Assume now that no player is suppressed.
Therefore, according to Lemma~\ref{lemma:NEchar:far_B:list}, player $1$
contributes only to the $2$-steep projects, and player $2$ contributes
to all the steep ones, and perhaps to a $2$-steep one as well.

First, we
assume that player $2$ does not contribute to non-steep projects and show
that it entails
condition~\ref{NE_charac_2players:conds:far_B:no_suppress:no_inter}.
Next, we assume that
$2$ does contribute to non-steep projects and show
that this entails
condition~\ref{NE_charac_2players:conds:far_B:no_suppress:inter}.

First, assume that player $2$ does not contribute to non-steep projects.
Since
player $1$ does not prefer to deviate by contributing exactly the threshold
to a steep project, $B_1 < \theta \frac{B_2}{k}$ is true.
In an \NE, player $2$ would not profit from suppressing player $1$ at
a $2$-steep project, and therefore

\begin{eqnarray*}
\alpha_m (B_1 / \theta) \geq \alpha_{m - k}(B_1(1 + 1 / \theta))
\iff \frac{\alpha_{m - k}}{\alpha_m} \leq \frac{1}{\theta + 1}.
\end{eqnarray*}

In addition, in an \NE, player $2$ would not profit from contributing exactly the
threshold at a $2$-steep project, and therefore
\begin{eqnarray*}
\alpha_m (\theta B_1) \geq \frac{\alpha_{m - k} (1 + \theta) B_1}{2}
\iff \frac{\alpha_{m - k}}{\alpha_m} \leq \frac{2\theta}{1 + \theta}.
\end{eqnarray*}
Thus, we have proved that
condition~\ref{NE_charac_2players:conds:far_B:no_suppress:no_inter} holds.

Assume now player $2$ contributes to non-steep projects.
By Lemma~\ref{lemma:NEchar:far_B},
$m - k$ is the single $2$-steep project where player $1$ contributes
all $B_1$, while player $2$ contributes $\theta B_1$ there, and he
splits the rest of his budget among all the steep projects,
yielding a positive contribution to each such project.

Assume that a steep project receives $y > 0$ (from player $2$, of course).
If player $1$ could achieve the threshold $\theta y$, it would
deviate, for the following reasons. We have
$\alpha_m (B_1 + y) / 2 > \alpha_{m - k}(B_1 + \theta B_1) / 2$,
unless, perhaps, if $y < \theta B_1$. In such a case, however, $1$
can suppress player $2$ and obtain $\alpha_m (B_1 + y)$, which is
larger than $\alpha_{m - k} (B_1 (1 + \theta)) / 2$.
Consequently, from the profile being an \NE, we conclude that $1$ is not able to achieve the threshold
$\theta y$, and therefore
\begin{eqnarray*}
B_1 < \theta \frac{B_2 - \theta B_1}{k}
\iff B_1 < \frac{\theta B_2}{k + \theta^2}.
\end{eqnarray*}
In addition, since player $1$ does not prefer to contribute to $m - k$
only the threshold $\theta^2 B_1$ and move the rest to $m - k - 1$,
it must hold that
\begin{eqnarray*}
\frac{\alpha_{m - k} (B_1 (1 + \theta))}{2}
& \geq & \frac{\alpha_{m - k} (B_1 (\theta^2 + \theta))}{2} + \alpha_{m - k - 1} (B_1(1 - \theta^2))\\
\iff \alpha_{m - k} & \geq & 2 \alpha_{m - k - 1}.
\end{eqnarray*}
Since player~$2$ does not want to suppress $1$ at $m - k$, we conclude
that
\begin{eqnarray*}
\alpha_m ((B_1) / \theta - \theta B_1) + \frac{\alpha_{m - k} (B_1 (1 + \theta))}{2}
& \geq & \alpha_{m - k}(B_1 (1 + 1 / \theta))\\
\iff \frac{\alpha_{m - k}}{\alpha_m} & \leq & \frac{2 (1 - \theta)}{2 - \theta}.
\end{eqnarray*}
Finally, since player $2$ does not prefer moving $\theta B_1$ to
a steep project over leaving it at $m - k$, it holds that
\begin{eqnarray*}
\frac{\alpha_{m - k (B_1 (1 + \theta))}}{2} \geq \alpha_m (\theta B_1)
\iff \frac{\alpha_{m - k}}{\alpha_m} \geq \frac{2\theta}{1 + \theta}.
\end{eqnarray*}
Therefore,
condition~\ref{NE_charac_2players:conds:far_B:no_suppress:inter} holds.
To conclude, at least one of the
sub-conditions of~\ref{NE_charac_2players:conds:far_B} holds, 
finalizing the proof of the other direction of the theorem.
}
\end{proof}

We prove Theorem~\ref{the:NE_effic_2players} now.

\begin{proof}%
\ifthenelse{\equal{\version}{AAMAS_02_2014}}{
According to the proof of
Theorem~\ref{the:NE_charac_2players},
equally dividing all the budgets
among the steep projects is an \NE. Therefore, $\pos = 1$.
}
{
We first  prove case~\ref{NE_effic_2players:conds:close_B}.
}%
Consider any \NE. By Lemma~\ref{lemma:NEchar:close_B}, each player
contributes to all the steep projects and if it contributes to a non-steep
project, then it is the only contributor there.
Take any non-steep project~$p$, where someone, say player~$i$,
contributes a positive amount $y$.
If $\alpha_p < 0.5 \alpha_m$,
consider moving all what player $i$ contributes to $p$ to a steep project.
According to part~\ref{lemma:NE_charac_2players:move_delta:list:prof_to_steep}
of Lemma~\ref{lemma:NE_charac_2players:move_delta},
this move is profitable, contradicting the assumption of an equilibrium.
Therefore, $\alpha_p = 0.5 \alpha_m$.
If the other player~$j$ could move $\theta y$ to project~$p$ from a
steep project without losing its half at the steep project, then $j$ would
strictly profit from this, because
\begin{eqnarray*}
\frac{\alpha_m}{2} (\theta y) < \frac{1}{2} \frac{\alpha}{2} ((\theta + 1) y)
\iff \theta < \frac{1}{2} (\theta + 1)
\iff \theta < 1,
\end{eqnarray*}
which is always the case. Since we assume an \NE, this cannot happen,
implying that at any steep project, the part above the minimum required
to get its half is strictly less than $\theta y$. Therefore, $i$ can
move $y$ to a steep project and suppress $j$ there.
Then,
according to part~\ref{lemma:NE_charac_2players:move_delta:list:prof_to_steep}
of Lemma~\ref{lemma:NE_charac_2players:move_delta},
this move is profitable, contradicting the \NE. Therefore, there
can be no contribution to $p$. This means that only the steep projects
obtain contributions,
and therefore, $\poa = 1$.
We have fully proven%
\ifthenelse{\equal{\version}{AAMAS_02_2014}}{
the theorem.}
{
case~\ref{NE_effic_2players:conds:close_B}.

Consider case~\ref{NE_effic_2players:conds:far_B:no_suppress:no_inter}
now. From the proof of the existence on an \NE{} in this case, we know
that $1$ investing all its budget in $m - k$ and $2$ dividing its budget
equally among the steep projects constitute an \NE. Thus,
$\pos \geq \frac{\alpha_m B_2 + \alpha_{m - k} B_1}{\alpha_m (B_1 + B_2)}$.
Since $\frac{\alpha_{m - k}}{\alpha_m} \leq \frac{1}{1 + \theta}$,
not all the projects have equal value functions. Therefore, in an \NE{}
no player is suppressed, since if $2$ dominated $1$, then $2$ would have
to invest more than $B_1 / \theta$ in each project, and $2$ would like to
deviate to contribute to the steep projects more. Since no player is
suppressed, we conclude from part~\ref{lemma:NEchar:far_B:list:contr_1} of Lemma~\ref{lemma:NEchar:far_B} that player
$1$ never contributes to a steep project in an \NE, and thus
$\pos = \frac{\alpha_m B_2 + \alpha_{m - k} B_1}{\alpha_m (B_1 + B_2)}$.

Next, let us approach the price of anarchy. According to
Lemma~\ref{lemma:NEchar:far_B}, the only way to reduce the efficiency
relatively to the price of stability is for player~$2$ to invest a $2$-steep
project. If this happens, then we obtain that
case~\ref{NE_effic_2players:conds:far_B:no_suppress:inter} must hold,
exactly as it is done in the proof of the other direction of
Theorem~\ref{the:NE_charac_2players}. Therefore, if this case does not
hold, then $\pos = \pos$. If it does, then we have the \NE{} when $2$
invests $\theta B_1$ in project $m - k$
(and $1$ invests all its budget there, and $2$ equally divides the rest
of its budget among the steep projects), which yields
the price of anarchy of
$\frac{\alpha_m (B_2 - \theta B_1) + \alpha_{m - k} (B_1 (1 + \theta))}{\alpha_m (B_1 + B_2)}$.

We prove case~\ref{NE_effic_2players:conds:far_B:no_suppress:inter} now.
We show in the proof of
case~\ref{NE_charac_2players:conds:far_B:no_suppress:inter} of
Theorem~\ref{the:NE_charac_2players} that player $1$ investing all its budget
in $m - k$ and $2$ investing $\theta  B_1$ in $m - k$ and uniformly
dividing the rest among the steep projects is an \NE. Thus,
$\pos \geq \frac{\alpha_m (B_2 - \theta B_1) + \alpha_{m - k} (B_1 (1 + \theta))}{\alpha_m (B_1 + B_2)}$.
Since $\frac{\alpha_{m - k}}{\alpha_m} \leq \frac{2(1 - \theta)}{2 - \theta}$,
we conclude analogously to what we did in the proof of the previous case
that no player is suppressed. Thus, Lemma~\ref{lemma:NEchar:far_B}
implies that the only way to achieve a more efficient \NE{} is for $2$
to contribute only to the steep projects, while player~$1$ contributes only to the two-steep projects. If this is an \NE, then we
obtain that case~\ref{NE_effic_2players:conds:far_B:no_suppress:no_inter}
must hold, exactly as it is done in the proof of the other direction of
Theorem~\ref{the:NE_charac_2players}. Therefore, if this does not hold,
we have
$\pos = \frac{\alpha_m (B_2 - \theta B_1) + \alpha_{m - k} (B_1 (1 + \theta))}{\alpha_m (B_1 + B_2)}$.
If case~\ref{NE_effic_2players:conds:far_B:no_suppress:no_inter}
does hold, then we know that the profile where $1$ invests all its budget
in $m - k$ and $2$ divides its budget among the steep projects is an \NE,
and thus
$\pos = \frac{\alpha_m B_2 + \alpha_{m - k} B_1}{\alpha_m (B_1 + B_2)}$.

We turn to the price of anarchy now. According to
Lemma~\ref{lemma:NEchar:far_B}, the \NE{} with
player $1$ investing all its budget
in $m - k$ and $2$ investing $\theta  B_1$ in $m - k$ and uniformly
dividing the rest of $b_2$ among the steep projects is the worst possible \NE,
and thus
$\poa = \frac{\alpha_m (B_2 - \theta B_1) + \alpha_{m - k} (B_1 (1 + \theta))}{\alpha_m (B_1 + B_2)}$.

Finally, in the case~\ref{NE_effic_2players:conds:far_B:suppress} we know
that $2$ dividing its budget equally and $1$ contributing all its budget
is an \NE, and therefore $\pos = 1$.
To find the price of anarchy, recall that if $2$ does as before while $1$
invests nothing at all, it still is an \NE, and thus $\poa \leq \frac{\alpha_m B_2}{\alpha_m (B_1 + B_2)}$.
Since $2$ always gets at least $\alpha_m B_2$ in any \NE, the price of
anarchy cannot decrease below it, and thus
$\poa = \frac{\alpha_m B_2}{\alpha_m (B_1 + B_2)}$.
}
\end{proof}

Let us now prove Corollary~\ref{cor:NE_effic_2players_tight_bound}.

\begin{proof}
The maxima are attained in case~\ref{NE_effic_2players:conds:close_B}
of Theorem~\ref{the:NE_effic_2players}.

To find the infima, find the infimum in every case, substituting the
extreme values in the expressions for $\pos$ and $\poa$. We begin
with the $\pos$.
In case~\ref{NE_effic_2players:conds:far_B:no_suppress:no_inter},
the infimum of the $\pos$ is $\frac{k}{k + \theta}$, attained for $\alpha_{m - k} = 0$ and
$B_1 = \frac{\theta B_2}{k}$.
In case~\ref{NE_effic_2players:conds:far_B:no_suppress:inter},
the infimum of the $\pos$ is the minimum of these two expressions
\begin{enumerate}
\item	$\frac{\alpha_m B_2 + \alpha_{m - k} B_1}{\alpha_m (B_1 + B_2)}$
when $\alpha_{m -k} = \frac{2 \theta}{1 + \theta} \alpha_m$ and
$B_1 = \frac{\theta B_2}{k + \theta^2}$, which is
$\frac{k + \theta^2 + \frac{2 \theta^2}{1 + \theta}}{k + \theta + \theta^2}$.

\item	$\frac{\alpha_m (B_2 - \theta B_1) + \alpha_{m - k} (B_1 (1 + \theta))}{\alpha_m (B_1 + B_2)}$
when $\alpha_{m - k} = \frac{2 \theta}{1 + \theta} \alpha_m$ and
$B_1 = \frac{\theta B_2}{k + \theta^2}$, which is
$\frac{k + 2 \theta^2}{k + \theta + \theta^2}$.
\end{enumerate}
The minimum of these expressions is
$\frac{k + 2 \theta^2}{k + \theta + \theta^2}$.
Finally, the infimum of the price of stability in
case~\ref{NE_effic_2players:conds:far_B:suppress} is $1$.
The absolute infimum is the minimum of these three expressions, which
is $\frac{k}{k + \theta}$.

We consider now the infimum of the  price of anarchy.
In case~\ref{NE_effic_2players:conds:far_B:no_suppress:no_inter},
the infimum of the $\poa$ is attained as follows:
\begin{enumerate}
\item If also case~\ref{NE_effic_2players:conds:far_B:no_suppress:inter}
holds, then it is the value of
$\frac{\alpha_m (B_2 - \theta B_1) + \alpha_{m - k} (B_1 (1 + \theta))}{\alpha_m (B_1 + B_2)}$
when $\alpha_{m - k} = \frac{2 \theta}{1 + \theta} \alpha_m$ and
$B_1 = \frac{\theta B_2}{k + \theta^2}$, which is
$\frac{k + 2 \theta^2}{k + \theta + \theta^2}$.

\item	Otherwise, $\poa = \pos$, and so the infimum is $\frac{k}{k + \theta}$.
\end{enumerate}
The minimum of these two expressions is $\frac{k}{k + \theta}$.
In case~\ref{NE_effic_2players:conds:far_B:no_suppress:inter}, the infimum
of the $\poa$ is $\frac{\alpha_m (B_2 - \theta B_1) + \alpha_{m - k} (B_1 (1 + \theta))}{\alpha_m (B_1 + B_2)}$
when $\alpha_{m - k} = \frac{2 \theta}{1 + \theta} \alpha_m$ and
$B_1 = \frac{\theta B_2}{k + \theta^2}$, which is
$\frac{k + 2 \theta^2}{k + \theta + \theta^2}$.
In case~\ref{NE_effic_2players:conds:far_B:suppress}, it is
$\frac{B_2}{B_1 + B_2}$ when $B_1 = \frac{\theta}{m}B_2$, which is
$\frac{m}{m + \theta}$.
Therefore, the infimum of the price of anarchy is $\frac{k}{k + \theta}$.
\end{proof}

We now present the proof of Theorem~\ref{the:NE_effic_nplayers}.

\begin{proof}%
\ifthenelse{\equal{\version}{AAMAS_02_2014}}{
According to proof of
Theorem~\ref{the:NE_charac_nplayers},
equally dividing all the budgets
among the steep projects is an \NE. Therefore, $\pos = 1$.}
{%
We first prove case~\ref{NE_effic_nplayers:conds:close_B}.
According to proof of case~\ref{NE_charac_nplayers:conds:close_B} in
Theorem~\ref{the:NE_charac_nplayers},
equally dividing all the budgets
among the steep projects is an \NE. Therefore, $\pos = 1$.

In order to bound the price of anarchy, we notice that since player~$n$
can always obtain at least
$$\frac{\alpha_m B_n (1 + (n - 1)\theta)}{n}$$ by investing all her budget
in a steep project, this is a lower bound on what she obtains in any \NE.
Since $B_{n - 1} \geq \theta B_n$, we analogically conclude that
player~$n - 1$ receives at least
$$\frac{\alpha_m B_{n - 1} (1 + (n - 1)\theta)}{n}$$ in any Nash equilibrium.
Therefore, the social welfare in an \NE{} is at least
$$\frac{\alpha_m (B_{n - 1} + B_n) (1 + (n - 1)\theta)}{n},$$ which,
in turn, implies the lower bound.

We consider case~\ref{NE_effic_nplayers:conds:far_B:suppress} now.
The lower bound on the price of anarchy stems from the fact that
player~$n$ always receives at least $\alpha_m B_n$ in any \NE, by
investing $B_n$ in a steep project.

Assume now that $B_{n - 1} < \frac{\theta}{\abs{\Omega}} B_n$ and all the
	project functions are equal. We know
that $n$ dividing its budget equally and all the other players contributing all their budgets
is an \NE, and therefore $\pos = 1$.
To find the price of anarchy, recall that if $n$ does as before while all the other players
invest nothing at all, it still is an \NE, and thus $\poa \leq \frac{\alpha_m B_n}{\alpha_m (\sum_{i\in \set{1, 2, \ldots, n}}{B_i})}$.
Since $n$ always gets at least $\alpha_m B_n$ in any \NE, the price of
anarchy cannot decrease below it, and thus
$\poa = \frac{\alpha_m B_n}{\alpha_m (\sum_{i\in \set{1, 2, \ldots, n}}{B_i})}$.
}%
\end{proof}

We now prove Corollary~\ref{cor:NE_effic_n_smooth_simpler}.

\begin{proof}
Theorem~\ref{the:NE_effic_n_smooth} implies
the lower bound of $$\frac{1 + (n - 1)\theta}{n} \frac{\sum_{i = l}^{n - 1}{B_i}}{\sum_{i = 1}^{n}{B_i}}
	+ \frac{1 + (n - l) \theta}{n - l + 1} \frac{B_n}{\sum_{i = 1}^{n}{B_i}}$$
	on the price of anarchy. Since $\theta \leq 1$, this is at least
\begin{eqnarray*}
\frac{\theta + (n - 1)\theta}{n} \frac{\sum_{i = l}^{n - 1}{B_i}}{\sum_{i = 1}^{n}{B_i}}
	+ \frac{\theta + (n - l) \theta}{n - l + 1} \frac{B_n}{\sum_{i = 1}^{n}{B_i}}
	= \theta \frac{\sum_{i = l}^{n - 1}{B_i}}{\sum_{i = 1}^{n}{B_i}}
	+ \theta \frac{B_n}{\sum_{i = 1}^{n}{B_i}}
	= \theta \frac{\sum_{i = l}^{n}{B_i}}{\sum_{i = 1}^{n}{B_i}},
\end{eqnarray*}
providing the lower bound.
\end{proof}

We shall now prove Theorem~\ref{the:mix_NE_exist}.
To remain self-contained, before proving the theorem, we bring here the necessary definitions used by%
~\cite{DasguptaMaskin1986}.
Given player~$i$ with the strategy set $A_i \subseteq \reals^m$, define
$A \defas A_1 \times \ldots \times A_n$.
\begin{defin}
For each pair of players $i, j \in {1, \ldots, n}$, let $D(i)$ be
a positive natural, and for a $d \in \set{1, \ldots, D(i)}$, let
$f^d_{i, j} \colon \reals \to \reals$ be continuous, such that
$(f^d_{i, j})^{-1} = f^d_{j, i}$. For every player~$i$, we define
\begin{eqnarray}
A^*(i) \defas
\{(a_1, \ldots, a_n) \in A \mid \exists j \neq i, \exists k \in \set{1, \ldots, m}, \exists d \in \set{1, \ldots, D(i)}, \nonumber\\
\text{ such that } a_{j, k} = f^d_{i, j}(a_{i, k})\}.
\label{eq:A_star_i}
\end{eqnarray}
\end{defin}

We now define weakly lower semi-continuity, which intuitively means that
there is a set of directions, such that approaching a point from any of
these directions gives values at least equal to the function at the point. 
\begin{defin}
Let $B^m \defas \set{z \in \reals^m \mid \sum_{i = 1}^m{ {z_l}^2} = 1}$,
i.e.\ the surface of the unit sphere centered at zero. Let
$e \in B^m$ and $\theta > 0$. Function $g_i(a_i, a_{-i})$ is
\defined{weakly lower semi-continuous in the coordinates of $a_i$} if for all
$\hat{a_i}$ there exists an absolutely continuous measure $\nu$ on $B^m$,
such that for all $a_{-i}$, we have
\begin{equation*}
\int_{B^m} \set{\liminf_{\theta \to 0} g_i(\hat{a_i} + \theta e, a_{-i}) d\nu(e)} \geq g_i(\hat{a_i}, a_{-i}).
\end{equation*}
\end{defin}

Finally, we are ready to prove Theorem~\ref{the:mix_NE_exist}.

\begin{proof}
We show now that all the conditions of Theorem~$5^*$
from \cite{DasguptaMaskin1986} hold. First, the strategy set
of player~$i$ is simplex, and as such, it is non-empty, convex and
compact. The utility function of player~$i$ is discontinuous only
at a threshold of one of the projects. These points belong to the set
$A^*(i)$, defined in \formsref{eq:A_star_i}, if we take
\begin{eqnarray*}
D(i) \defas 2;\\
f^1_{i,j}(y) \defas y
\begin{cases}
\theta & \text{if } i < j, \\
1 / \theta & \text{if } i > j;
\end{cases}\\
f^2_{i,j}(y) \defas y
\begin{cases}
\theta & \text{if } j < i, \\
1 / \theta & \text{if } j > i.
\end{cases}
\end{eqnarray*}
The sum of all the utilities is a continuous function.
In addition, the utility of player~$i$ is bounded by the
largest project's value when all the players contribute their budgets
there. It is also weakly lower semi-continuous in $i$'s contribution,
since if we take the measure $\nu$ to be
$\nu(S) \defas \lambda(S \cap B^{m}+)$, where $\lambda$ is
the Lebesgue measure on $B^m$ and
$B^m+ \defas \set{z \in B^m \mid z_l \geq 0, \forall l = 1, \ldots, m}$,
we obtain an absolutely continuous measure $\nu$, such that the
integral sums up only the convergences to a point from the positive
directions, and such convergences will never become less than the
function at the point. 

Finally, Theorem~$5^*$ implies
that a mixed \NE{} exists.
\end{proof}

%% file: Simulation.tex
\section{Simulations}\label{Sec:Sim}

The
theory provides only sufficiency results for shared effort games with
more than $2$ players. 
To better explore the existence and efficiency of \NE{} in these games, we simulate a variation of
fictitious play~\cite{Robinson1951}; intuitively, we try fictitious play because of its well-known
properties of converging  to an \NE{} in some finite cases (see~\sectnref{fict_play}).
The main value of this section is in the methodology; the concrete simulations
are best done based on the studied case at hand.
%

%
We first adapt the fictitious play~\cite{Brown1951} to our infinite game.
Danskin~\cite{Danskin1981} defines
the best response to maximise the average utility against all the previous
strategies of the other players, while we, as well as the original
fictitious play, best respond to the averaged (cumulative) strategy of the others.

%

Aspiring to implement the adapted
fictitious play,
we then suggest an algorithm for finding a best response, if it exists, for the
case of two projects. For more projects, we prove that even best responding
is already \NPH.

\subsection{Infinite-Strategy Fictitious Play for Shared Effort Games}

In fictitious play, a player best responds to the averaged strategies
of the other players.
Since the game has convex strategy spaces (simplexes), we do not need mixing to average the
strategies. Each player's strategy is averaged separately, and each player best responds to
the product distribution of the other players.
Denote the set of all the best responses of
player $i$ to profile $x^{-i}$ of the others by
$\BR(x^{-i})$.
\begin{defin}\label{def:ISFP}
Given a shared effort game with players $N$, budget-defined strategies
$S^i = \set{x^i = (x^i_\omega)_{\omega \in \Omega} \in \reals_{+}^{\abs{\Omega}} \mid \sum_{\omega \in \Omega}{x^i_\omega} \leq B_i}$
and utilities
$u^i(x) \defas \sum_{\omega \in \Omega}{\phi^i_\omega(x_\omega)}$,
define an \defined{Infinite-Strategy Fictitious Play (\ISFP)} as
the following set of sequences. Consider a (pure) strategy in this game at time~$1$,
i.e.~%
$(x^i(1))^{i\in N} = ((x^i(1)_\omega)_{\omega \in \Omega})^{i\in N}$,
define $X^i(1) \defas \set{x^i(1)}$,
and define recursively, for each $i\in N$ and $t \geq 0$, the set
of the possible strategic histories at time~$t + 1$ to be all the possible
combinations of the history with the best responses to the previous strategic histories:
\begin{eqnarray}
X^{i}(t+1) \defas \set{\frac{t x^i(t) + \br(x^{-i}(t))}{t + 1} \mid x^i(t) \in X^i(t), \br(x^{-i}(t)) \in \BR(x^{-i}(t))}.
\end{eqnarray}

We say that an \ISFP{} \defined{converges} to $x^* \in \realsP^n$
if at least one of its sequences converges to $x^*$ in every coordinate.
Please note that if at a given time the sequence plays an \NE, it can stay there forever.
\end{defin}
Since $\BR(x^{-i}(t))$ is a set, there may be multiple \ISFP{} sequences.
For an \ISFP{} to be defined, we need that $\BR(x^{-i}(t)) \neq \emptyset$, meaning that the
utility functions attain a maximum. 
However, the functions are, generally speaking, not upper semi-continuous,
and may sometimes not attain a maximum. 

In \ISFP, all the plays have equal weights in the averaging. In the other extreme, a
player just best-responds to the previous strategy profile of other
players, thereby attributing the last play with the weight of $1$ and
all the other plays with $0$. In general, we define, for an $\alpha \in [0, \infty]$, an
\defined{$\alpha-\ISFP$} play as in Definition~\ref{def:ISFP},
but with the following formula instead\footnote{For $\alpha = \infty$, we just obtain a constant sequence.}
\begin{eqnarray}
X^{i}(t+1) \defas \set{\frac{\alpha t x^i(t) + \br(x^{-i}(t))}{\alpha t + 1} \mid x^i(t) \in X^i(t), \br(x^{-i}(t)) \in \BR(x^{-i}(t))}.
\end{eqnarray}
Here, the last play's weight is~$\frac{1}{\alpha t + 1}$.

We do not know whether and when any convergence property
can be proven for the generalised fictitious play in shared effort games.
In simulations, our generalised fictitious play
often converges to an \NE, 
bolstering the importance of
equilibrium efficiency, as stated by
Roughgarden and Tardos~\cite[\chapt{17}]{Nisanetal07}.

Next, we solve the algorithmic problem of finding whether a best response
exists, and if it does, what it is.
\subsection{Best Response} 

First, we show that finding a best response is \NPH, even when the game
has only two players, and then,
we do find it in polynomial time for a $2$-project game.
\begin{theorem}\label{The:BR_2_player_nph}
Best-responding (if a best response exists and otherwise returning that none exists) in
shared effort games with $\theta$-equal
sharing is \NPH{} already for two players, even if all the project
functions have the same coefficients.
\end{theorem}
\begin{proof}
We reduce the following \NPH{}
problem~\cite[\sectn{1.3}]{MartelloToth1990} to best-responding by,
w.l.o.g., player~$1$.
\begin{defin}
The \defined{subset sum} problem receives items $\set{1, \ldots, n}$ of
sizes $s_1, \ldots, s_n$ and a cap $C$, and returns a subset $S$ of
$\set{s_1, \ldots, s_n}$ such that $\sum_{i \in S}{s_i}$ is maximum
possible that is at most $C$.
\end{defin}

The reduction proceeds as follows.
Let $B_1$ be $C$. For each item~$i$, create project~$i$, where
player~$2$ contributes $s_i / \theta$, where $\theta$ is set small
enough so that player $1$ will not be able to suppress $2$, i.e.,
$C \leq s_i / (\theta^2)$, for each $i$.
First, the inability to suppress makes player $1$'s utility 
upper semi-continuous, so a best response exists.
All the equal project coefficients $\alpha$ are such
that $\alpha C < s_i, \forall i = 1, \ldots, n$,
so that it is always better to achieve the threshold at yet another
project than to contribute more to a project where the threshold is
already achieved.
We finally require that $\alpha (1 + \theta) = 2 \theta$, which
can be achieved by making $\alpha$ and/or $\theta$ smaller.

Fix any best response in the obtained game.
The projects where $1$ contributes above the threshold constitute
an optimal solution to the
subset sum problem, because player~$1$ needs to provide at least
$\theta (s_i / \theta) = s_i$ in order to be above the threshold
at project~$i$, and such a contribution provides it with the utility of
$\frac{\alpha (s_i / \theta + s_i)}{2}$, which is, because of
the requirement $\alpha (1 + \theta) = 2 \theta$, equal to $s_i$.
The equivalence is completed by the fact that contributing more to
such a project is less important than achieving the threshold at another
project, because $\alpha C < s_i$.
\end{proof}

In order to execute the infinite-strategy fictitious play, 
from now on we assume that the game has only
two projects, $\Omega = \set{\psi, \omega}$.
We would like to find a best response for a player $i \in N$, all the other players' strategies
$x^{-i} \in S^{-i}$ being fixed. From the weak
monotonicity of the sharing functions $\phi^i_\omega$,
we may assume w.l.o.g.\ that a best responding player contributes all her budget.
Then, a strategy is uniquely determined by the contribution to project
$\psi$ and we shall write $x^i$ for $x^i_\psi$, meaning that
$x^i_\omega = B_i - x^i$.

We now state the conditions for the following theorem. Consider $\thetaEq$ sharing
and convex project value functions. 
Let $D^i_0 < D^i_1 < \ldots < D^i_m$
and $W^i_0 < W^i_1 < \ldots < W^i_l$
be the jumps of $\phi^i_{\psi}$ and $\phi^i_{\omega}$ respectively.
(The first points in each list are the minimum contributions to projects $\psi, \omega$,
respectively, required for $i$ to obtain a share.
The other points are the points at which another
player becomes suppressed at the respective project.) The possible discontinuity
points of the total utility of $i$ are thus
$D^i_0 < D^i_1 < \ldots < D^i_m$
and $B_i - W^i_l < \ldots <B_i - W^i_1 < B_i - W^i_0$. Denote the
distinct points of these lists merged in the increasing order by $L$.
%
%
Let $L_{B_i}$ denote the points of the list $L$ that are on $[0, B_i]$,
together with $0$ and $B_i$,
and let $M_{B_i}$ be $L_{B_i}$ with an arbitrary point added between
each two consecutive points.
\begin{theorem}\label{The:BR_2_proj_part_conv_weak_mon}
The maximum of the one-sided limits at
the points of $L_{B_i}$ and of the values at the points of $M_{B_i}$
yields the utility supremum\footnote{The supremum is the exact upper bound; it always exists.} of the responses of player $i$.
This supremum is a maximum (and in particular, a best response exists)
if and only if it is achieved at a point of $M_{B_i}$.
\end{theorem}
\begin{proof}
The utility of
$i$ is $u^i(x^i) = \phi^i_\psi(x^i) + \phi^i_\omega(B_i - x^i)$.
Consider the open intervals between the consecutive points of $L_{B_i}$.
On each of these intervals,
the function $\phi^i_\psi(x^i)$ is convex, being proportional to the
convex project value function,
and $\phi^i_\omega(B_i - x^i)$ is convex because the function $B_i - x$
is convex and concave and $\phi^i_\omega$ is convex and weakly monotone.
Therefore, the utility is also convex, as the sum of convex functions.

Therefore, the supremum of the utility on the closure of such a convexity interval is attained
as the one-sided limit of at least one of its edge points. This supremum
can be a maximum if and only if it is not larger than the maximum of the
utility at an interval edge point or at an internal point of an interval
(in the last case, the convexity implies that the utility is constant on this interval).
\end{proof}

When finding the one-sided limit at a point of $L_{B_i}$ takes constant
time, the resultant algorithm runs in $O(n \log n)$ time
and in linear space. We employ this algorithm in the simulations.

\subsection{The Simulation Method}

We consider the $\theta$-equal $2$-project case,
where 
Theorem~\ref{The:BR_2_proj_part_conv_weak_mon} provides an algorithm to best respond.
For each of the considered shared effort games, we run several $\alpha-\ISFP$s,
for several $\alpha$s.
If at least once in the
simulation process no best response exists,
we drop this attempt. 
Otherwise, we stop after
a predefined number of iterations ($50$), or if an \NE{} has been found. 
%

We choose an initial belief state about all the players and run the $\ISFP$ from this state on,
updating this common belief state at each step by finding a best response of
each player to the current belief state and averaging it with the history. 
To increase the chances of finding an existing \NE,
for each game, we generate $45$ fictitious plays by randomly and independently
generating the initial belief state on each player's actions, uniformly over the possible
histories (this number was experimentally found to be a good balance 
between run time and precision). 
While simulating, when a player has multiple
best responses, we choose a closest one to the current belief state of the
fictitious play, in the sense of minimising the maximum distance from
the last action's components.
For each found \NE, we calculate its efficiency
by dividing its total utility by the optimum possible total utility
and plot it using shades of gray. When no \NE{} is found, we
plot it black.

\subsection{Results and Conclusions}\label{Sec:Sim:res_conclud}


\begin{figure*}[h!tbp]
\centering
\subfloat
{
\includegraphics[trim = 10mm 5mm 0mm 10mm, clip=true,width=0.35\textwidth]{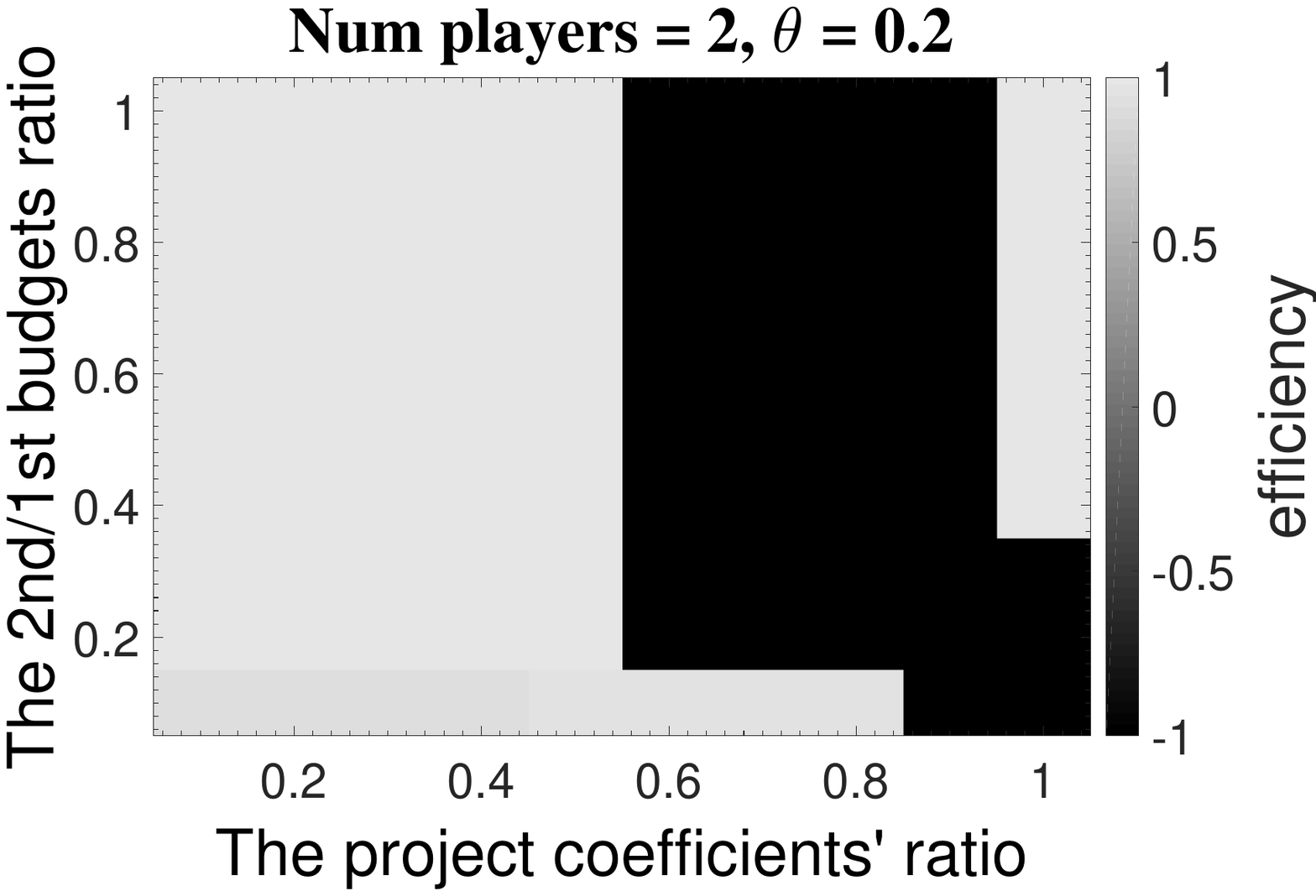}
}
\subfloat
{
\includegraphics[trim = 10mm 5mm 0mm 10mm, clip=true,width=0.35\textwidth]{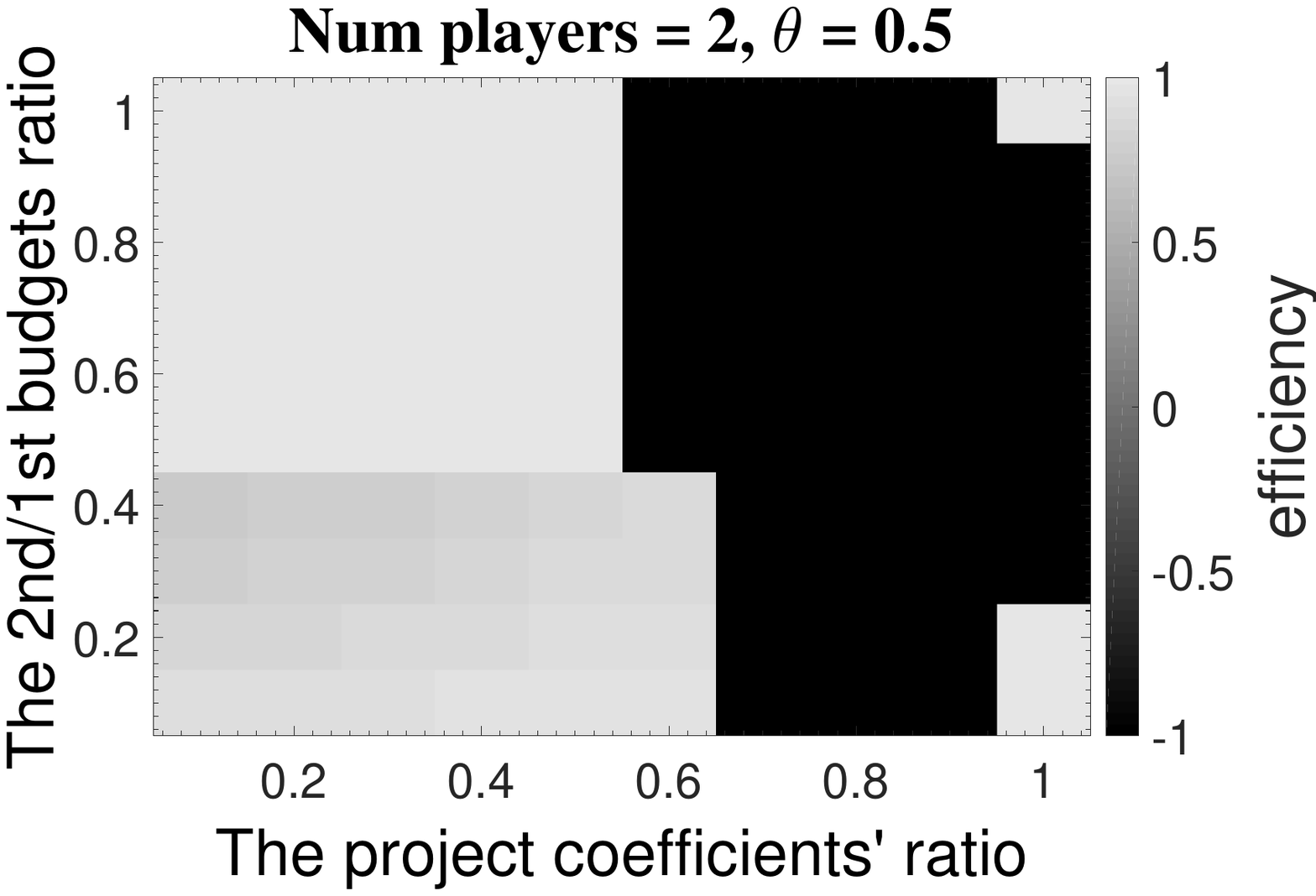}
}
\subfloat
{
\includegraphics[trim = 10mm 5mm 0mm 10mm, clip=true,width=0.35\textwidth]{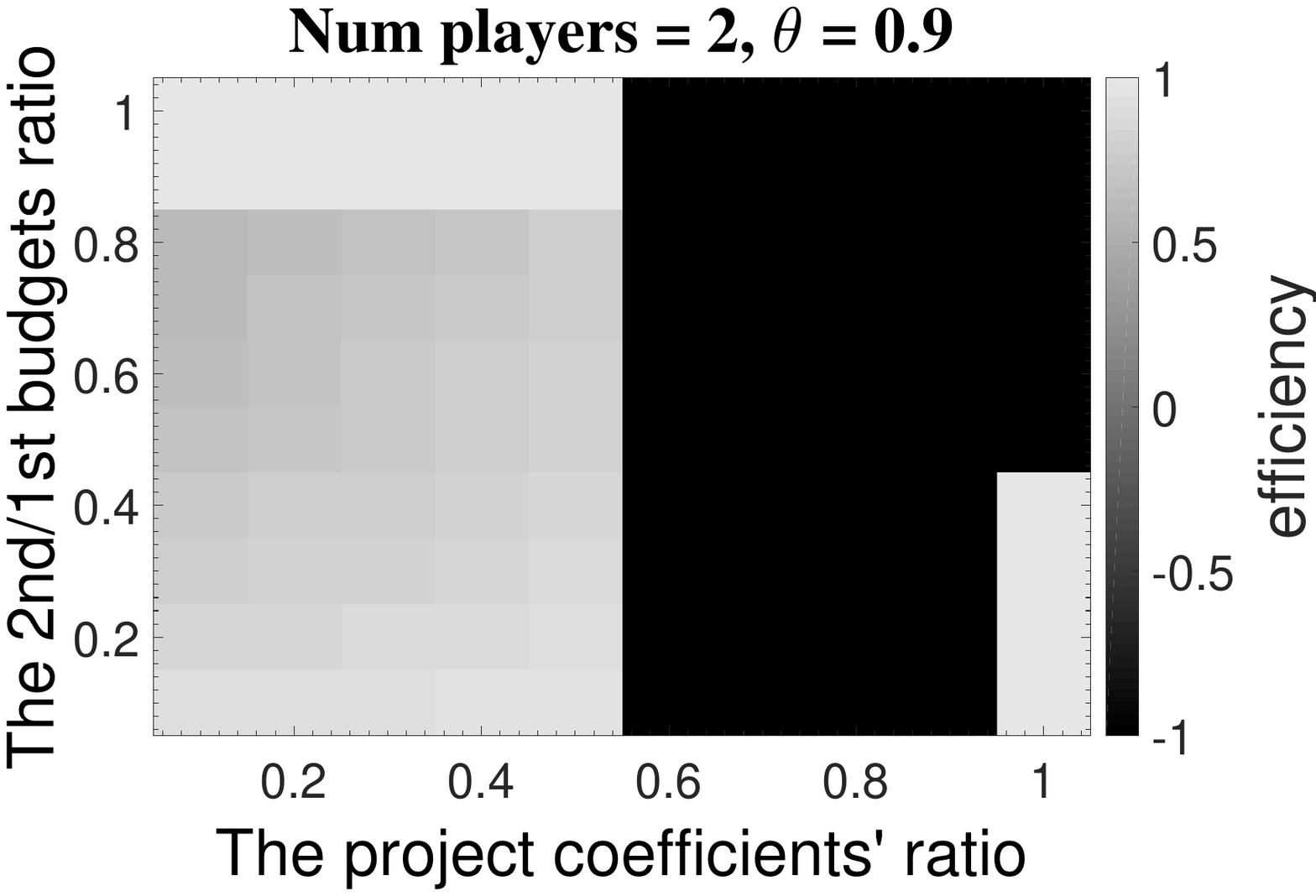}
}

\subfloat
{
\includegraphics[trim = 10mm 5mm 0mm 10mm, clip=true,width=0.35\textwidth]{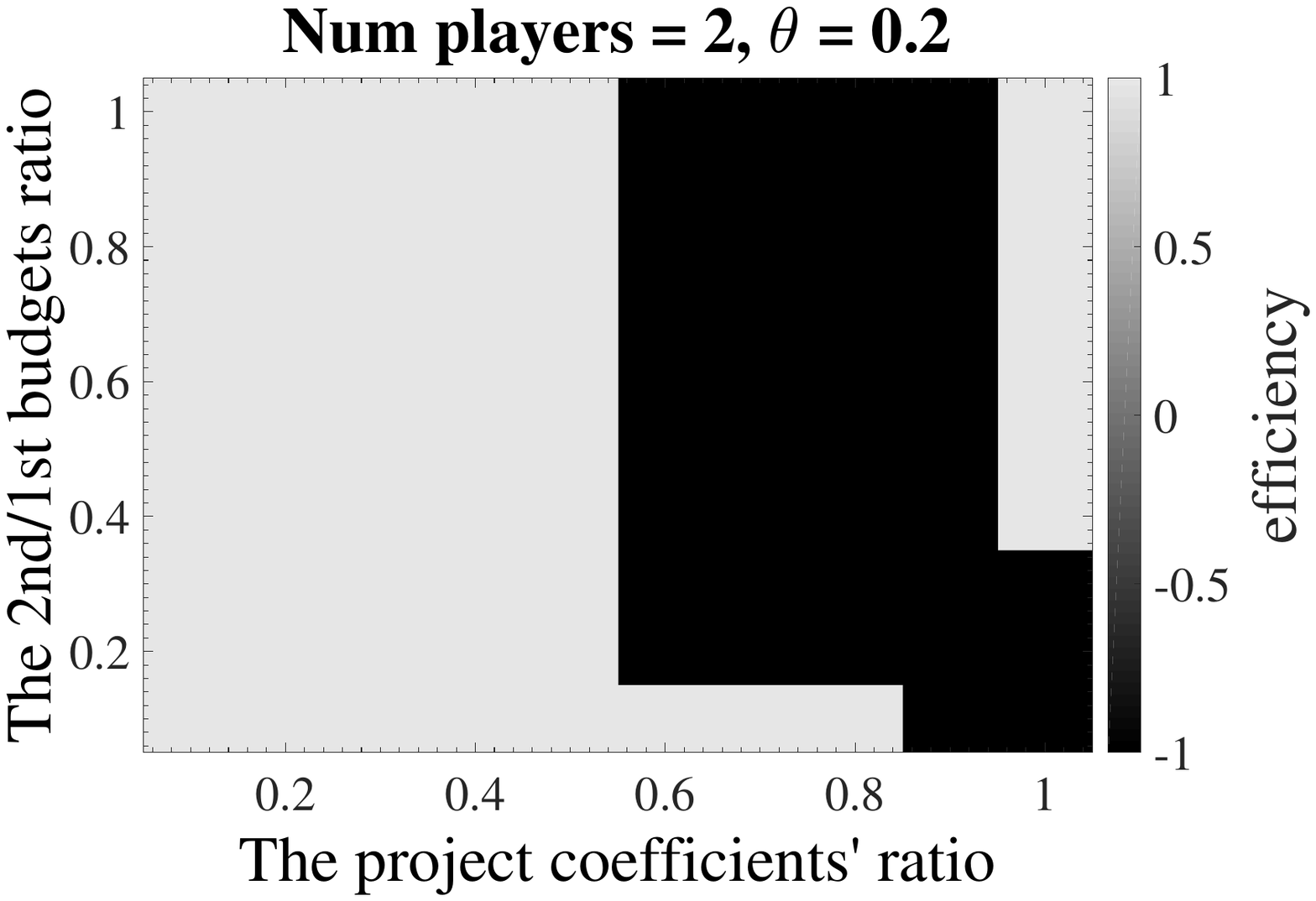}
}
\subfloat
{
\includegraphics[trim = 10mm 5mm 0mm 10mm, clip=true,width=0.35\textwidth]{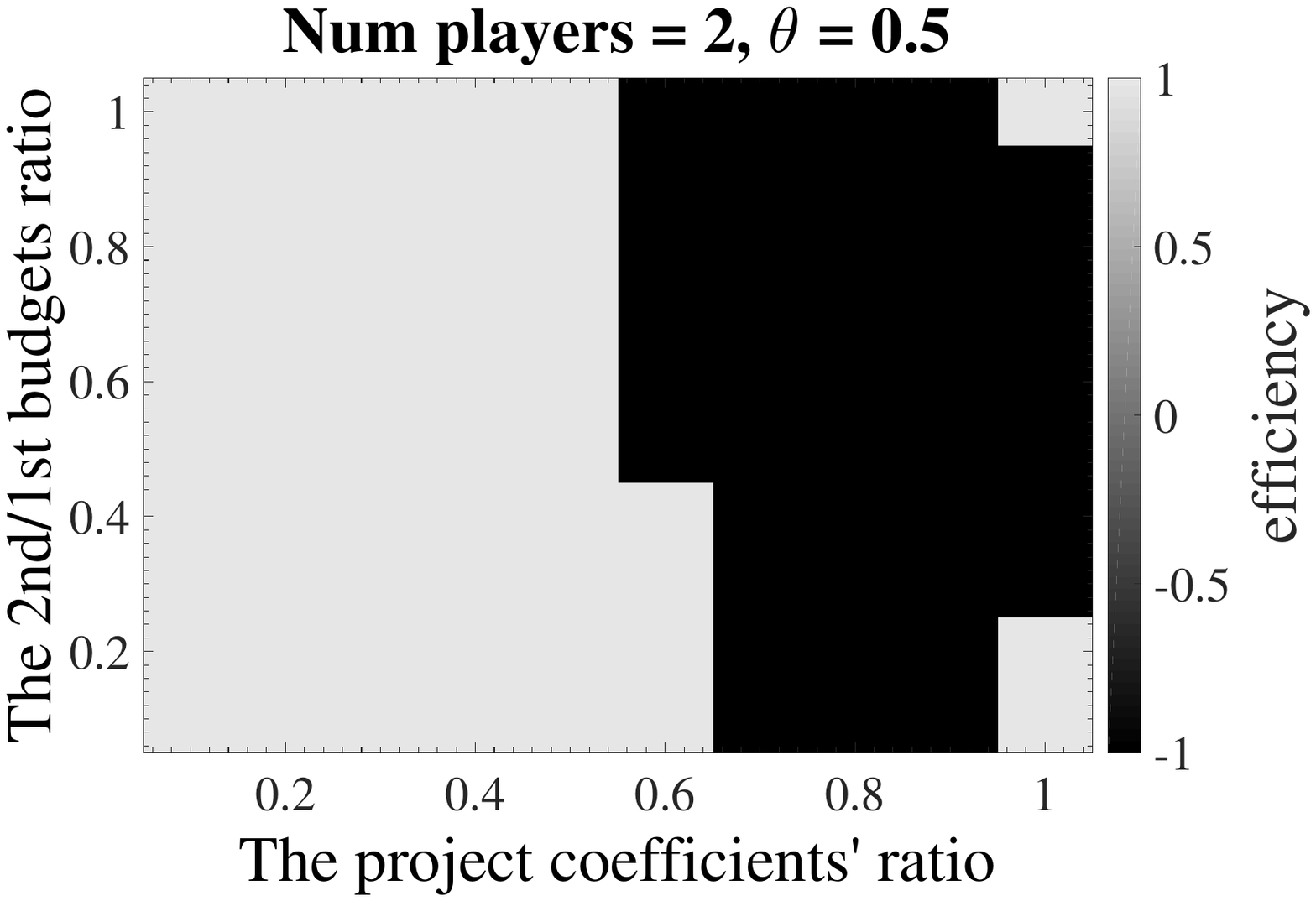}
}
\subfloat
{
\includegraphics[trim = 10mm 5mm 0mm 10mm, clip=true,width=0.35\textwidth]{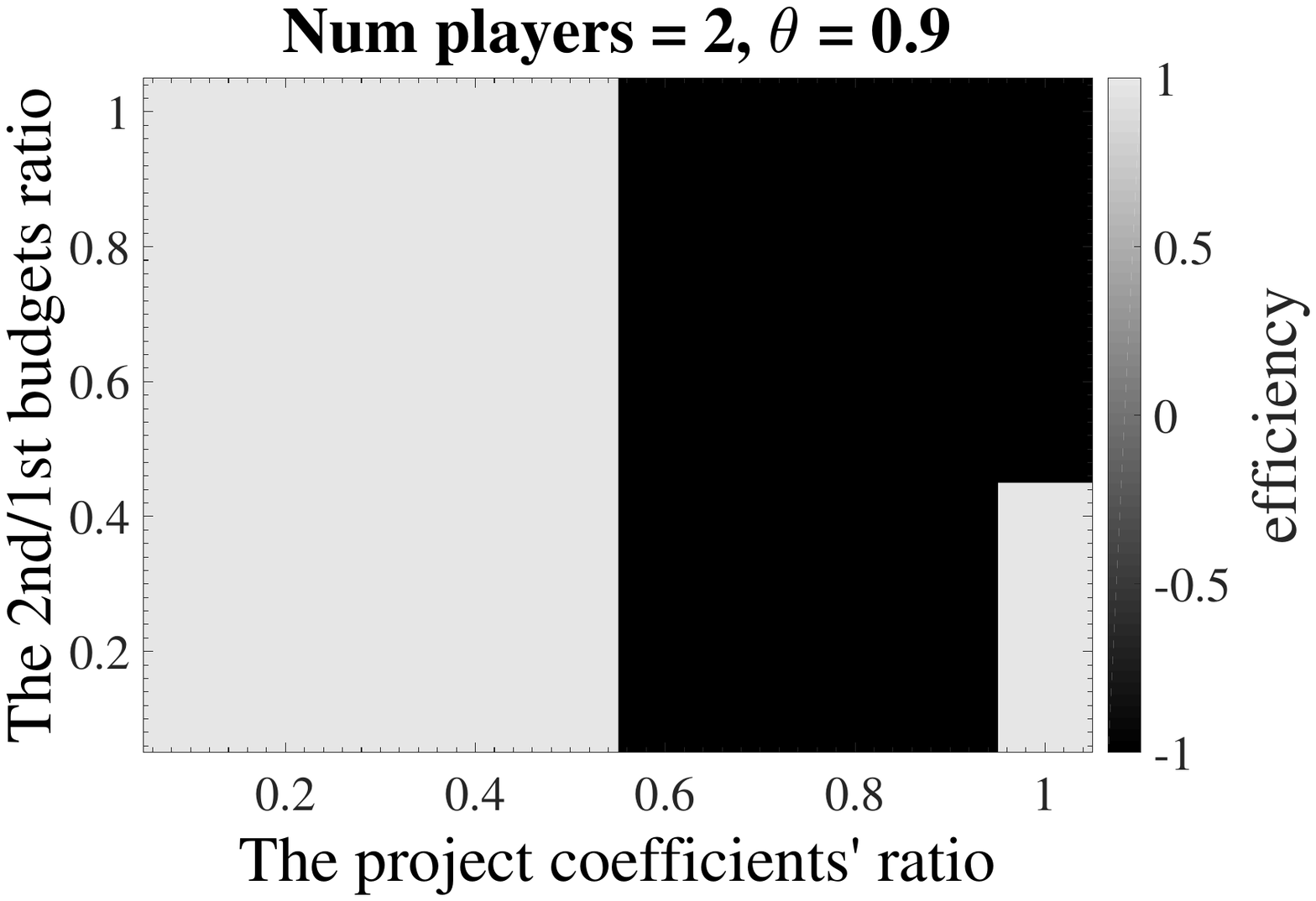}
}
\caption{The existence and efficiency of \NE{} for $2$ players as a function of
the ratio of the project functions coefficients and the ratio of the two largest budgets.
The first row plots the results of the simulations, and the second row
shows the theoretical predictions.
Black means that an equilibrium has not been found.
}%
\label{fig:NE_exist_effic_proj_rat_budg_rat_sim_theory_2}%
\end{figure*}

\ifthenelse{\equal{\version}{AAMAS_02_2014}}{
For two players with far away budgets ($B_1 < \theta B_2$),
an \NE{} exists when the project function coefficients are not too close
to one another. For $\theta = 0.5$, an equilibrium exists also when the
project functions are exactly the same, since player $2$ can just
dominate player $1$ everywhere.
When an \NE{} exists, the efficiency is quite high, it begins with $0.914$
and increases as the project functions become closer to each other.
}{
}%

We present the results representing the simulation trends; 
We have obtained similar results in more extensive simulations as well.

First, to validate our simulations, we compare the results of the simulations
to the theoretical predictions 
in \figref{fig:NE_exist_effic_proj_rat_budg_rat_sim_theory_2}, where
we go over the ratios of the project coefficients and budgets, because
the invariance to multiplication from Proposition~\ref{prop:inv_mult}
implies that only the ratios matter for \NE. This is not exactly the case here,
since the budgets are equally distanced, and multiplying the largest
budgets by the same factor may multiply the other budgets differently.
However, ratios constitute a good start.
%
In this case, the simulations are in complete agreement with the theory.

The difference between the areas
in \figref{fig:NE_exist_effic_proj_rat_budg_rat_sim_theory_2}
that correspond to
$B_1 \geq \theta B_2$ and to $B_1 < \theta B_2$ fits
Theorem~\ref{the:NE_charac_2players}.
For two players, 
except for the \NE{} when the two projects are equal, 
an \NE{} exists for a budget ratio
if and only if the value functions ratio is below a certain value,
fitting 
Corollary~\ref{cor:NE_exist_proj_rat_smaller_2players}.


\begin{figure*}[h!tbp]
\centering
\subfloat
{
\includegraphics[trim = 10mm 5mm 0mm 10mm, clip=true,width=0.30\textwidth]{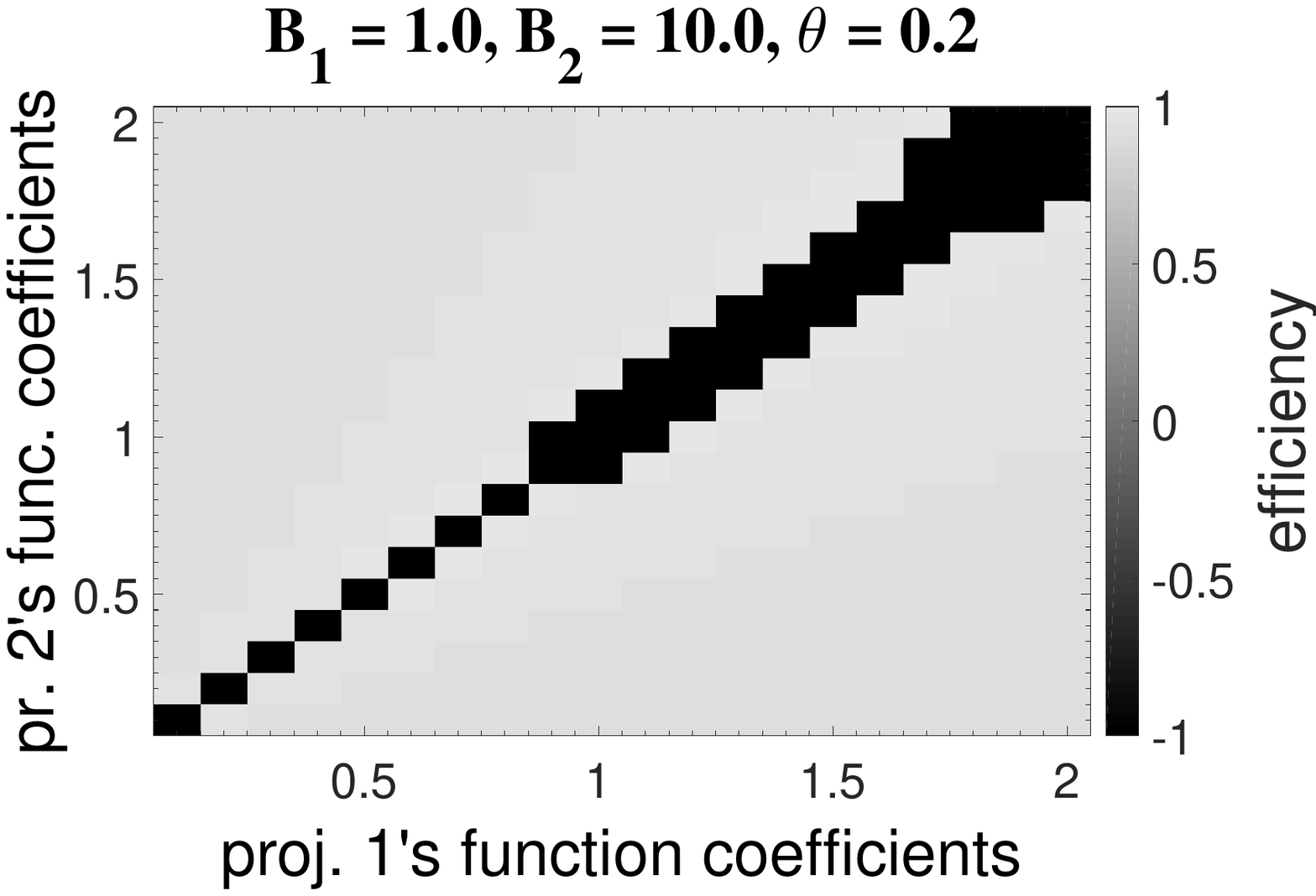} 
\label{fig:figure1}}
\subfloat
{
\includegraphics[trim = 10mm 5mm 0mm 10mm, clip=true,width=0.30\textwidth]{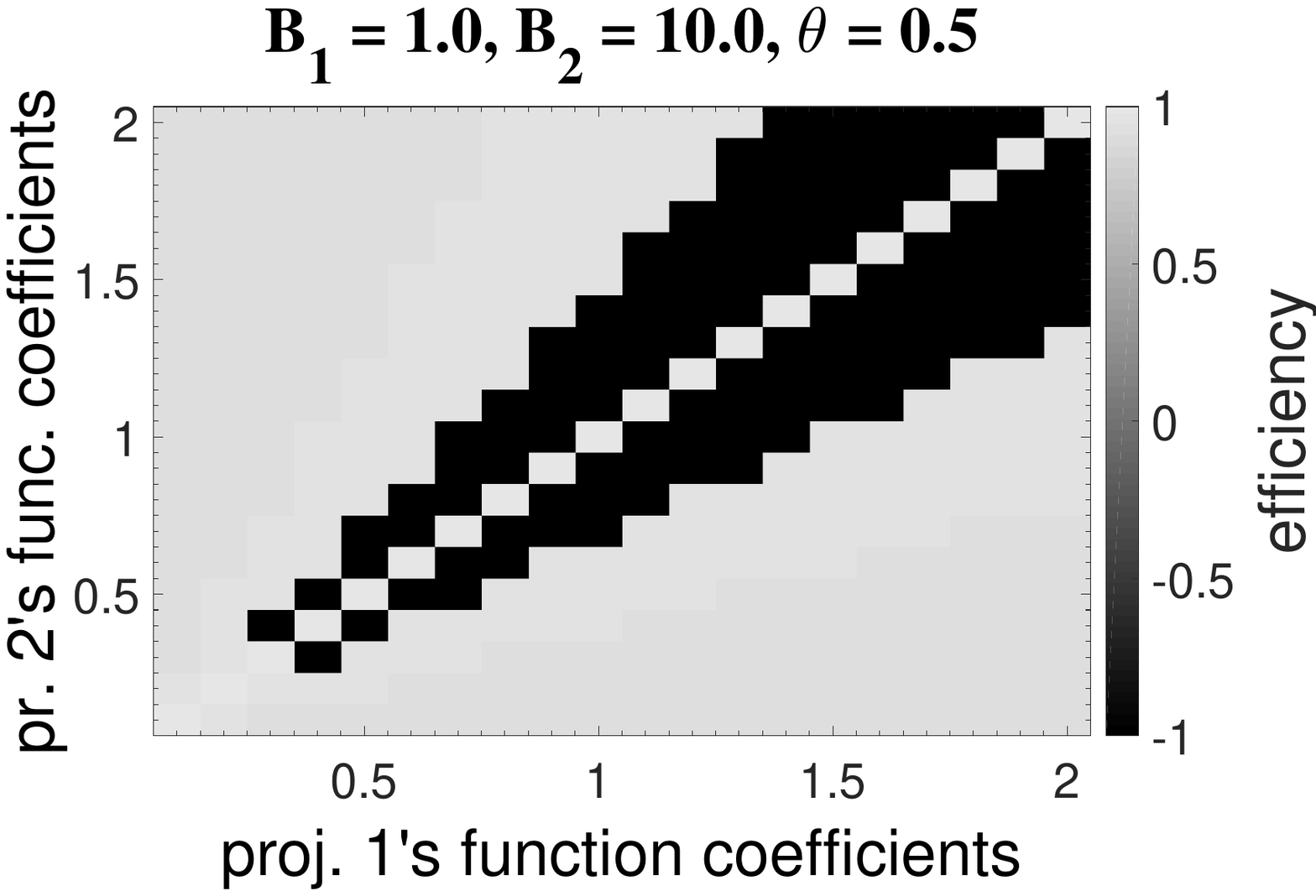} 
\label{fig:figure2}}

\subfloat
{
\includegraphics[trim = 10mm 5mm 0mm 10mm, clip=true,width=0.30\textwidth]{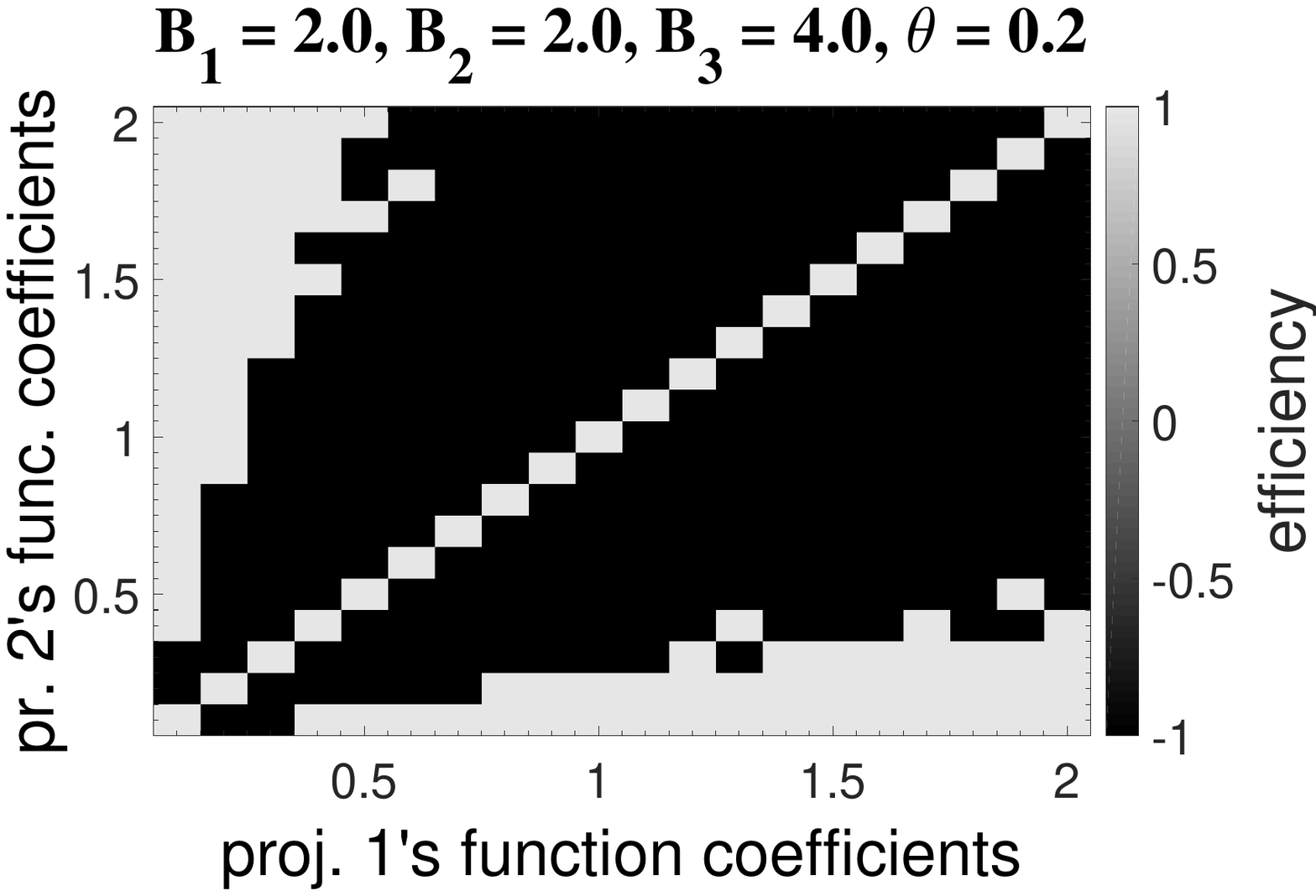} 
\label{fig:figure6}}
\subfloat
{
\includegraphics[trim = 10mm 5mm 0mm 10mm, clip=true,width=0.30\textwidth]{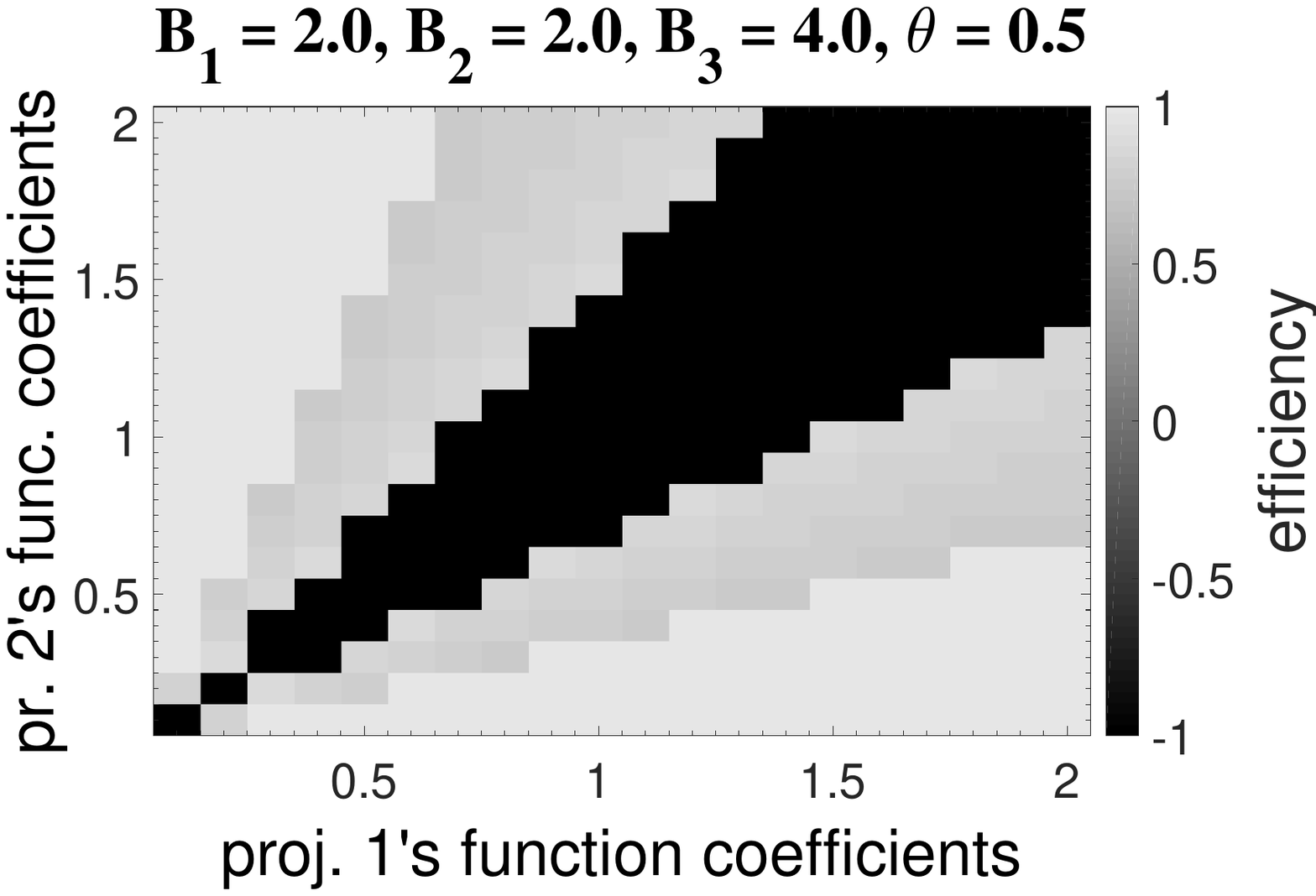}
\label{fig:figure3}}
\subfloat
{
\includegraphics[trim = 10mm 5mm 0mm 10mm, clip=true,width=0.30\textwidth]{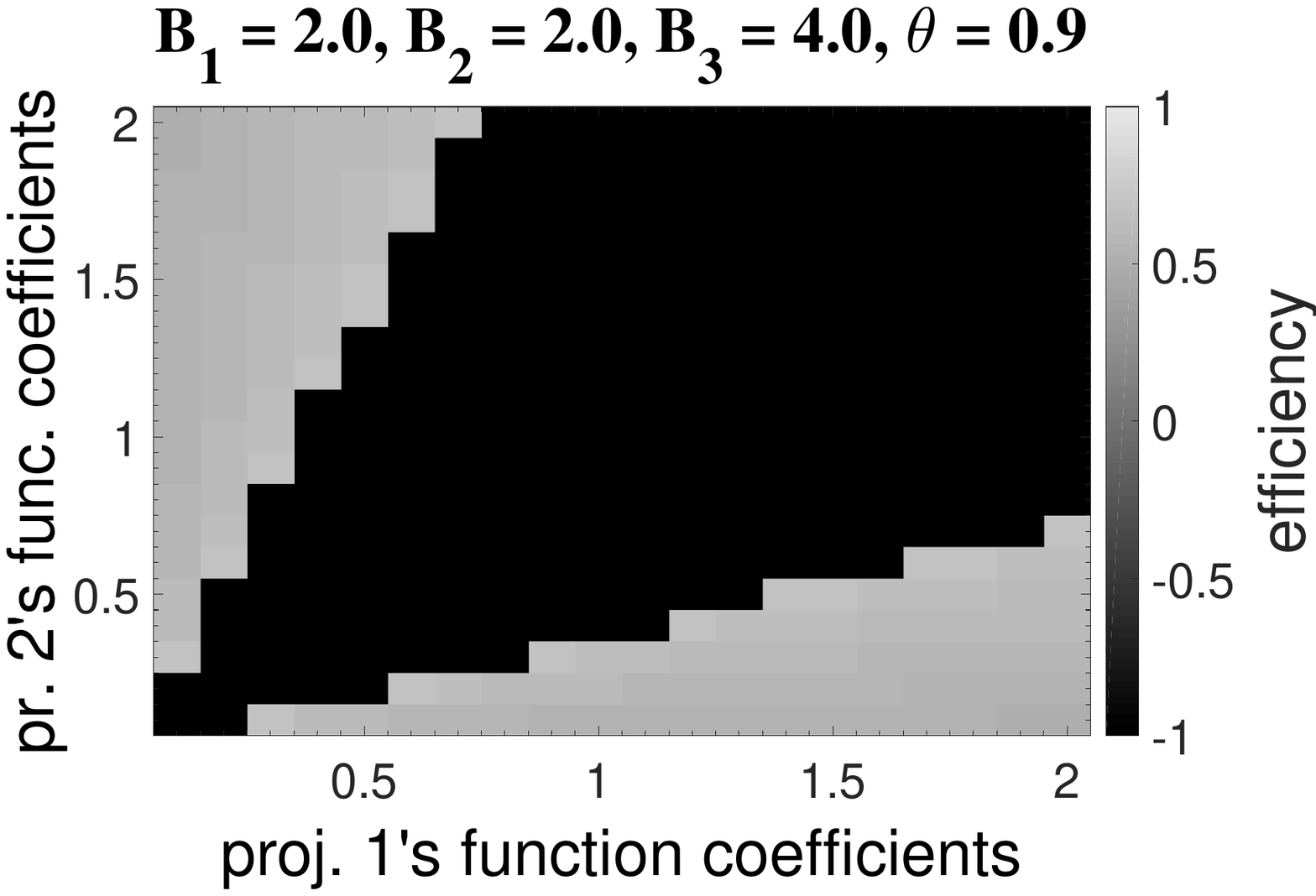} 
\label{fig:figure7}}

\subfloat%
{
\includegraphics[trim = 10mm 5mm 0mm 10mm, clip=true,width=0.30\textwidth]{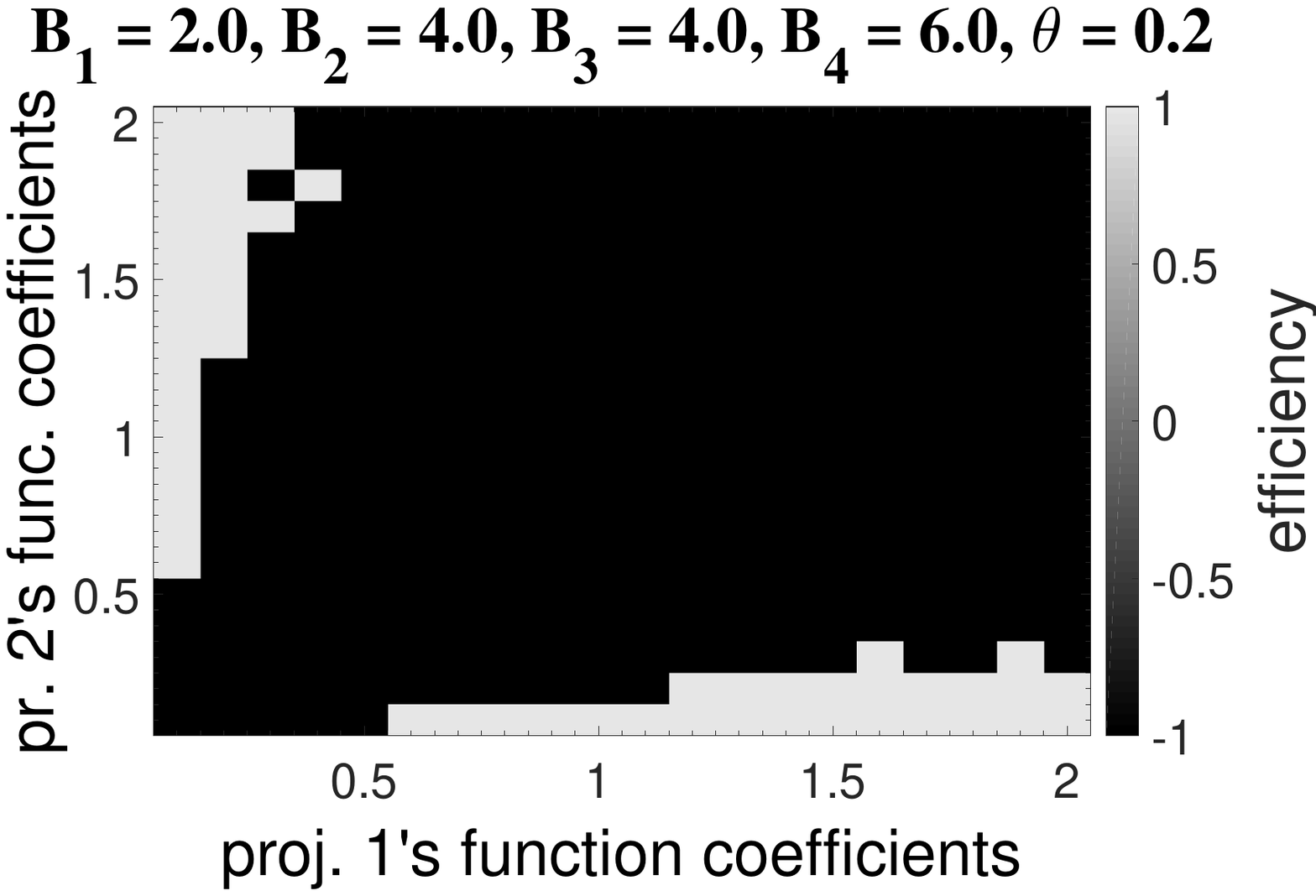}
\label{fig:figure9}}
\subfloat%
{
\includegraphics[trim = 10mm 5mm 0mm 10mm, clip=true,width=0.30\textwidth]{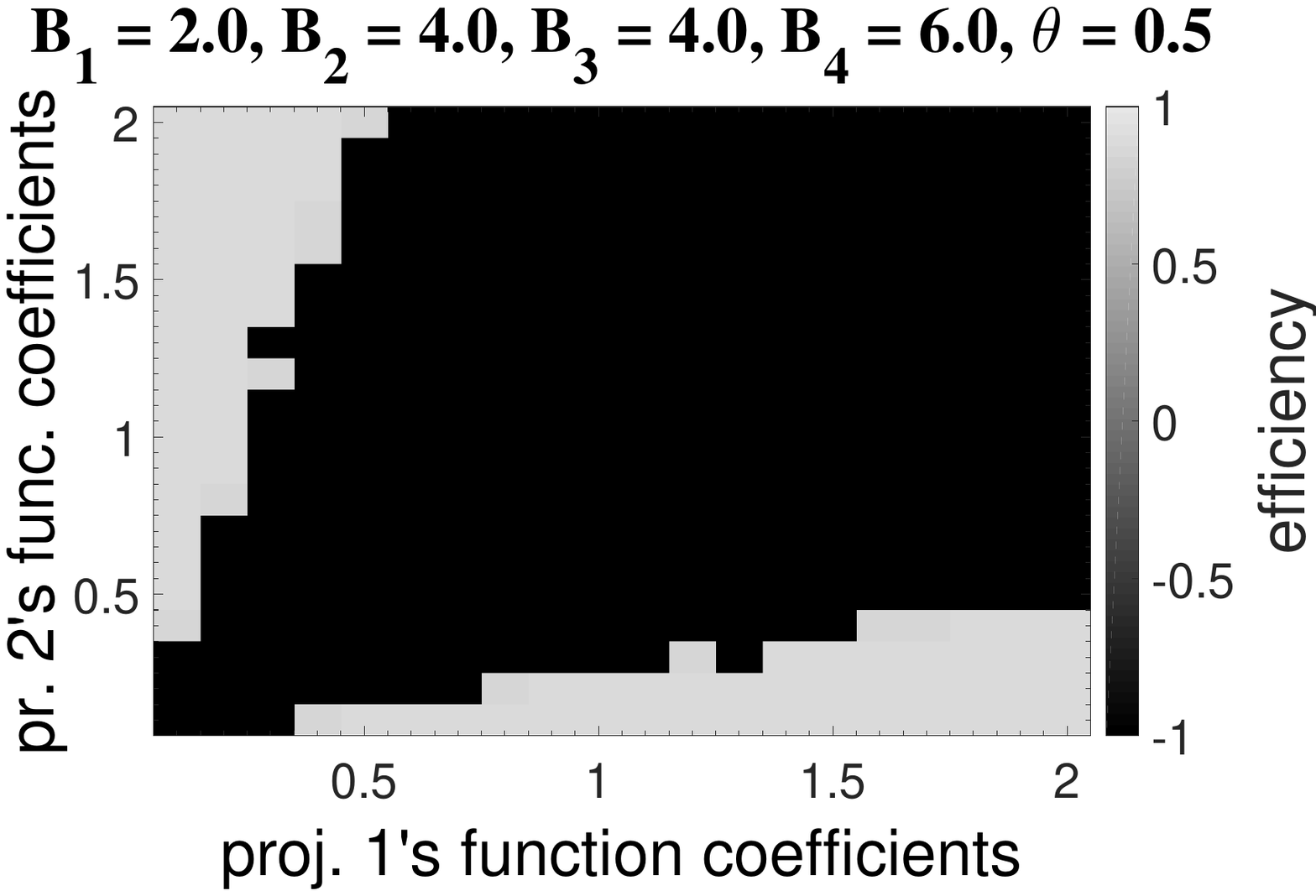}
\label{fig:figure10}}
\subfloat%
{
\includegraphics[trim = 10mm 5mm 0mm 10mm, clip=true,width=0.30\textwidth]{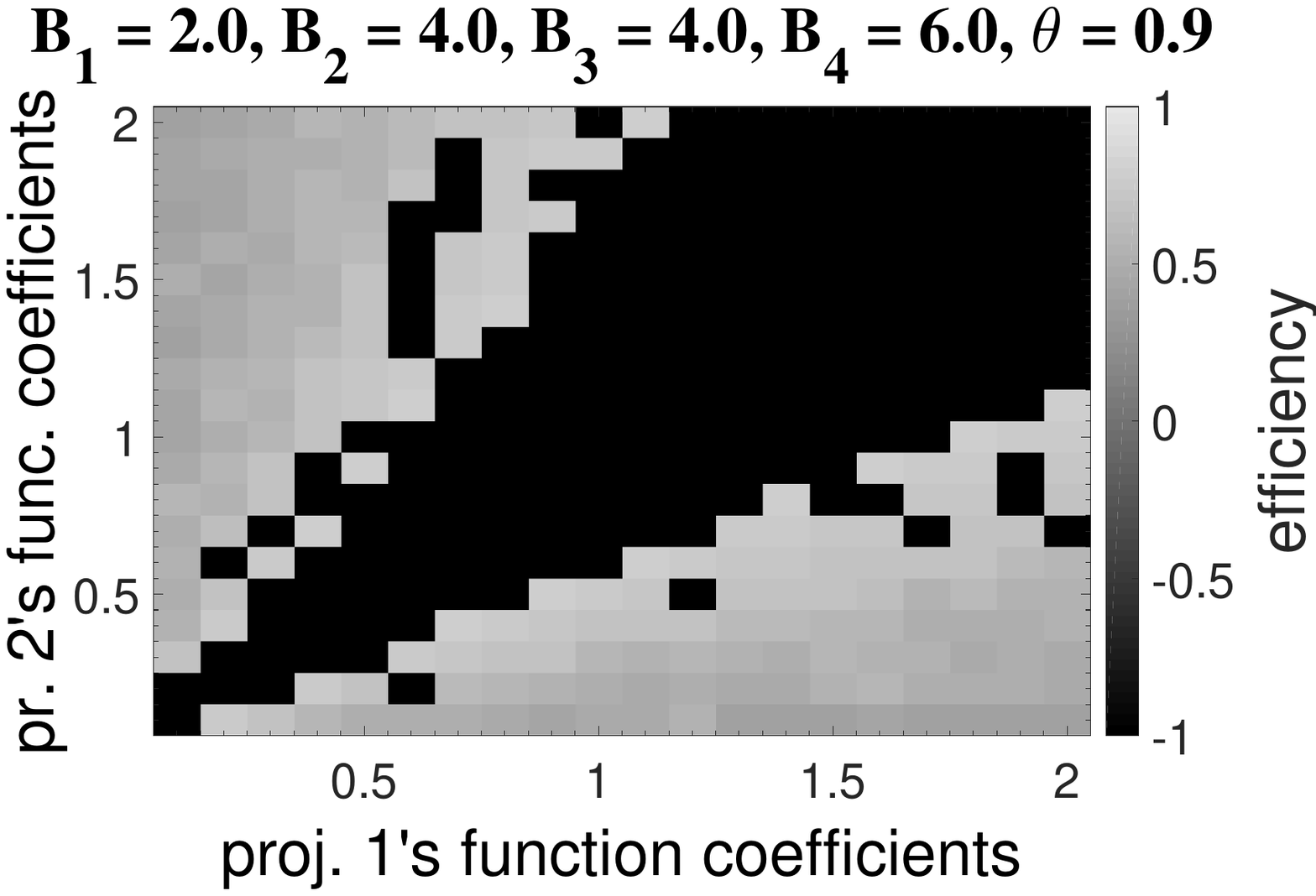}
\label{fig:figure11}}

\subfloat%
{
\includegraphics[trim = 10mm 5mm 0mm 5mm, clip=true,width=0.30\textwidth]{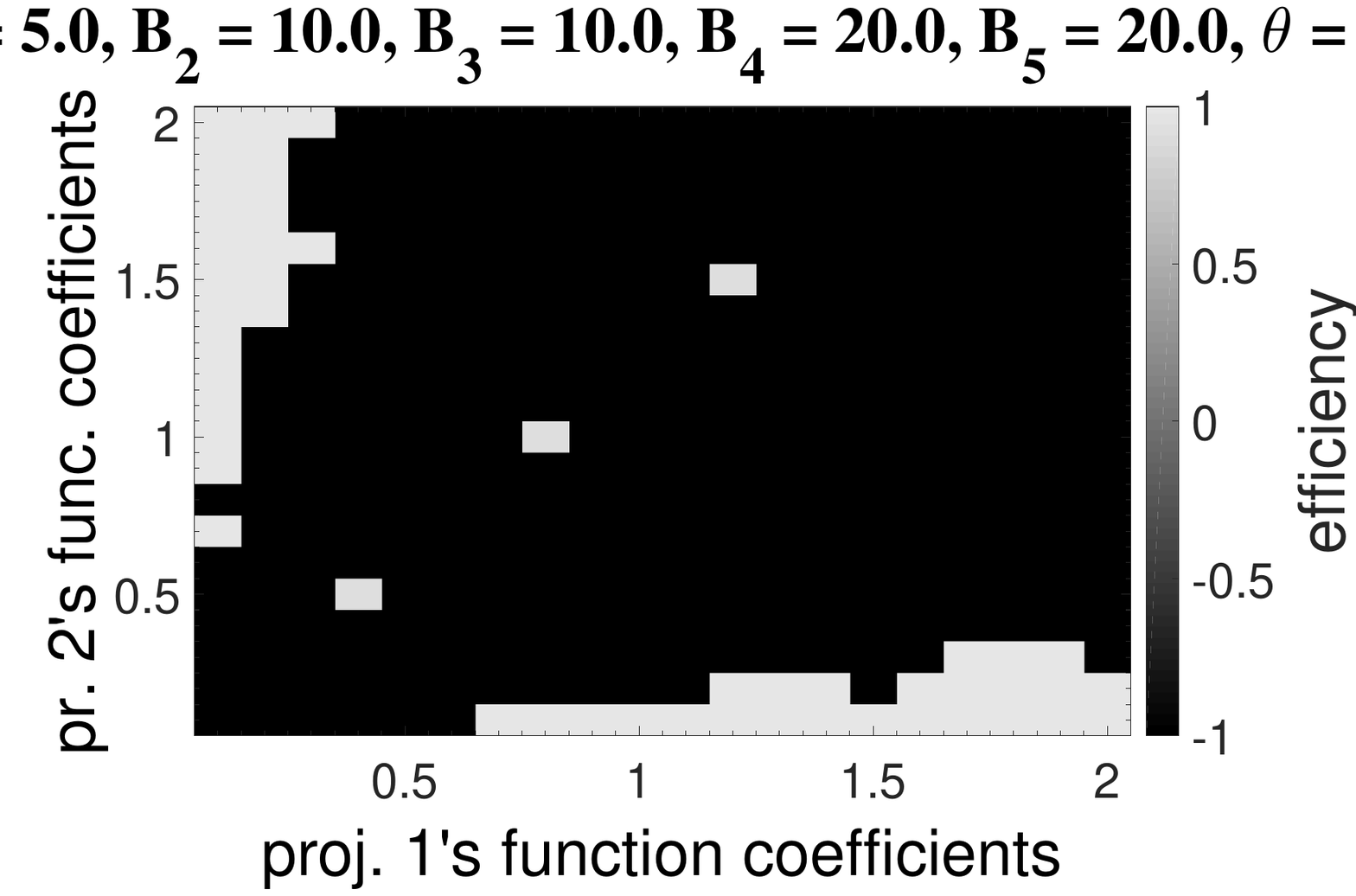}
\label{fig:figure13}}
\subfloat%
{
\includegraphics[trim = 10mm 5mm 0mm 5mm, clip=true,width=0.30\textwidth]{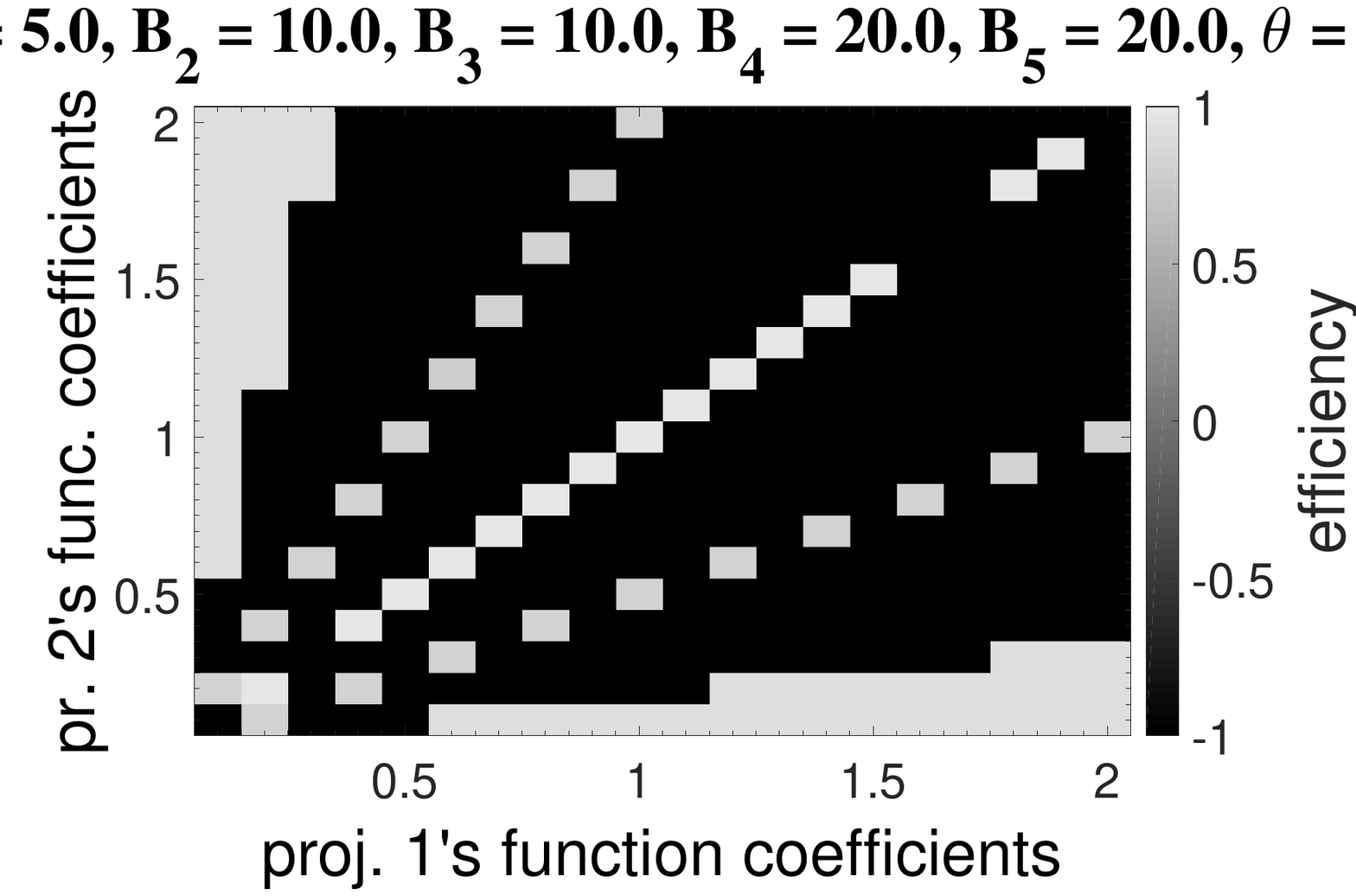}
\label{fig:figure28}}
\subfloat%
{
\includegraphics[trim = 10mm 5mm 0mm 5mm, clip=true,width=0.30\textwidth]{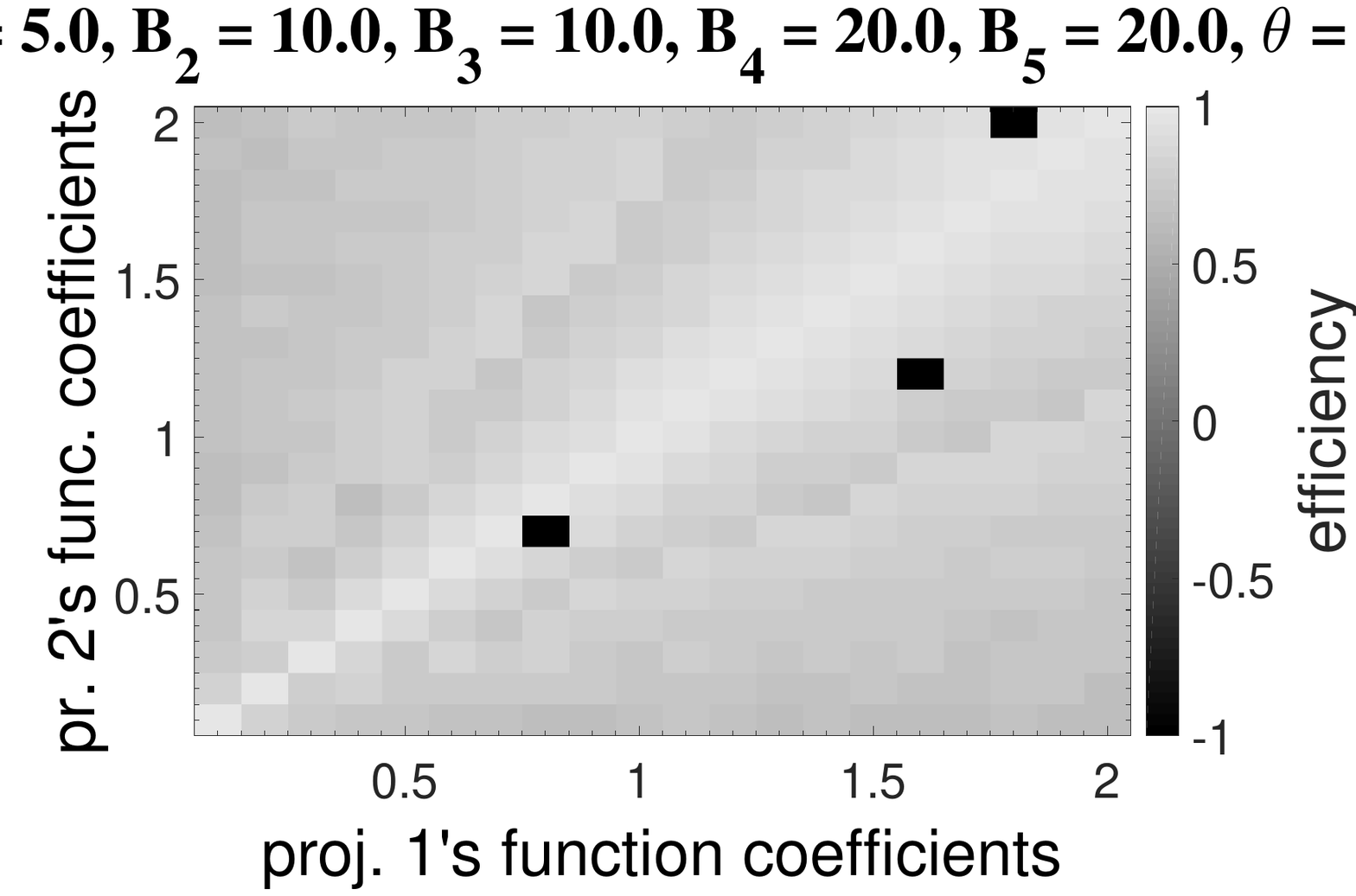}
\label{fig:figure29}}

\subfloat%
{
\includegraphics[trim = 10mm 5mm 0mm 5mm, clip=true,width=0.30\textwidth]{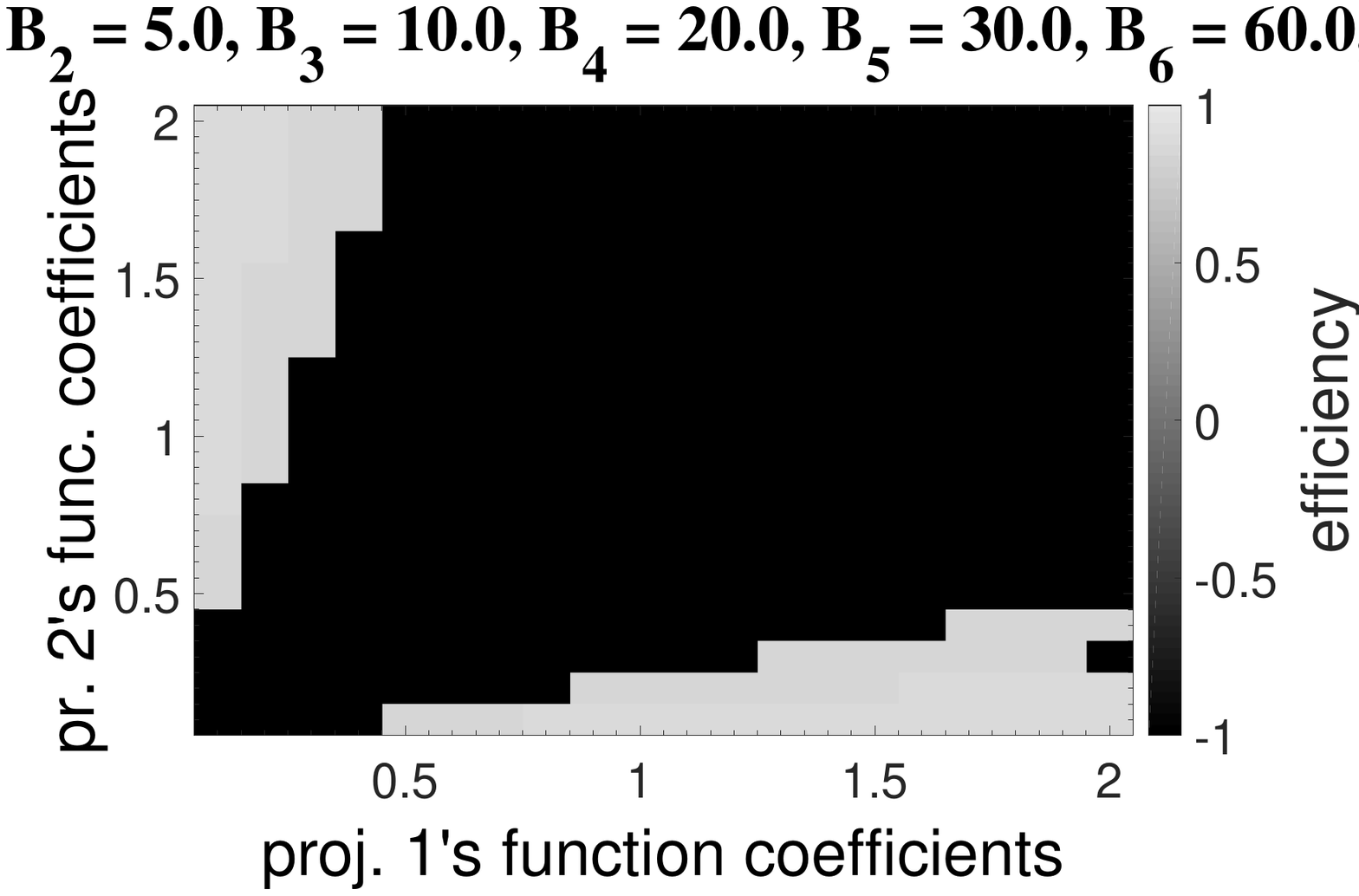}
\label{fig:figure17}}
\caption{The existence and efficiency of \NE{} as function of
project functions for $2, 3, 4, 5$ and $6$ players.
Black means that Nash Equilibrium has not been found.
}%
\label{fig:NE_exist_effic_proj_2_3_4_5_6}%
\end{figure*}

\begin{figure*}[h!tbp]
\centering

\subfloat%
{
\includegraphics[trim = 10mm 20mm 30mm 10mm, clip,width=0.30\textwidth]{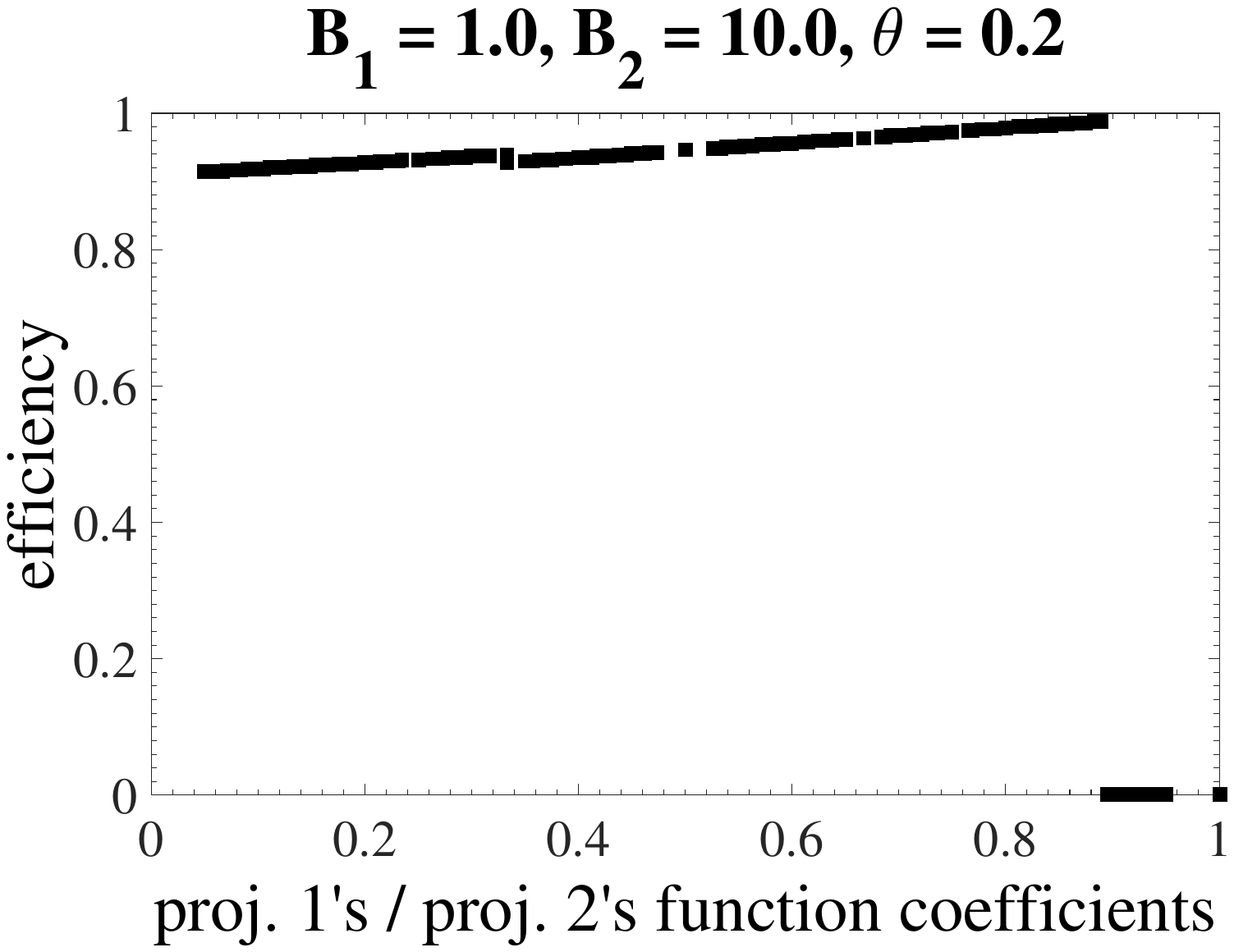}
\label{fig:plot1}}
\subfloat%
{
\includegraphics[trim = 10mm 20mm 30mm 10mm, clip,width=0.30\textwidth]{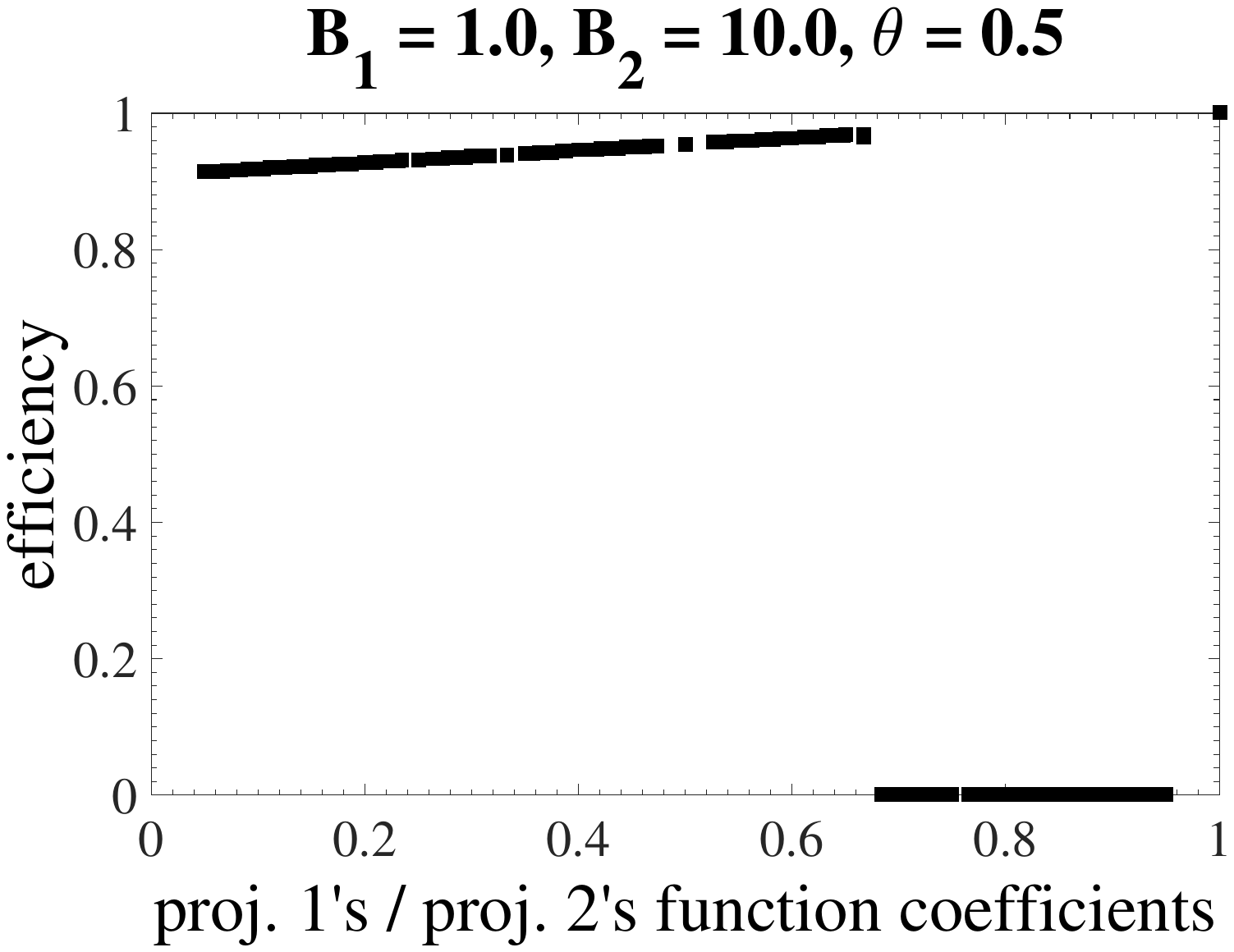}
\label{fig:plot2}}

\subfloat%
{
\includegraphics[trim = 10mm 20mm 30mm 10mm, clip,width=0.30\textwidth]{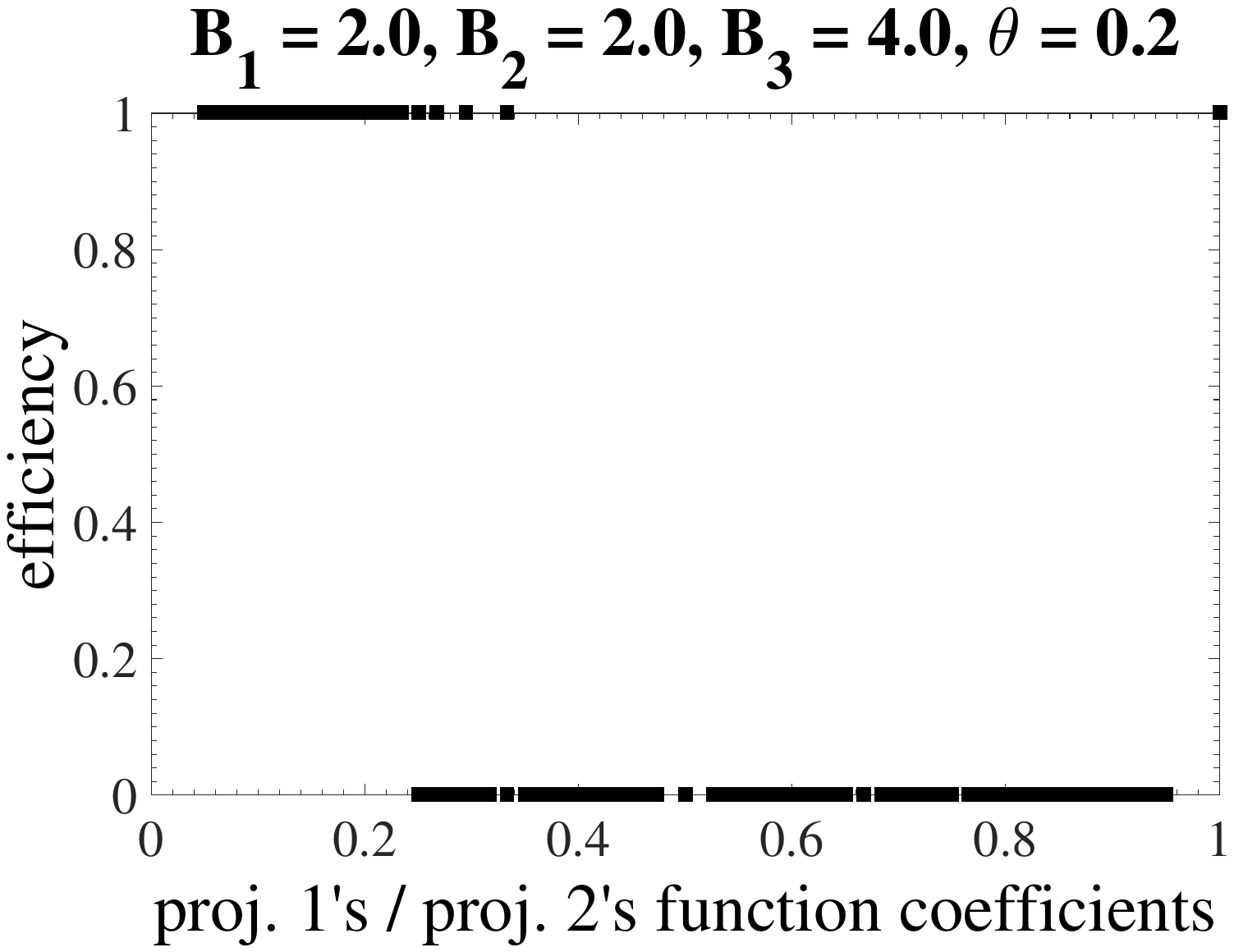}
\label{fig:plot6}}
\subfloat%
{
\includegraphics[trim = 10mm 20mm 30mm 10mm, clip,width=0.30\textwidth]{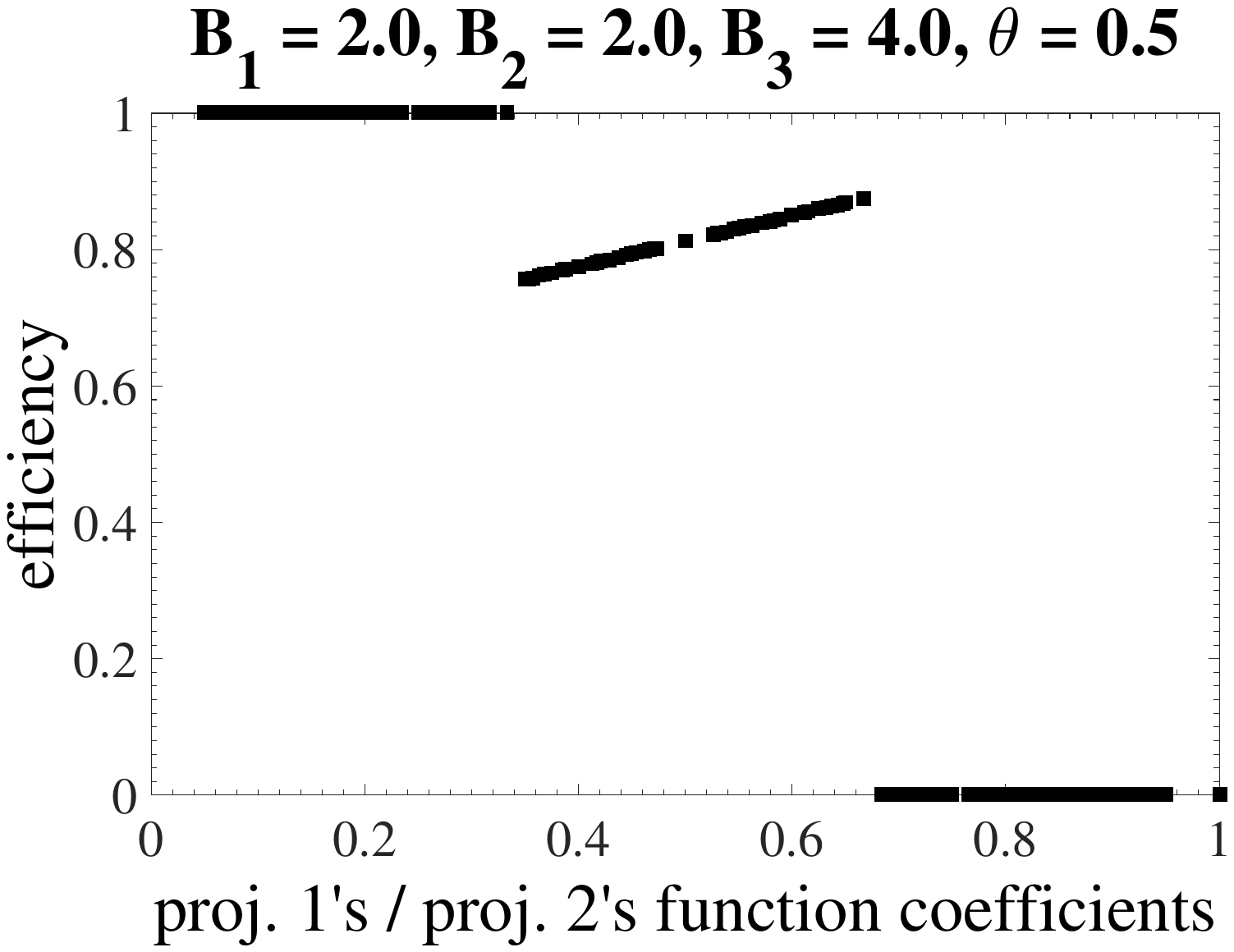}
\label{fig:plot3}}
\subfloat%
{
\includegraphics[trim = 10mm 20mm 30mm 10mm, clip,width=0.30\textwidth]{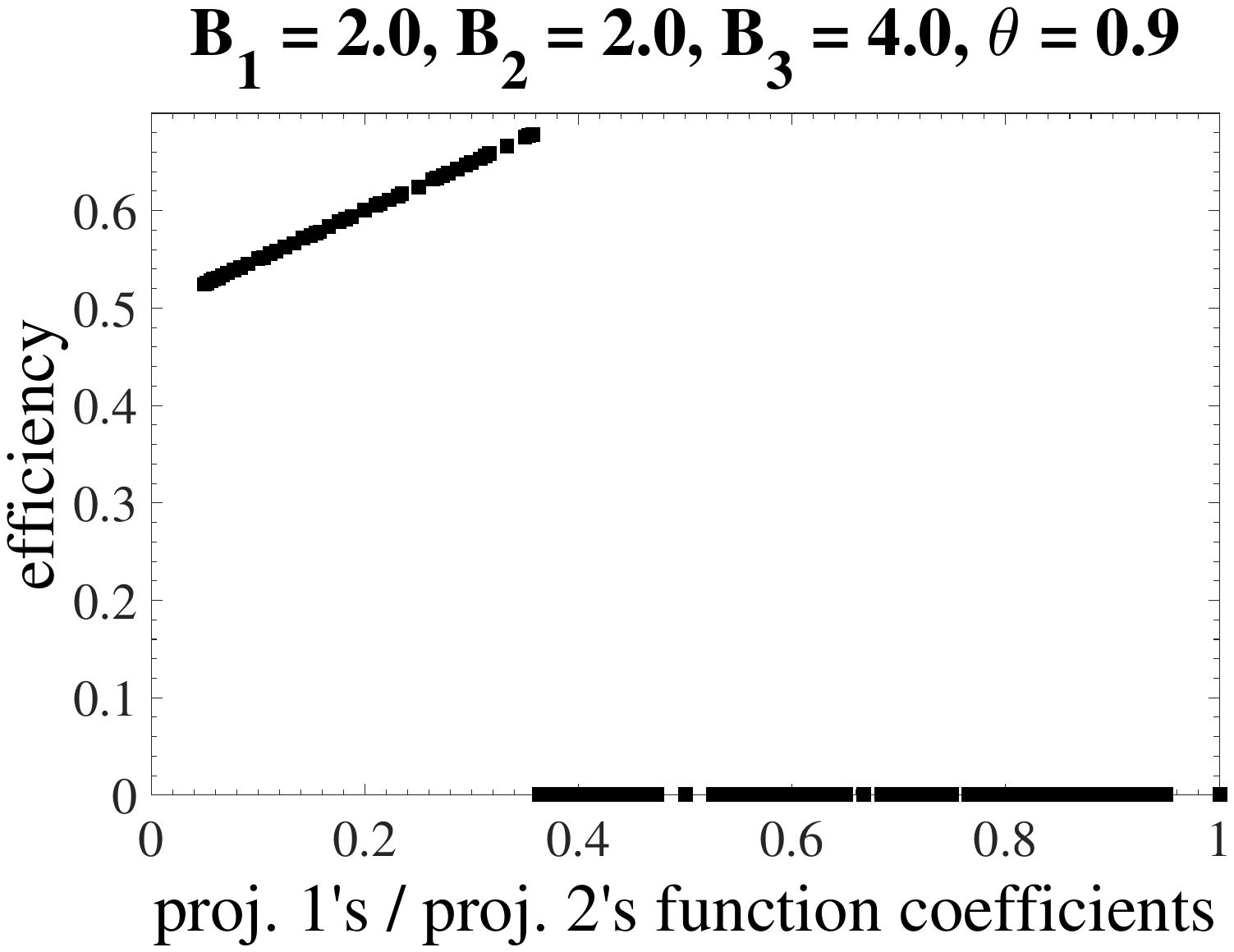}
\label{fig:plot7}}

\subfloat%
{
\includegraphics[trim = 10mm 20mm 10mm 10mm, clip,width=0.30\textwidth]{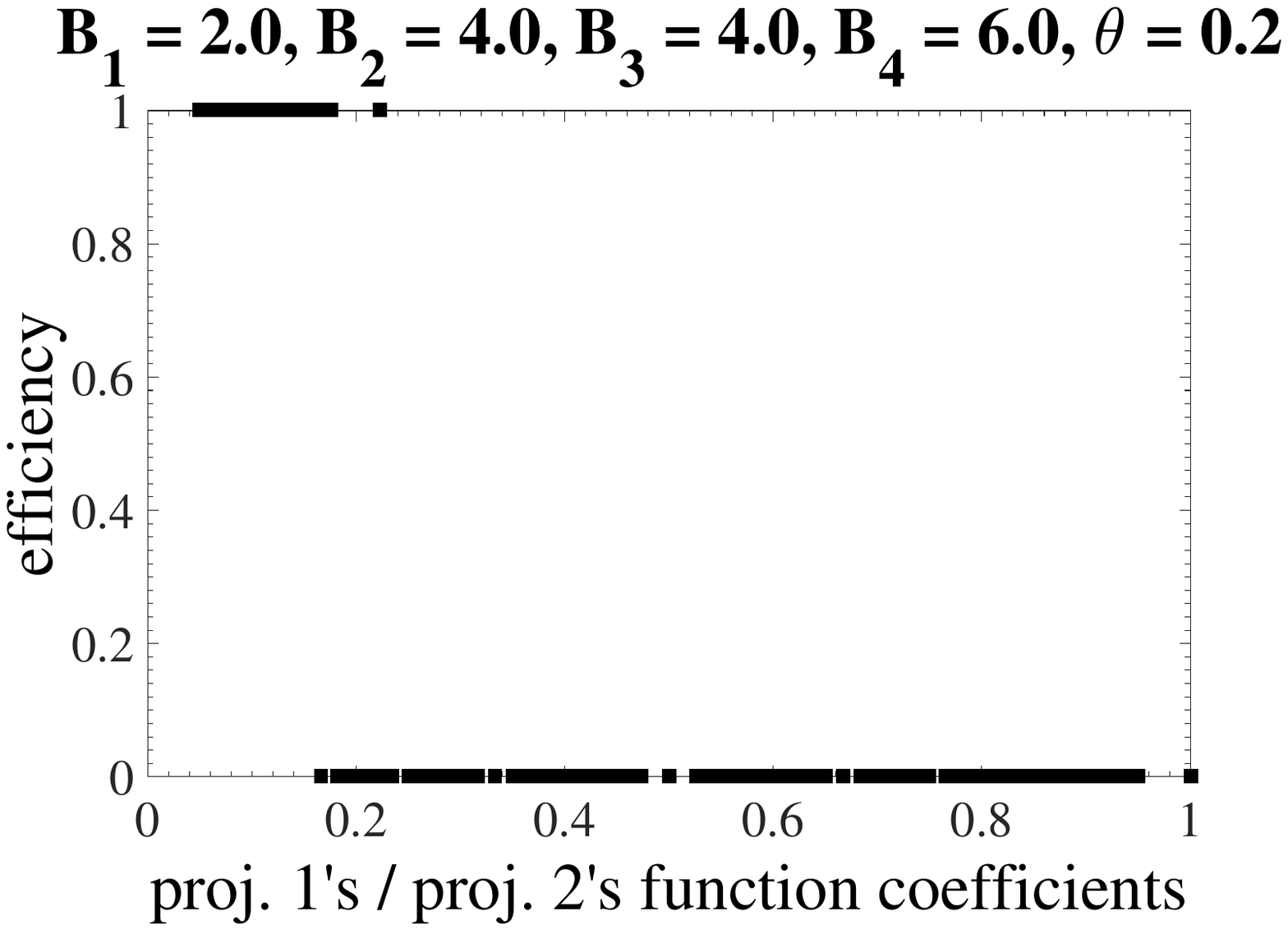}
\label{fig:plot9}}
\subfloat%
{
\includegraphics[trim = 10mm 20mm 10mm 10mm, clip,width=0.30\textwidth]{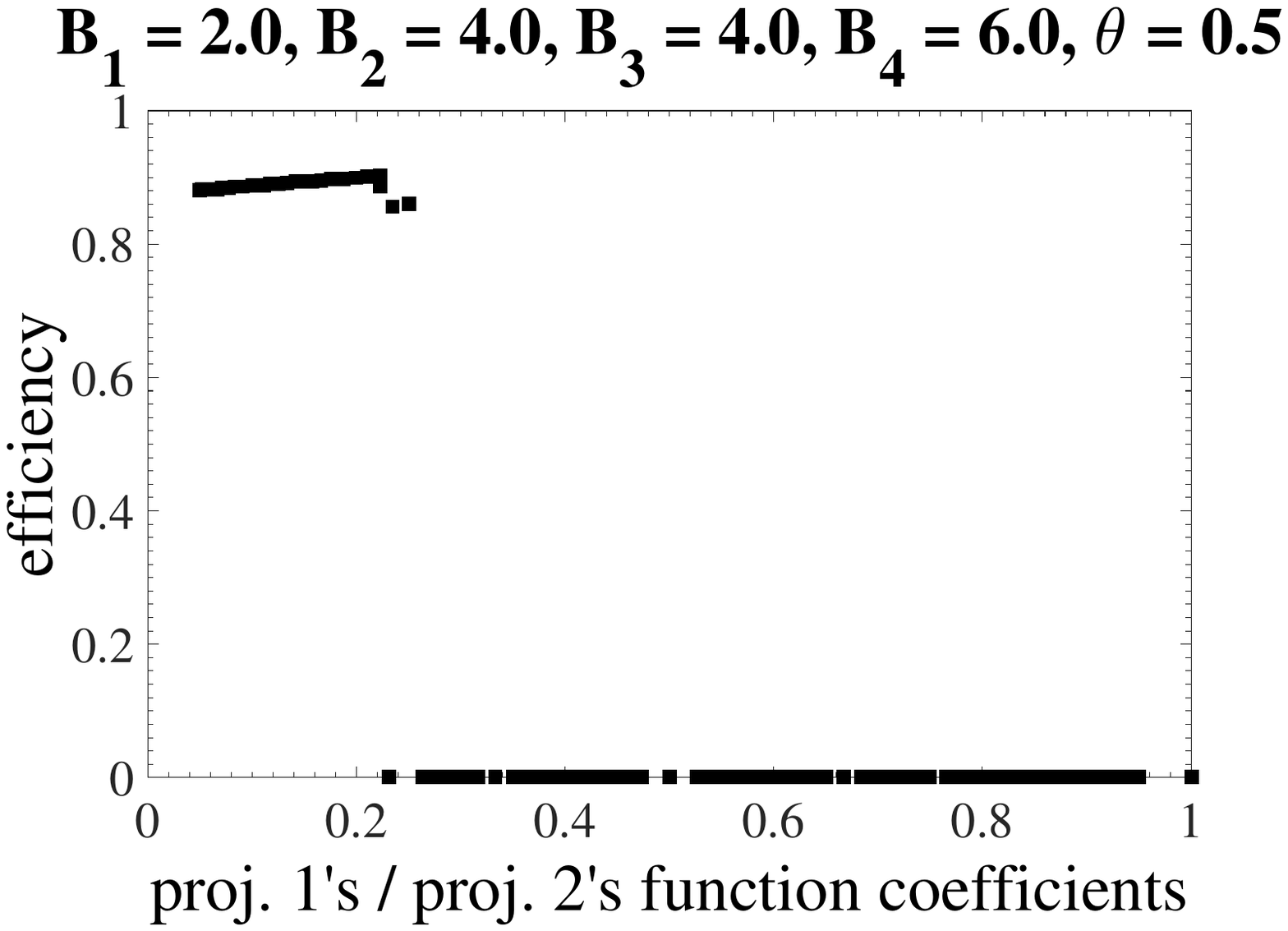}
\label{fig:plot10}}
\subfloat%
{
\includegraphics[trim = 10mm 20mm 10mm 10mm, clip,width=0.30\textwidth]{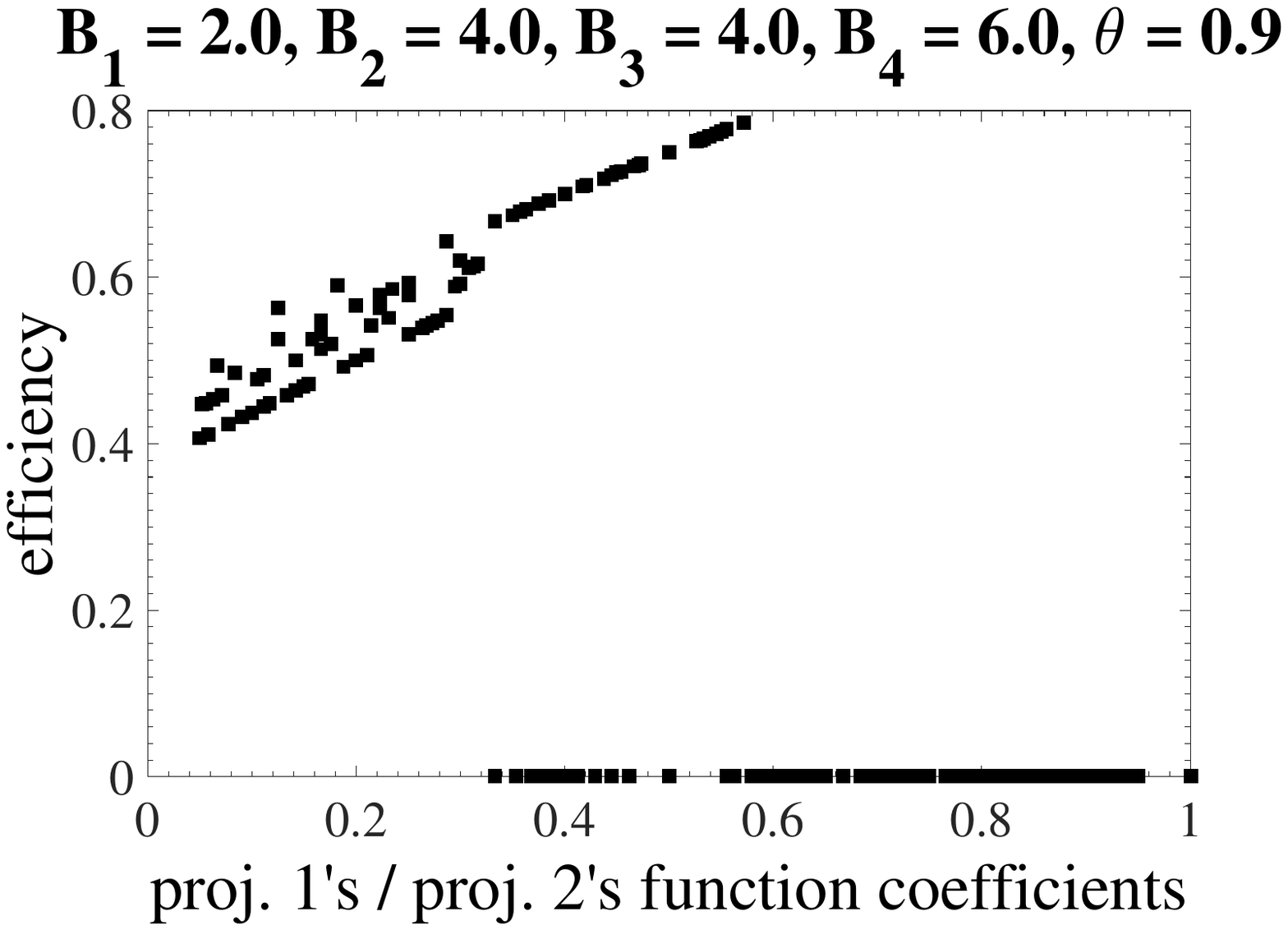}
\label{fig:plot11}}

\subfloat%
{
\includegraphics[trim = 10mm 20mm 15mm 5mm, clip,width=0.30\textwidth]{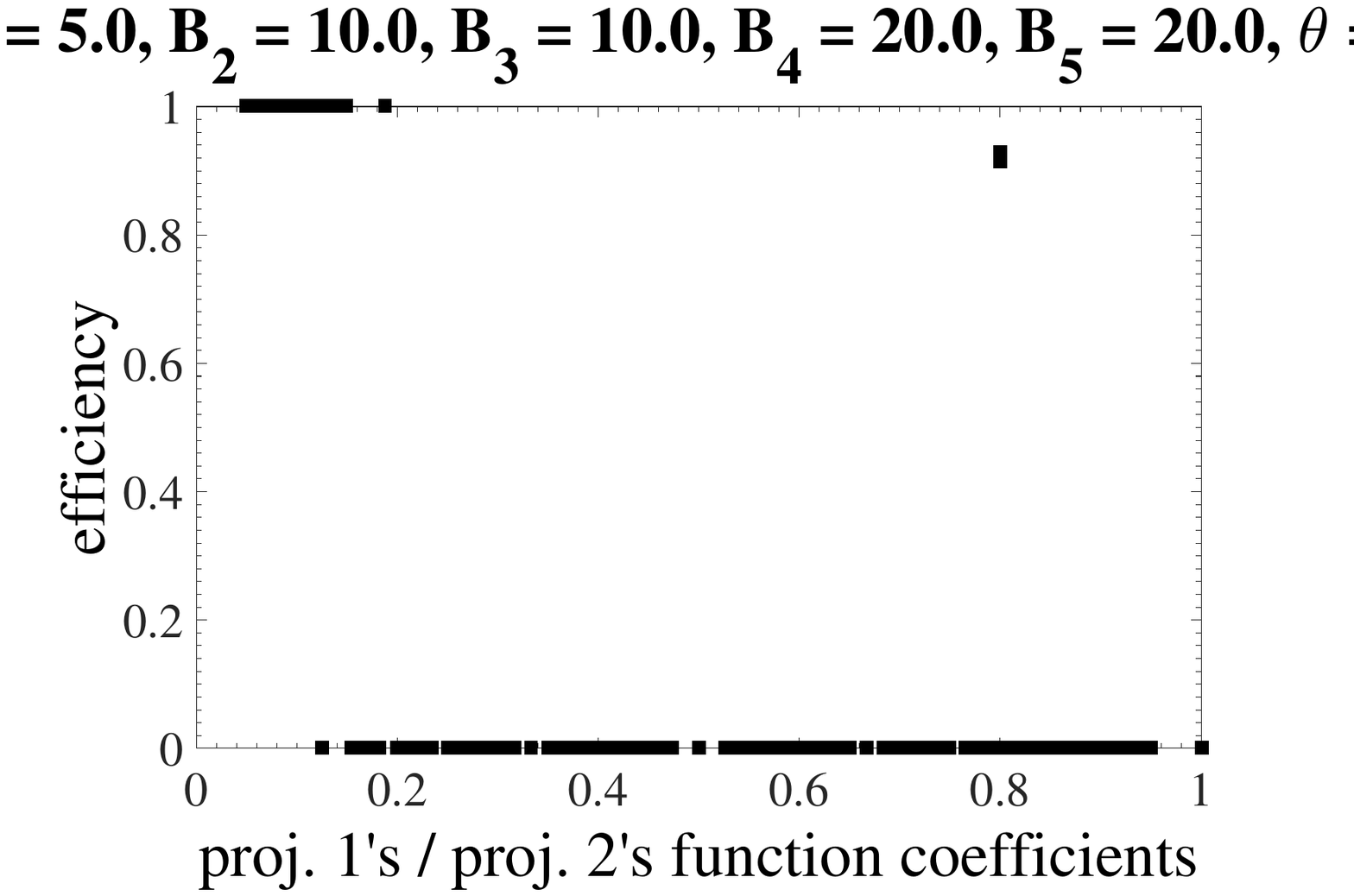}
\label{fig:plot13}}
\subfloat%
{
\includegraphics[trim = 10mm 20mm 15mm 5mm, clip,width=0.30\textwidth]{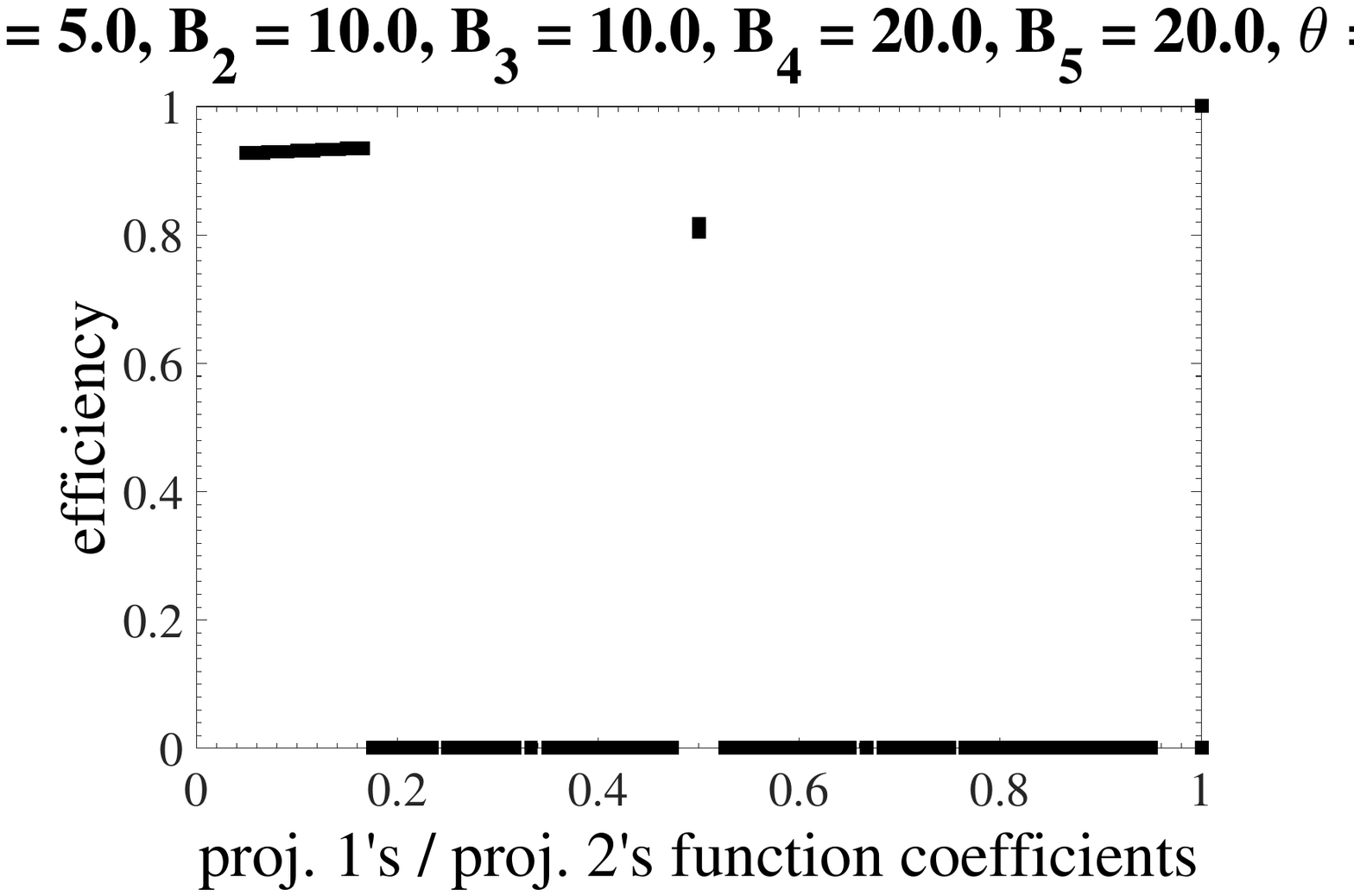}
\label{fig:plot28}}
\subfloat%
{
\includegraphics[trim = 10mm 20mm 15mm 5mm, clip,width=0.30\textwidth]{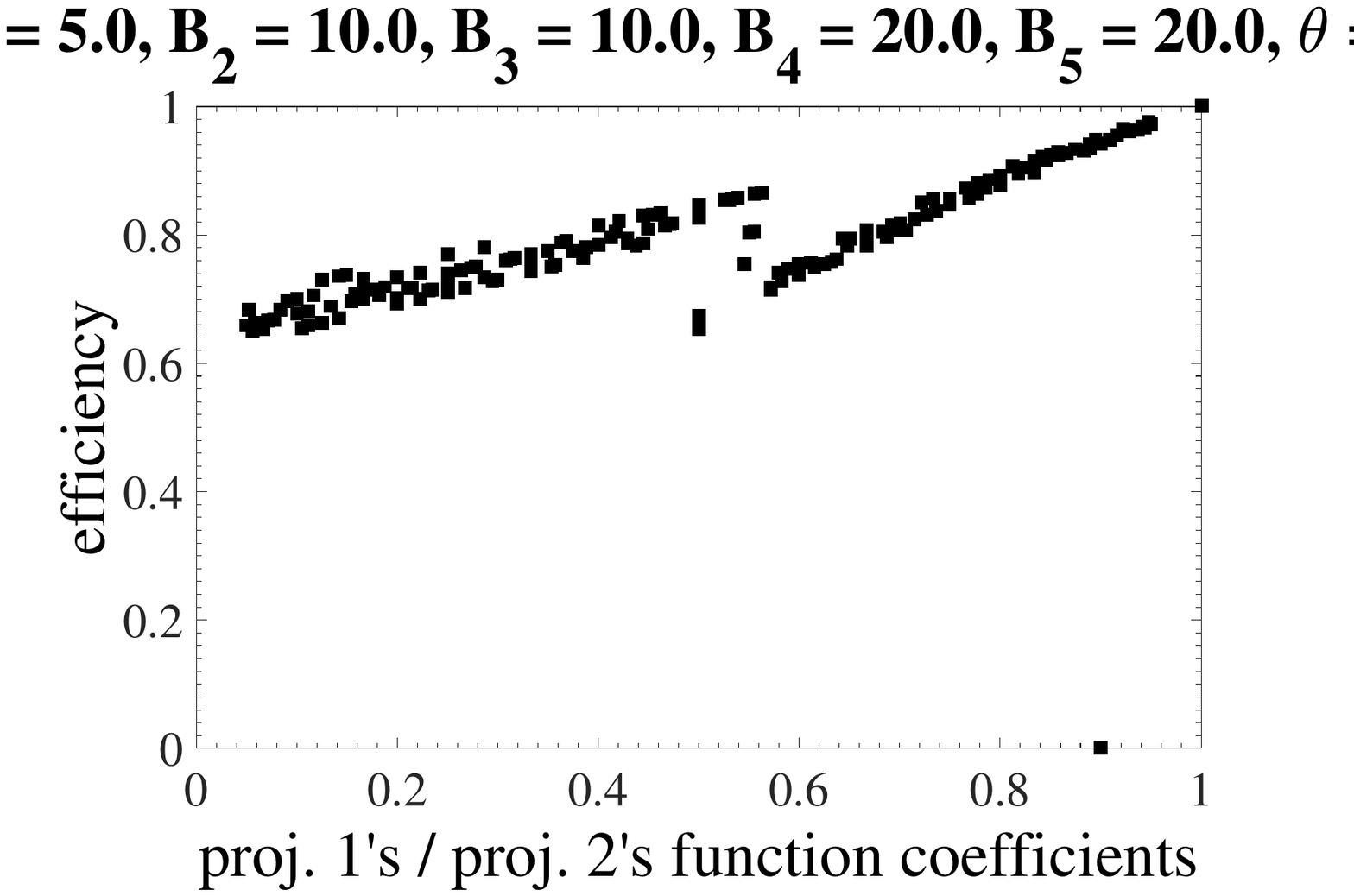}
\label{fig:plot29}}

\subfloat%
{
\includegraphics[trim = 10mm 20mm 20mm 5mm, clip,width=0.30\textwidth]{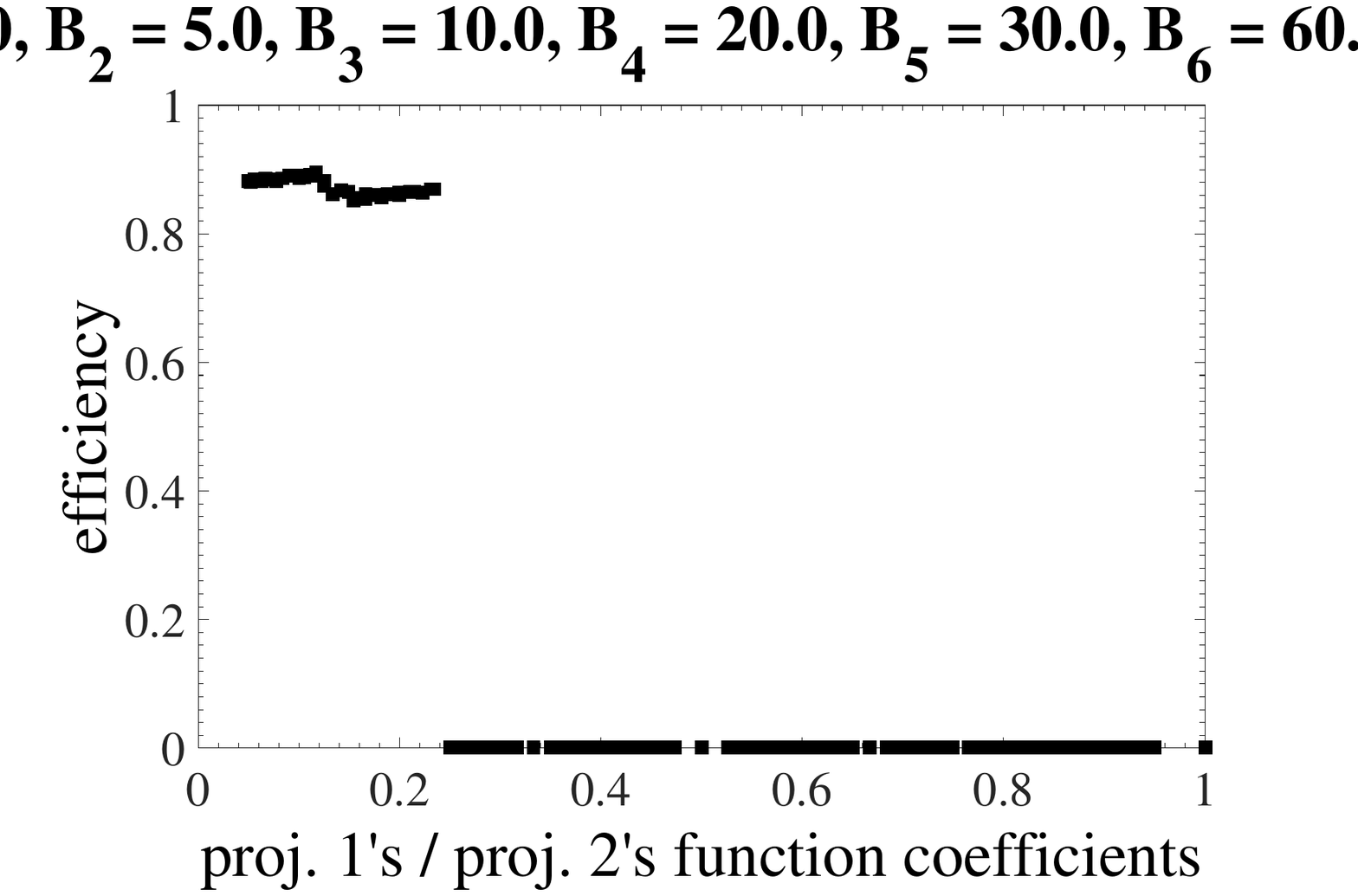}
\label{fig:plot17}}
\caption{The efficiency as a function of 
the ratio of the project value functions for $2, 3, 4, 5$ and $6$ players.
Efficiency of~$0$ means that Nash Equilibrium has not been found.
}%
\label{fig:NE_effic_proj_rat_2_3_4_5_6}%
\end{figure*}

When the project function coefficients
are the independent variables, \figref{fig:NE_exist_effic_proj_2_3_4_5_6} presents the \NE.
Each line of the simulation plots corresponds to a setting for a given number of players,
and within a line, the plots are generated for an increasing sequence of $\theta$.
Mostly, an \NE{} exists except for a cone where
the project functions are quite close to each other.
Interestingly, sometimes an \NE{} exists also when the project functions are
nearly the same (at ratio $1$), or at another constant ratio with each other. In all cases, the ratio of project functions
determines existence of an \NE, 
as Proposition~\ref{prop:inv_mult} implies.
Usually, the more players there are, the fewer settings
with an \NE{} we find.
%

\figref{fig:NE_effic_proj_rat_2_3_4_5_6} demonstrates
efficiency as a function of the ratio of the coefficients of the project functions.
This efficiency is uniquely determined by the ratio of the project
functions, complying with Proposition~\ref{prop:inv_mult}.
The dependency is piecewise linear and non-decreasing in each linear interval.
For two players, it is linear, in the spirit of Theorem~\ref{the:NE_effic_2players}.
(Though not directly predicted by it, since the theorem considers extremely
efficient or inefficient equilibria, and has several cases, which
imply piecewise linearity.)
The larger the $\theta$ is, the steeper the
piecewise linear dependency becomes.
%

\begin{figure*}[h!tbp]
\centering
\subfloat
{
\includegraphics[trim = 10mm 0mm 5mm 0mm, clip=true,width=0.35\textwidth,height=0.25\textwidth]{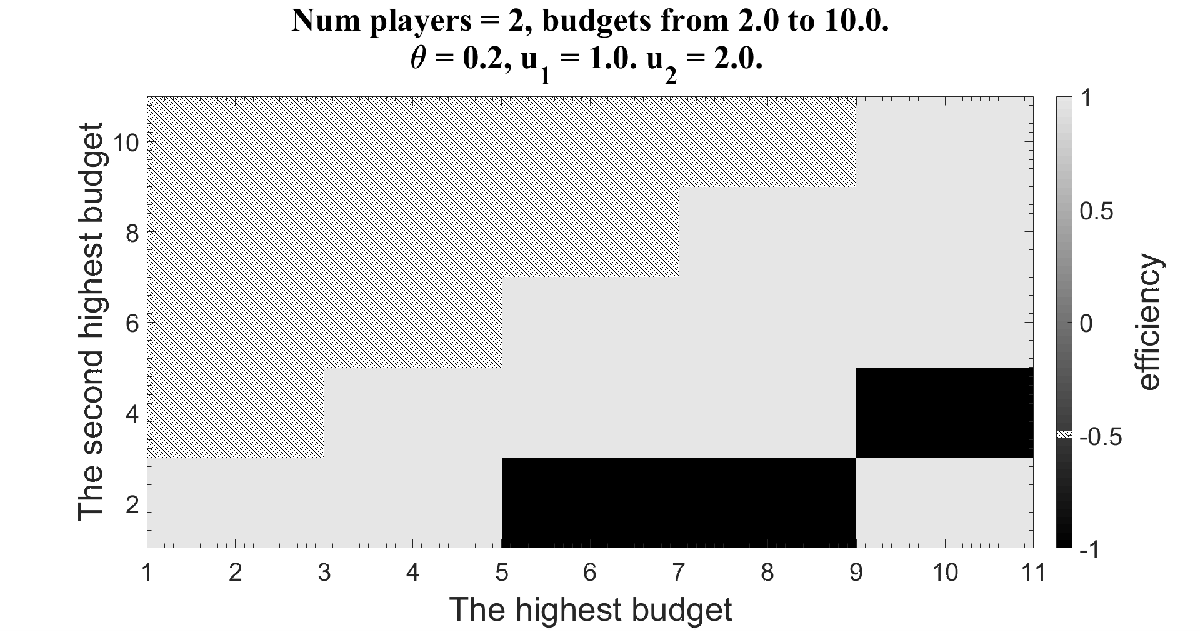}
}
\subfloat
{
\includegraphics[trim = 10mm 0mm 5mm 0mm, clip=true,width=0.35\textwidth,height=0.25\textwidth]{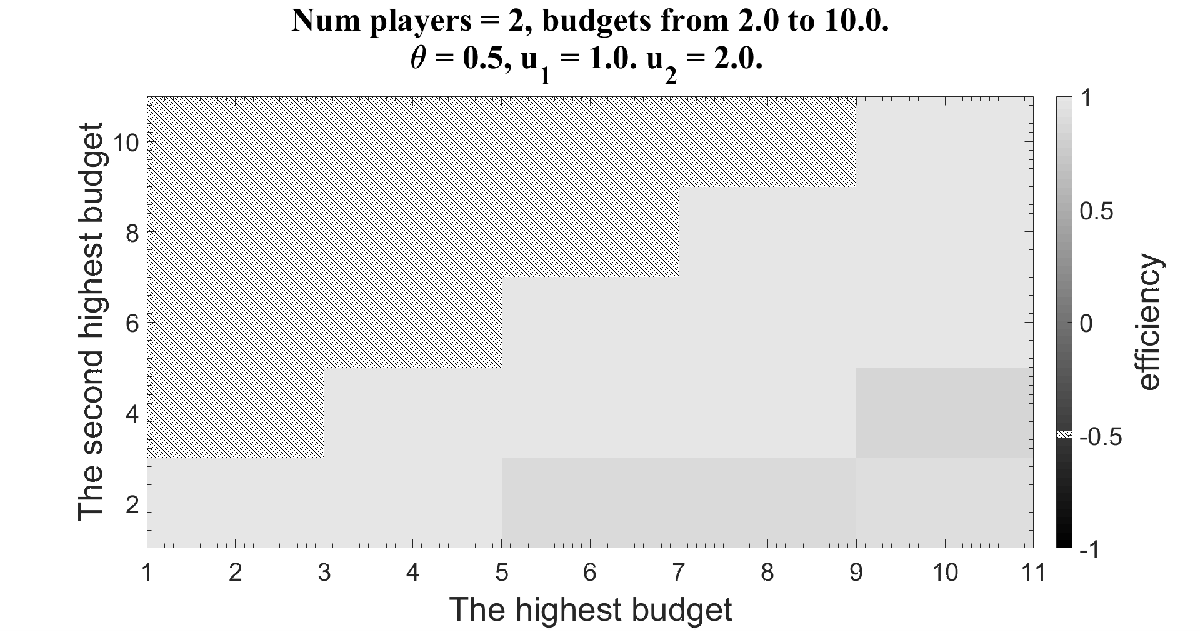}
}
\subfloat
{
\includegraphics[trim = 10mm 0mm 5mm 0mm, clip=true,width=0.35\textwidth,height=0.25\textwidth]{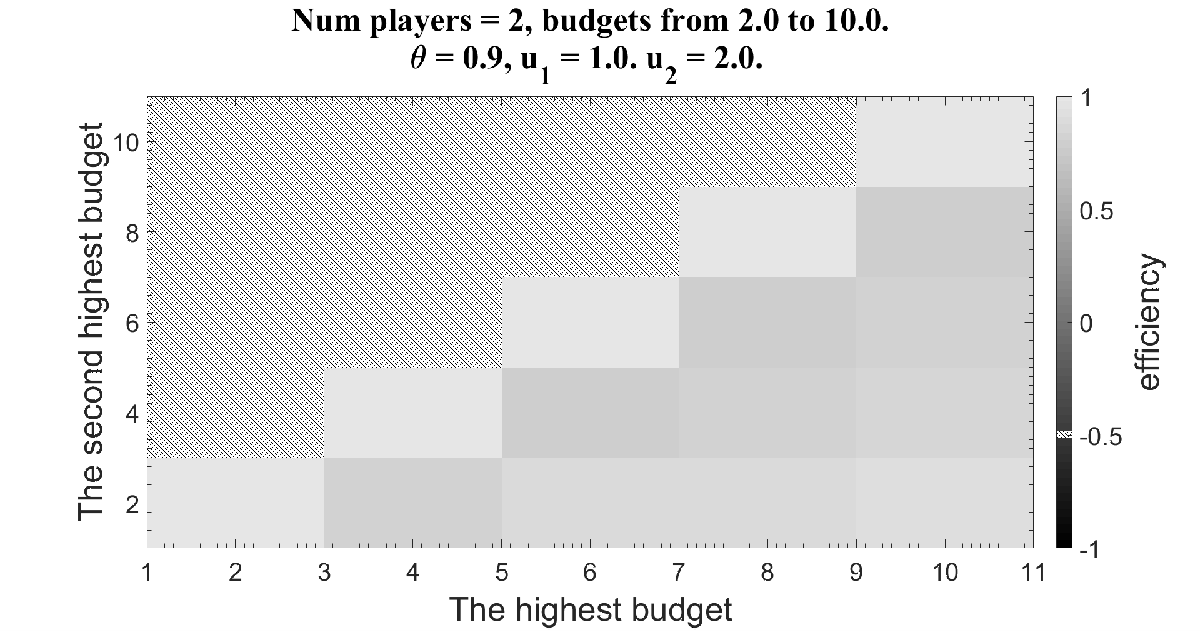}
}

\subfloat
{
\includegraphics[trim = 10mm 0mm 5mm 0mm, clip=true,width=0.35\textwidth,height=0.25\textwidth]{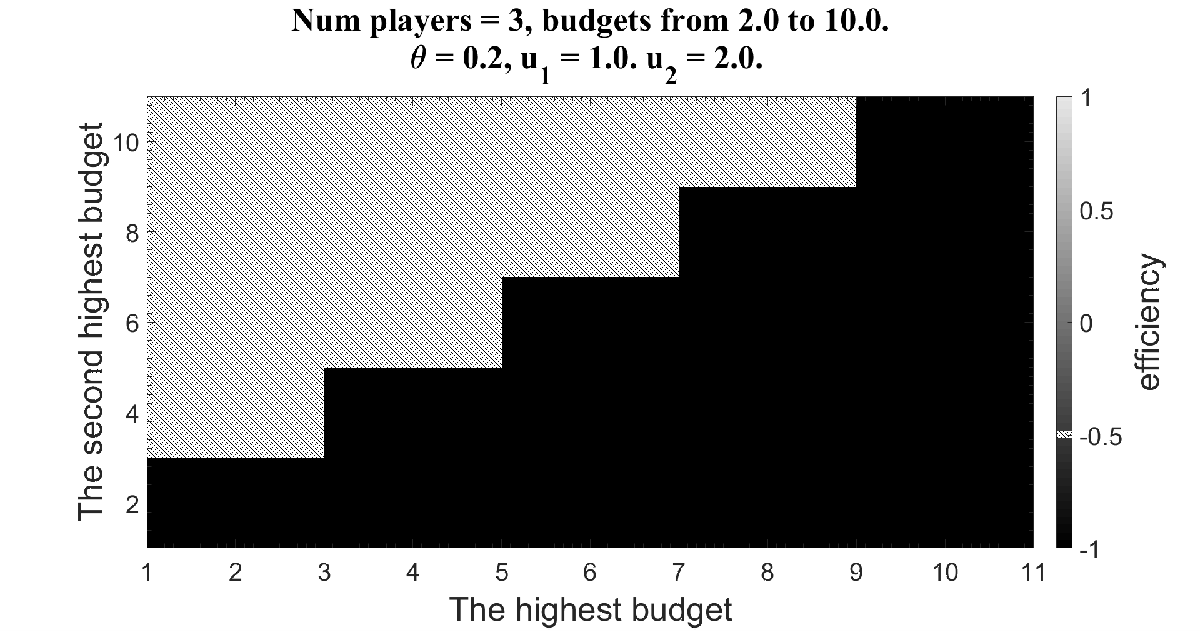}
}
\subfloat
{
\includegraphics[trim = 10mm 0mm 5mm 0mm, clip=true,width=0.35\textwidth,height=0.25\textwidth]{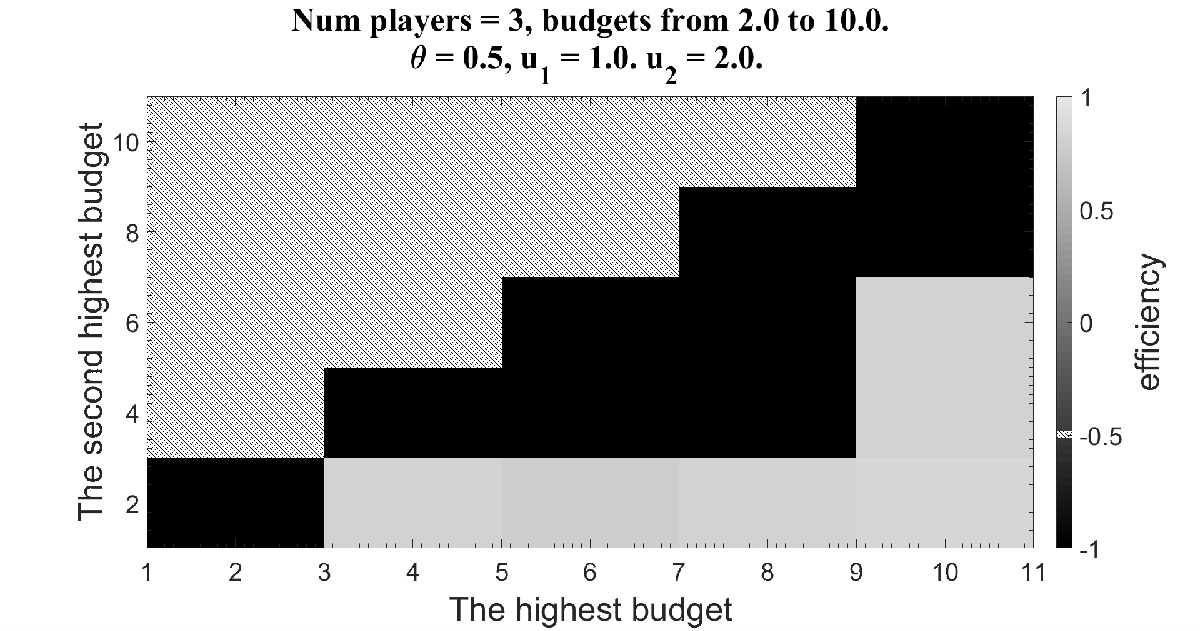}
}
\subfloat
{
\includegraphics[trim = 10mm 0mm 5mm 0mm, clip=true,width=0.35\textwidth,height=0.25\textwidth]{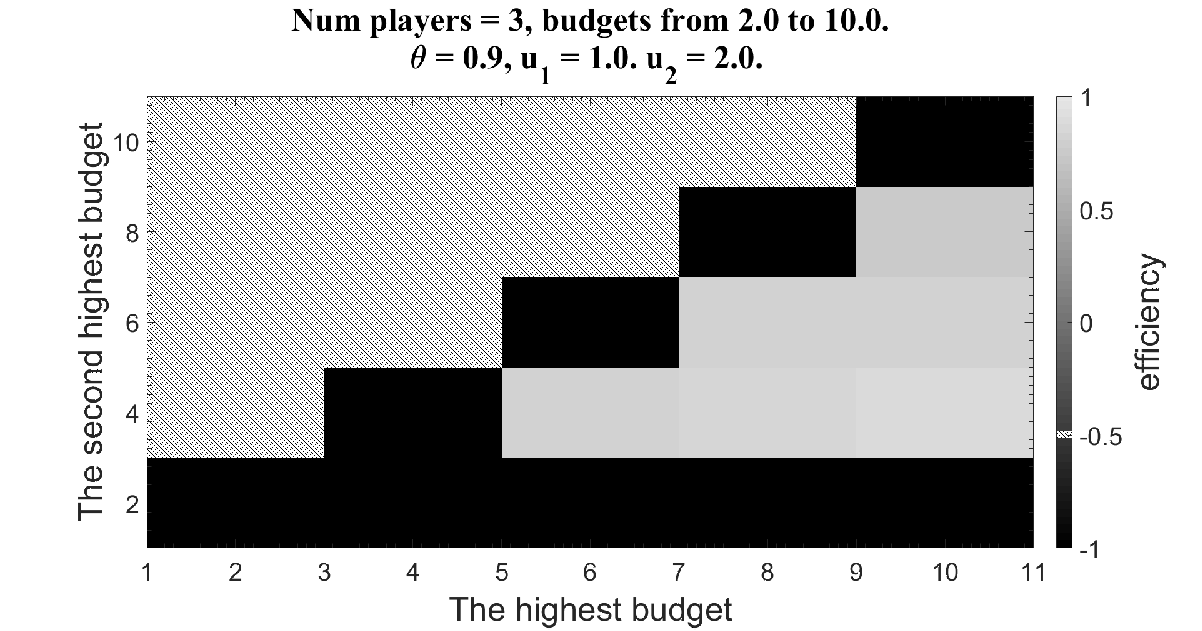}
}

\subfloat
{
\includegraphics[trim = 10mm 0mm 5mm 0mm, clip=true,width=0.35\textwidth,height=0.25\textwidth]{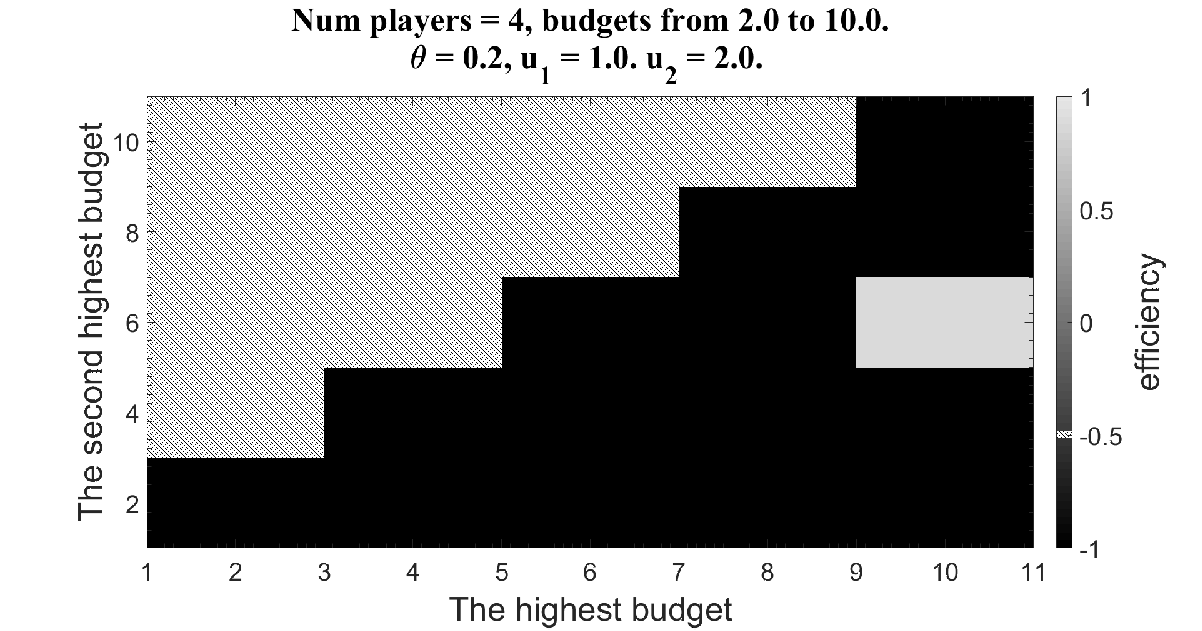}
}
\subfloat
{
\includegraphics[trim = 10mm 0mm 5mm 0mm, clip=true,width=0.35\textwidth,height=0.25\textwidth]{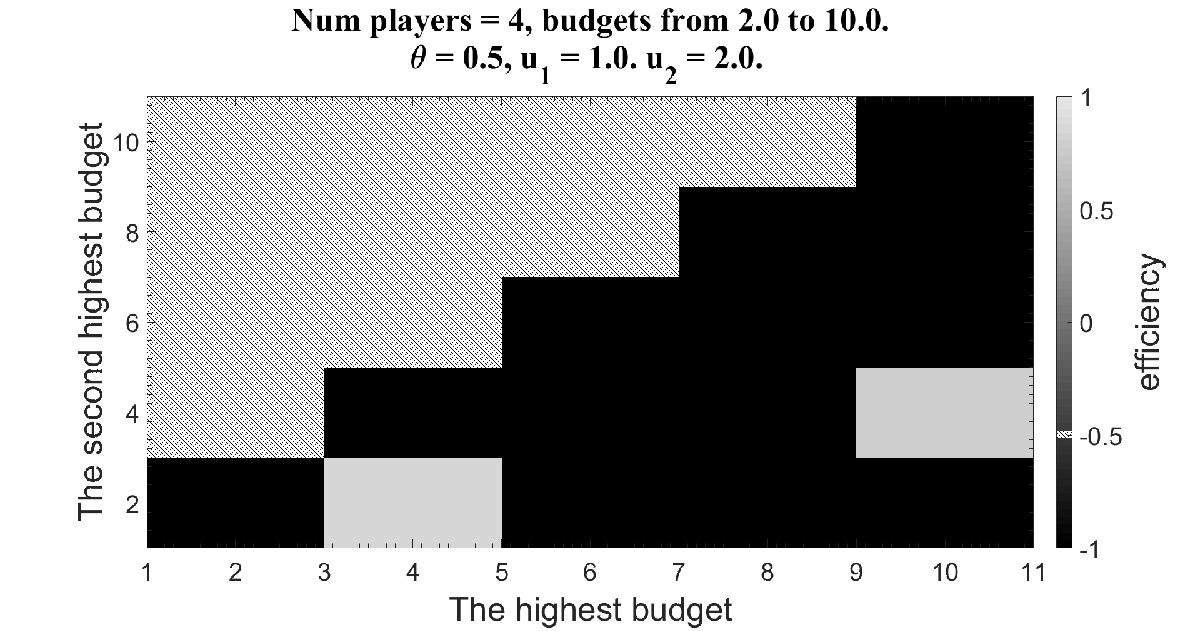}
}
\subfloat
{
\includegraphics[trim = 10mm 0mm 5mm 0mm, clip=true,width=0.35\textwidth,height=0.25\textwidth]{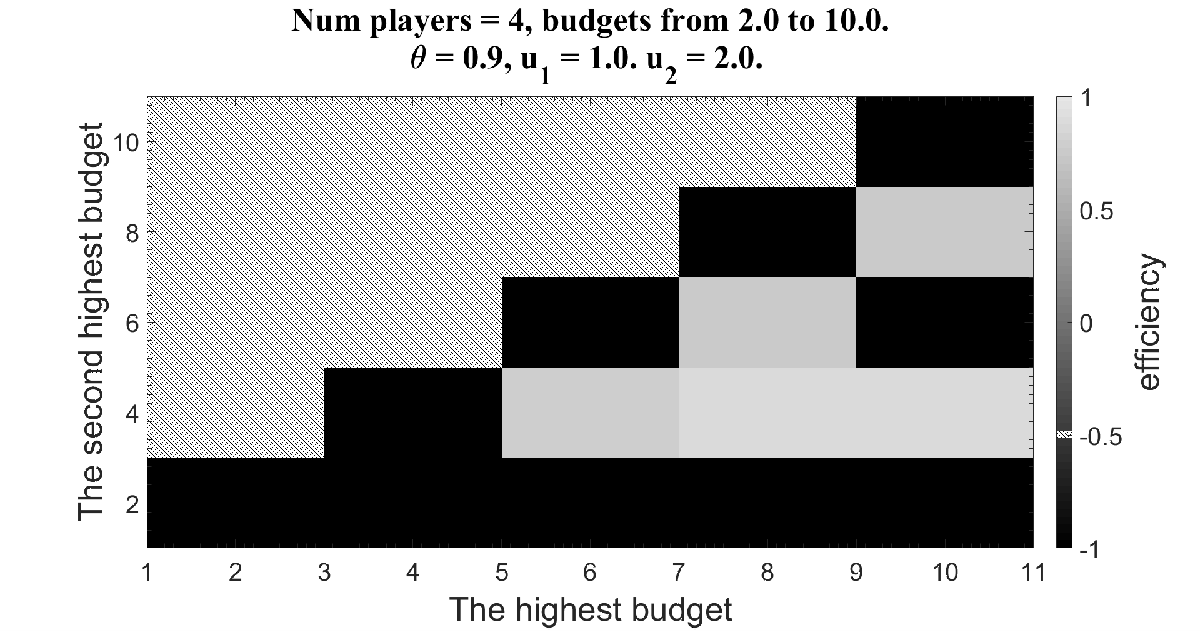}
}

\caption{The existence and efficiency of \NE{} as function of
the largest and the second largest budgets.
Black means that Nash Equilibrium has not been found, and
gray hatching indicates the non-defined area, since the second highest
budget may not be larger than the highest one.
}%
\label{fig:NE_exist_effic_budg}%
\end{figure*}

\begin{figure*}[h!tbp]
\centering
\subfloat
{
\includegraphics[trim = 10mm 5mm 15mm 10mm, clip=true,width=0.35\textwidth]{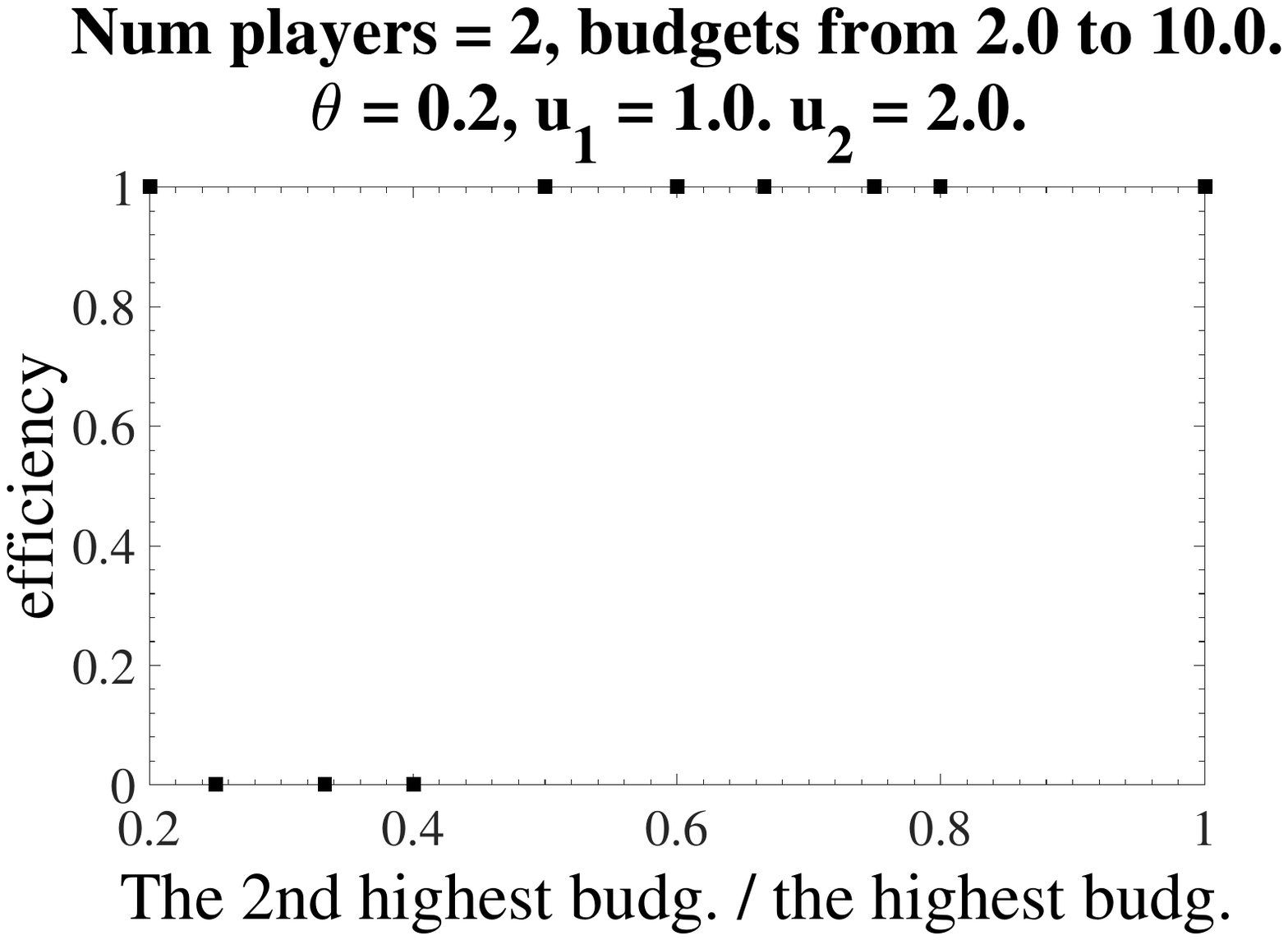}
}
\subfloat
{
\includegraphics[trim = 10mm 5mm 15mm 10mm, clip=true,width=0.35\textwidth]{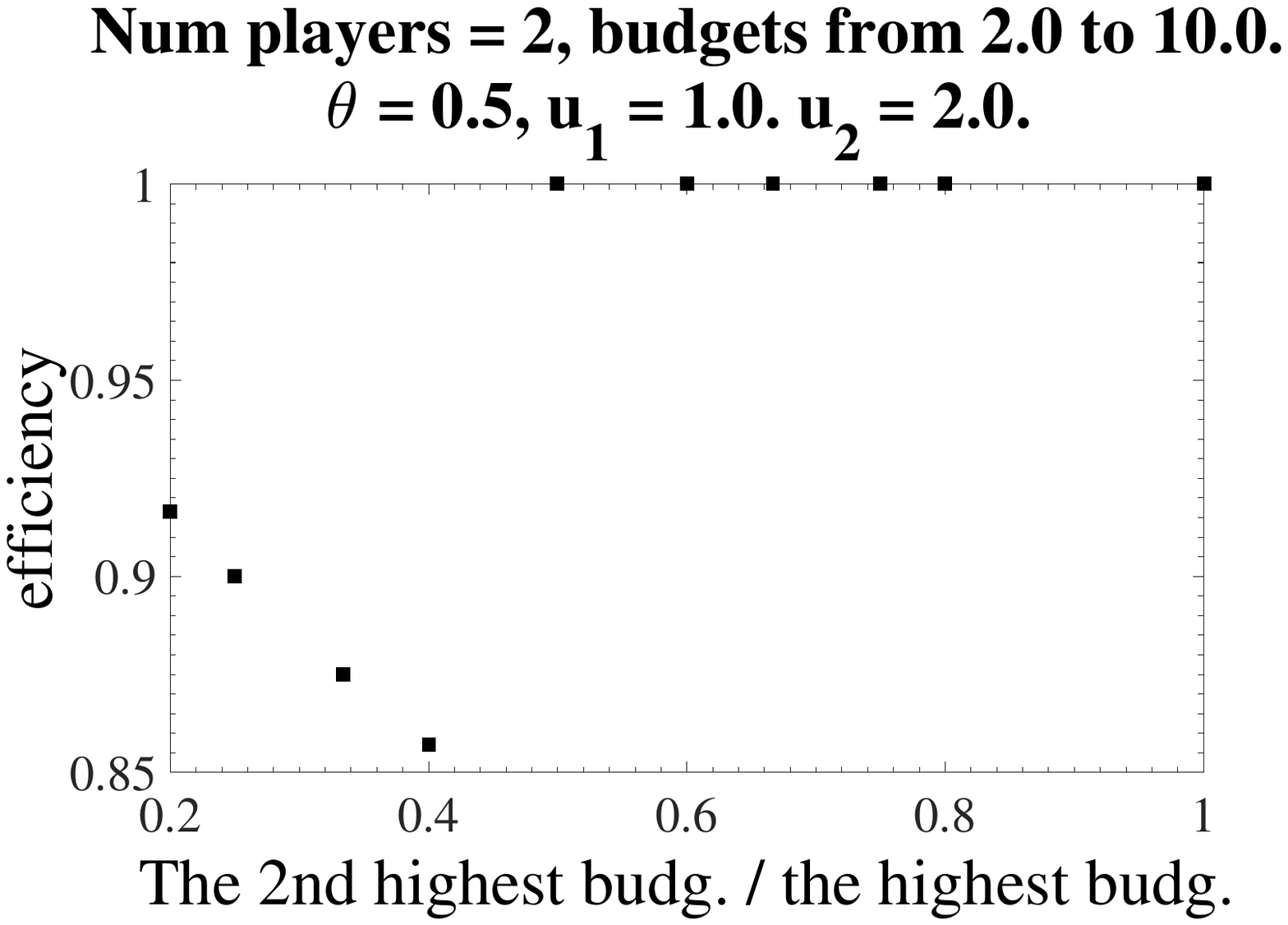}
}
\subfloat
{
\includegraphics[trim = 10mm 5mm 15mm 10mm, clip=true,width=0.35\textwidth]{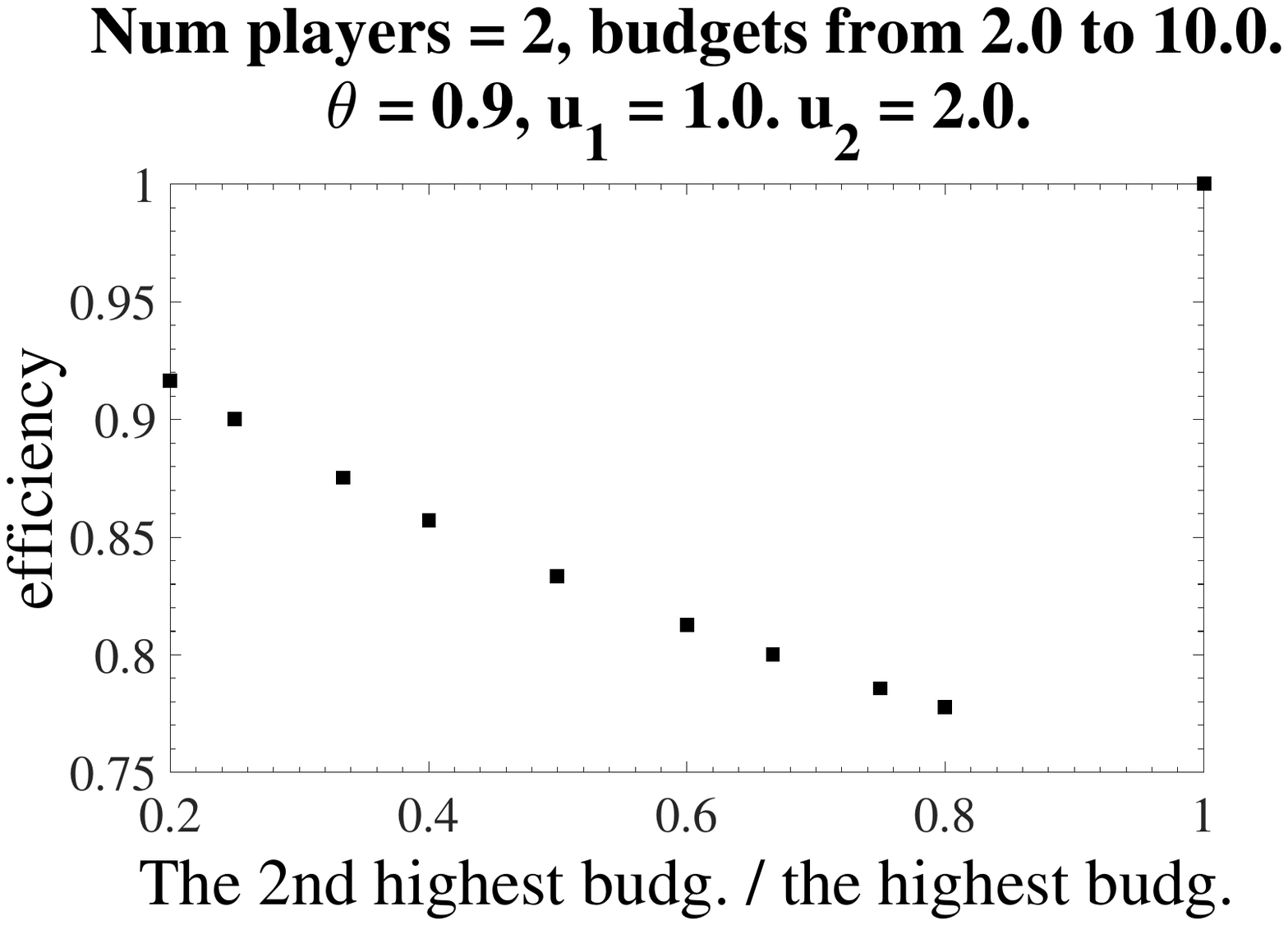}
}

\subfloat
{
\includegraphics[trim = 10mm 5mm 15mm 10mm, clip=true,width=0.35\textwidth]{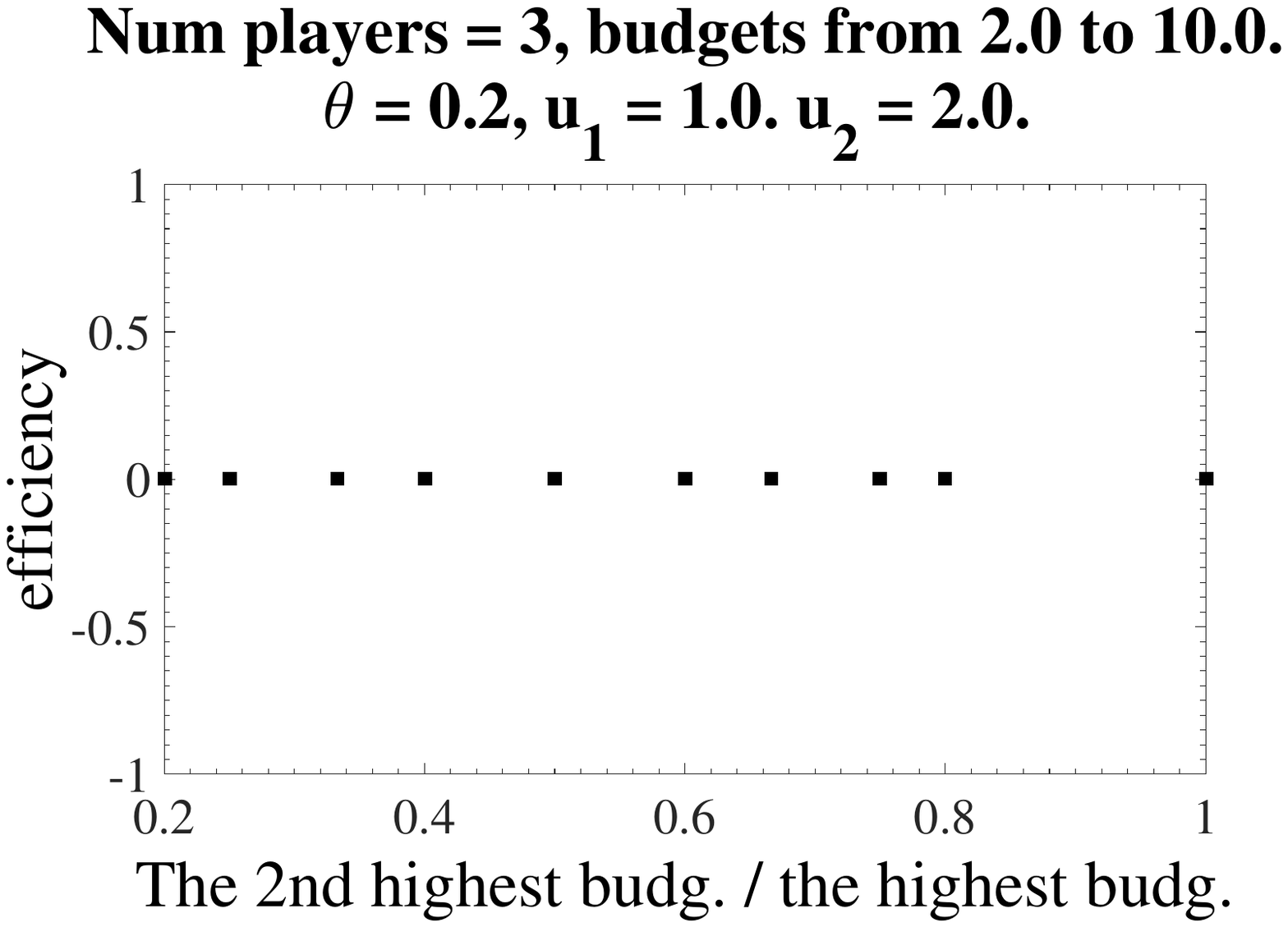}
}
\subfloat
{
\includegraphics[trim = 10mm 5mm 15mm 10mm, clip=true,width=0.35\textwidth]{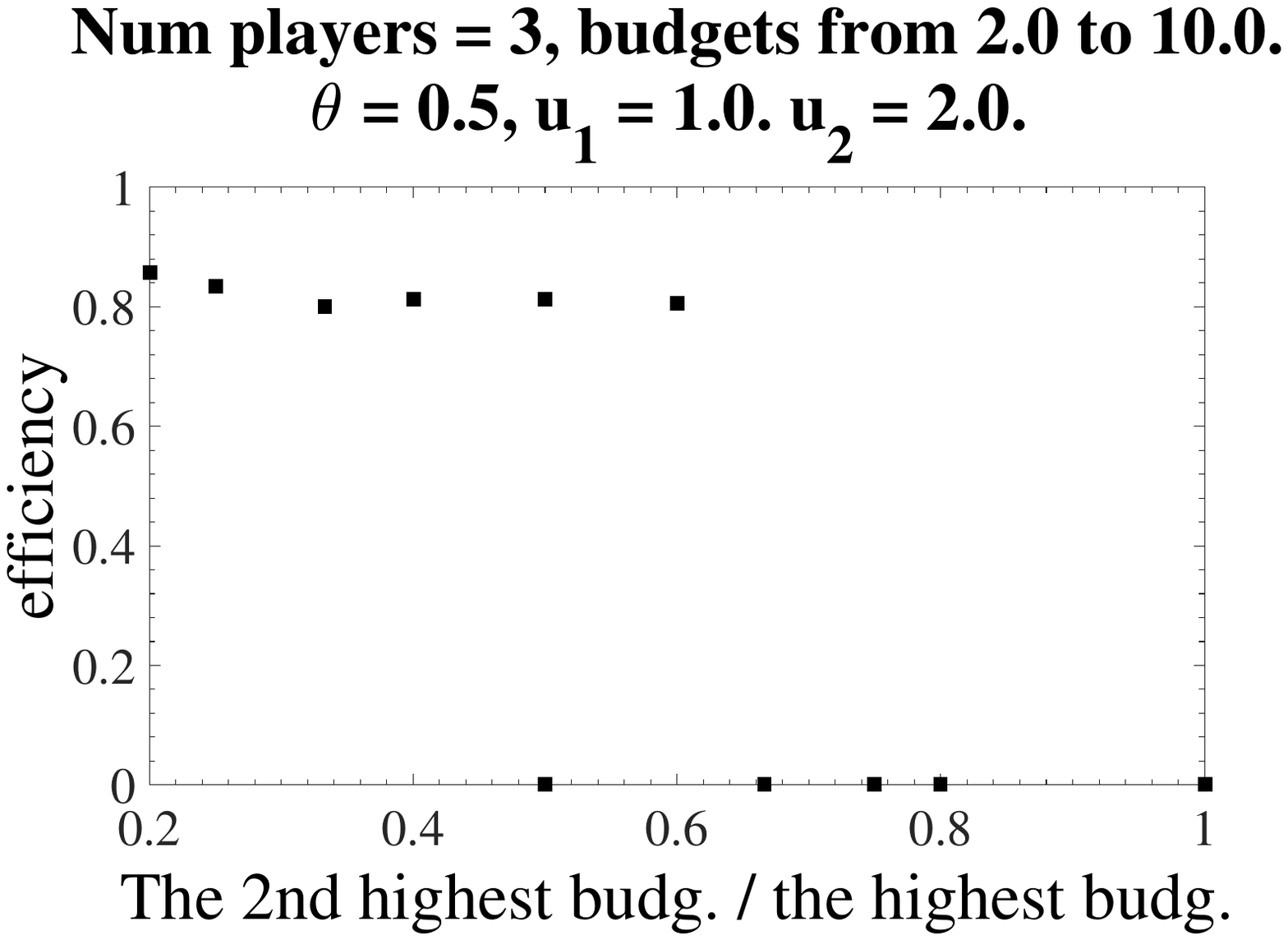}
}
\subfloat
{
\includegraphics[trim = 10mm 5mm 15mm 10mm, clip=true,width=0.35\textwidth]{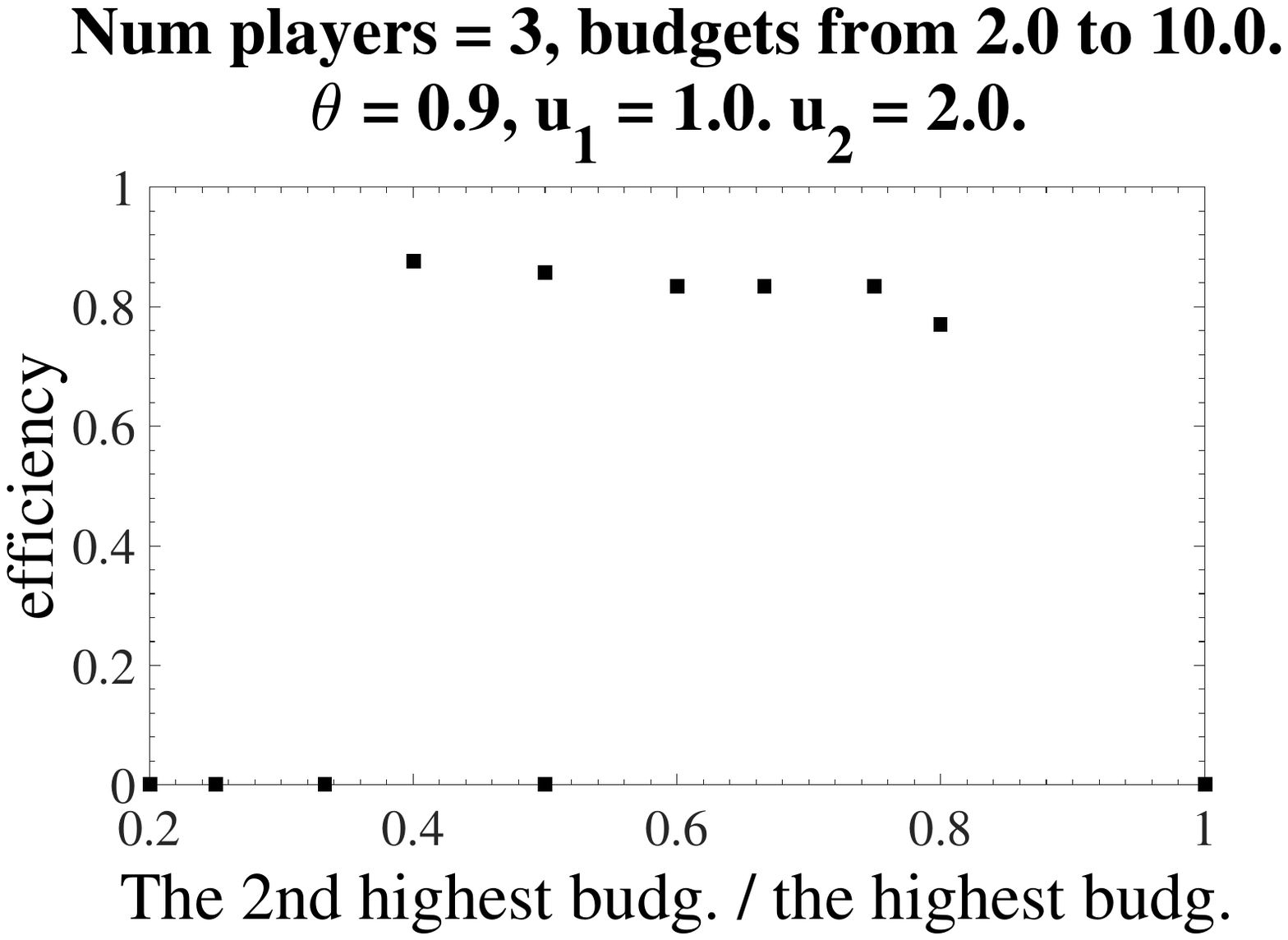}
}

\subfloat
{
\includegraphics[trim = 10mm 5mm 15mm 10mm, clip=true,width=0.35\textwidth]{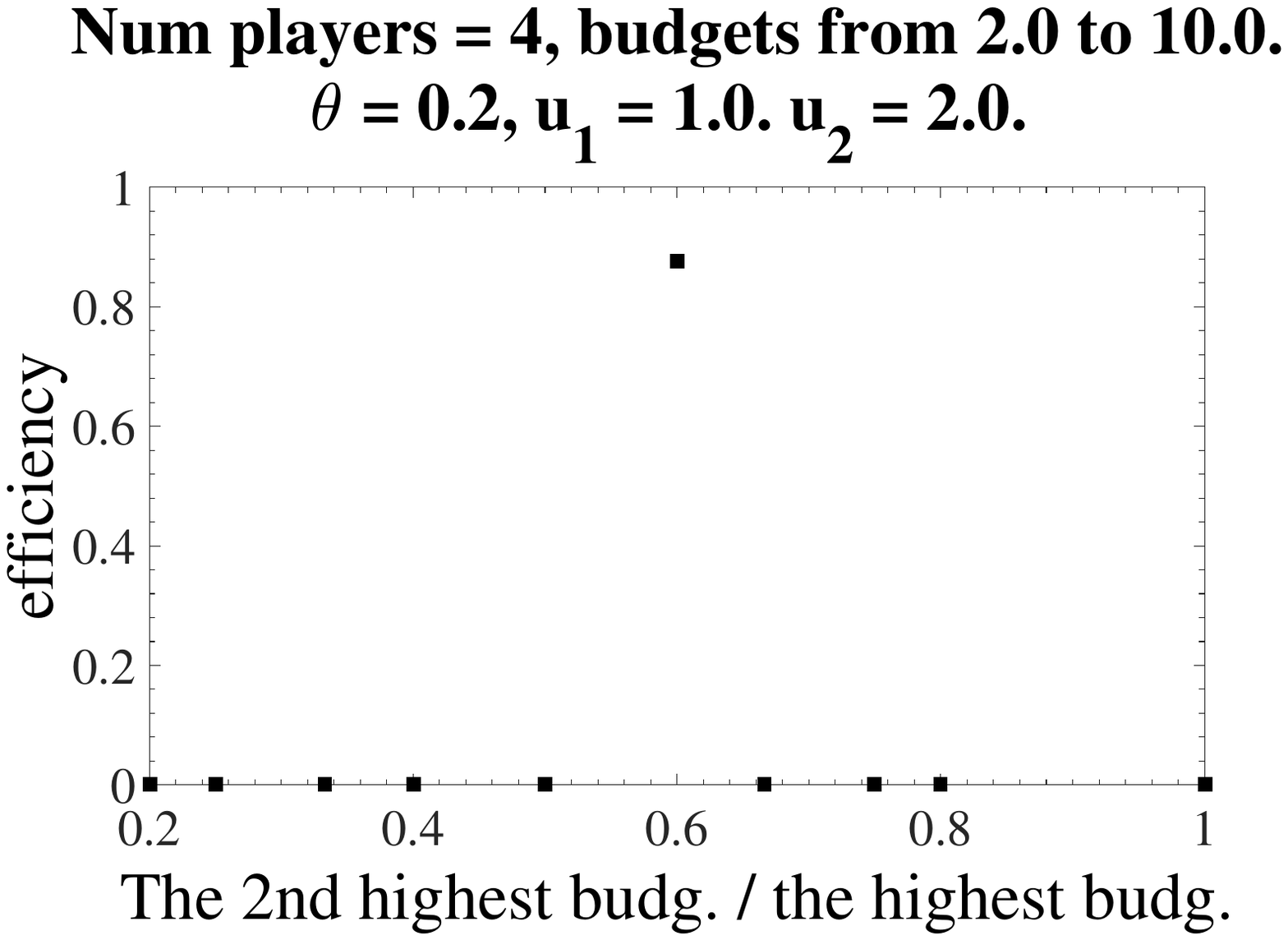}
}
\subfloat
{
\includegraphics[trim = 10mm 5mm 15mm 10mm, clip=true,width=0.35\textwidth]{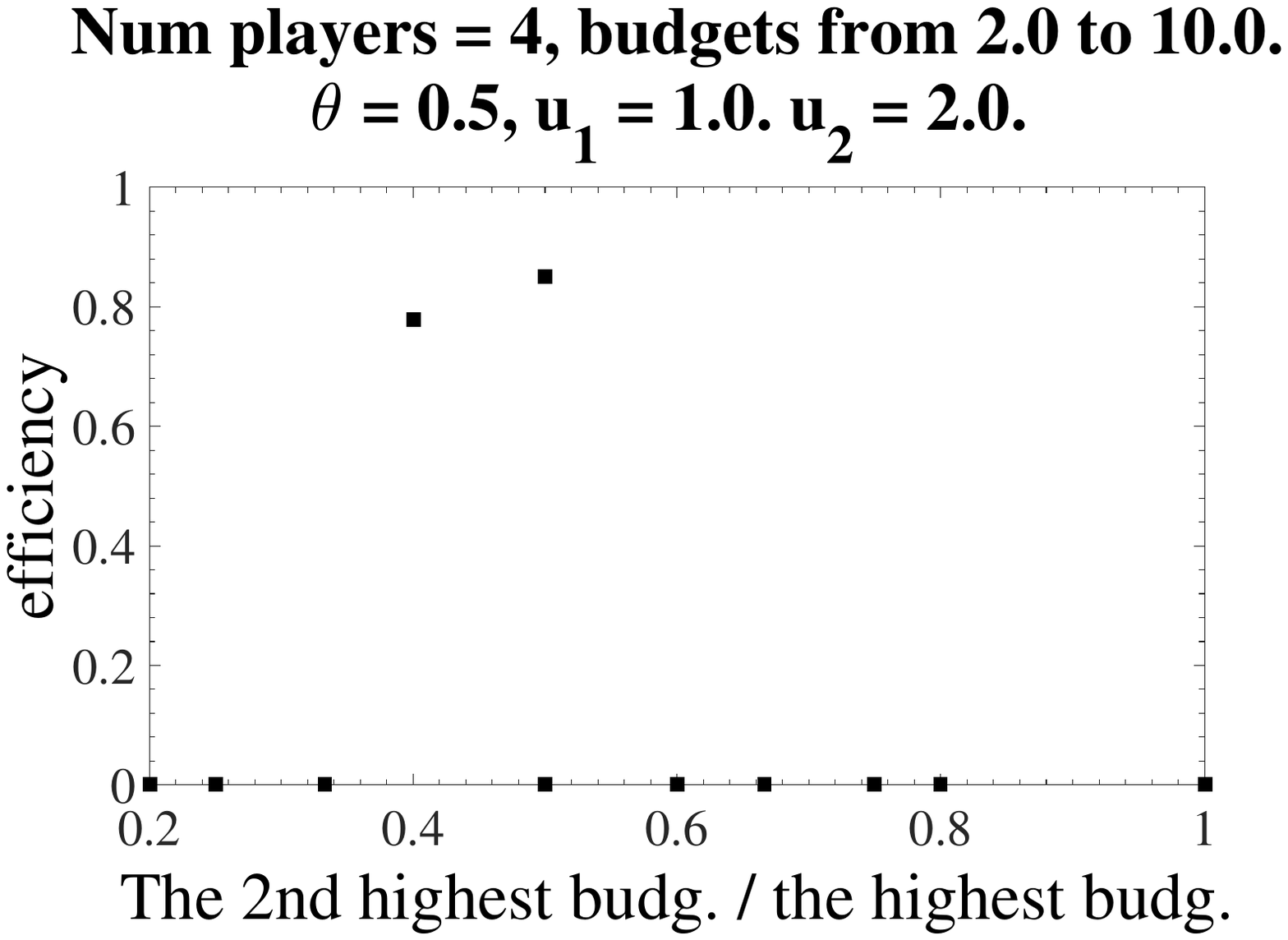}
}
\subfloat
{
\includegraphics[trim = 10mm 5mm 15mm 10mm, clip=true,width=0.35\textwidth]{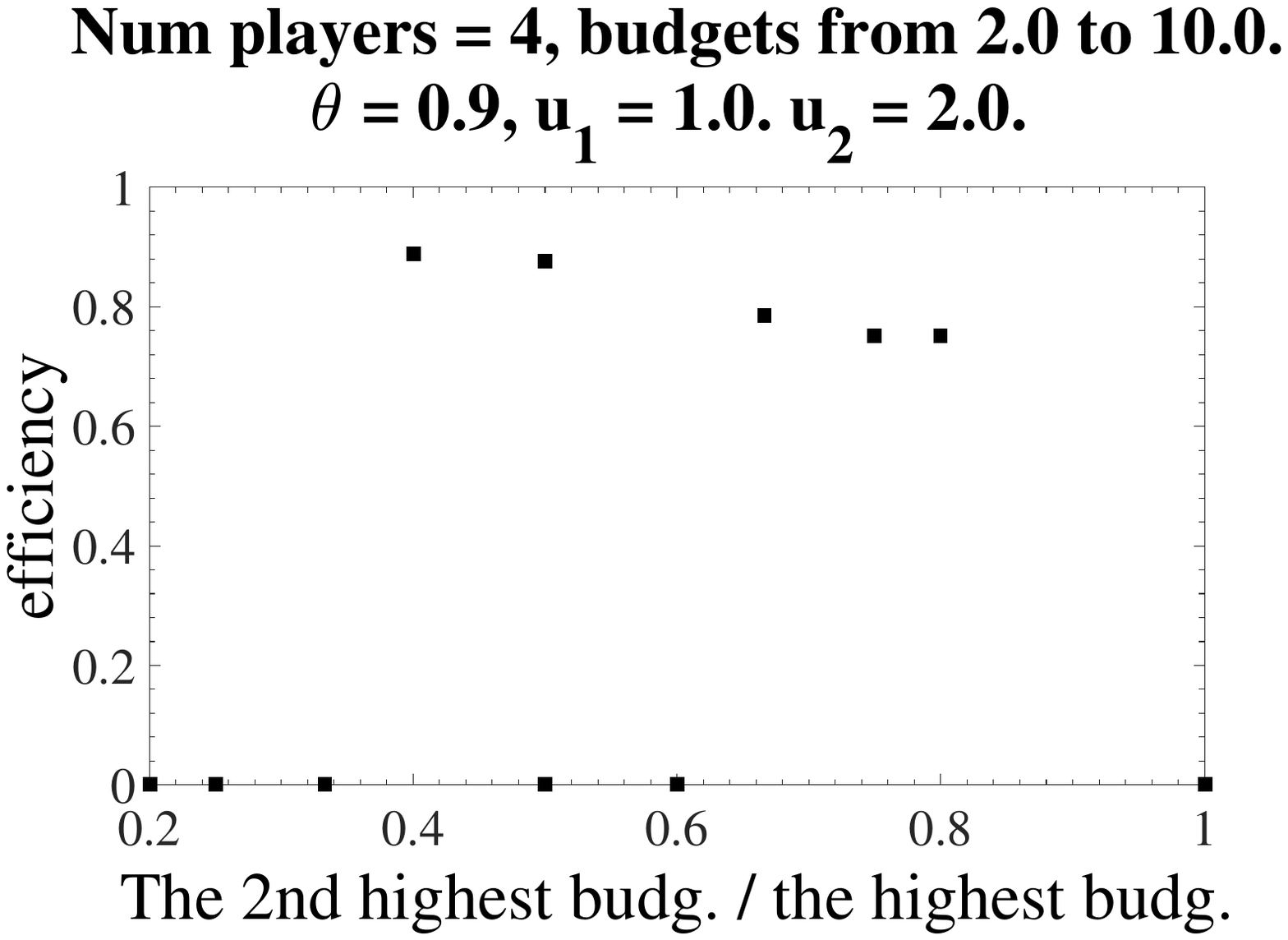}
}

\caption{The efficiency as a function of the ratio of the two largest budgets.
Efficiency of~$0$ means that Nash Equilibrium has not been found. 
}%
\label{fig:NE_effic_budg_rat}%
\end{figure*}

When the largest and the second largest budgets
are the independent variables, \figref{fig:NE_exist_effic_budg} presents the equilibria.
(The budgets of the other players are spread on equal intervals).
Each line of the plot corresponds to a setting of
project function coefficients for a given number of players, and
within a line, the plots are generated for an increasing sequence of $\theta$.
The more players we have, the fewer \NE{} we typically find.
%
%
\figref{fig:NE_effic_budg_rat} plots efficiency as a function of the ratio of the two largest budgets.
To summarise, the existence of \NE{} is related to the ratio of
project function coefficients
and the budget ratio being in some limits, limits that in particular
depend on the threshold. This is in the spirit of 
Theorem~\ref{the:one_threshold_char},
Proposition~\ref{prop:inv_mult} and
Theorems~\ref{the:NE_charac_2players} and~\ref{the:NE_charac_nplayers}.
Based on Theorem~\ref{the:one_threshold_char}, Proposition~\ref{prop:inv_mult},
Theorem~\ref{the:NE_charac_2players},
Corollary~\ref{cor:NE_exist_proj_rat_smaller_2players}, and the simulation
results,
we hypothesise that also for multiple projects and players, an \NE{} exists if and only if at least one of several
sets of conditions on the ratios of the budgets holds, and for every such
set, several conditions of being smaller or equal or exactly equal to a function of
the threshold (and not the budgets) on the ratios of project function
coefficients hold together. 

The efficiency of the \NE{} that we find depends on the ratio of
the project function coefficients and on the ratio of the budgets,
rather than on the project functions and budgets themselves,
in the spirit of Proposition~\ref{prop:one_threshold_eff},
Proposition~\ref{prop:inv_mult}. 
%
Based on Proposition~\ref{prop:one_threshold_eff}, Proposition~\ref{prop:inv_mult},
Theorem~\ref{the:NE_effic_2players} and the simulation results,
we hypothesise that also for multiple projects and players, the price of anarchy and stability of
a shared effort game depends piecewise linearly on the project function
coefficients ratios, 
and within each linear domain, this dependency is non-decreasing
with $\theta$.